\documentclass[aps,prd,amsmath,floats,floatfix,twocolumn,
  superscriptaddress,nofootinbib,showpacs]{revtex4-1}

\pdfoutput=1 
\usepackage{amsfonts}
\usepackage{amsmath}
\usepackage{amssymb}
\usepackage{amsthm}
\usepackage{array}
\usepackage{bm}
\usepackage{braket}
\usepackage{dcolumn}
\usepackage{diagbox}
\usepackage{epsfig}
\usepackage{float}
\usepackage{graphicx}
\usepackage{graphics}
\usepackage[latin1]{inputenc}
\usepackage{latexsym}
\usepackage{rotating}
\usepackage[dvipsnames]{xcolor}
\usepackage{mathrsfs}
\usepackage{mathtools}
\usepackage{microtype}
\usepackage{multirow}
\usepackage{siunitx}
\usepackage{url}
\usepackage{verbatim}
\usepackage{xspace}
\usepackage{yfonts}

\usepackage[caption=false]{subfig}
\usepackage[normalem]{ulem}
\usepackage[breaklinks=true]{hyperref}
\usepackage[toc,page]{appendix}
\hypersetup{
    colorlinks=true,
    linkcolor=NavyBlue,
    filecolor=Magenta,
    urlcolor=NavyBlue,
    citecolor=NavyBlue
}

\makeatletter
\newcommand*{\rom}[1]{\expandafter\@slowromancap\romannumeral #1@}
\newcommand{\Even}{{\mbox{\tiny E}}}
\newcommand{\Odd}{{\mbox{\tiny O}}}
\newcommand{\Hertz}{{\mbox{\tiny H}}}

\newcommand{\dCS}{{\mbox{\tiny dCS}}}

\newcommand{\GR}{{\mbox{\tiny GR}}}
\newcommand{\bGR}{{\mbox{\tiny bGR}}}

\newcommand{\geo}{{\mbox{\tiny geo}}}
\newcommand{\typeD}{{\mbox{\tiny D}}}
\newcommand{\nonD}{{\mbox{\tiny non-D}}}
\newcommand{\EdGB}{{\mbox{\tiny EdGB}}}

\newcommand{\matter}{{\mbox{\tiny matter}}}
\newcommand{\field}{{\mbox{\tiny field}}}

\makeatother
\allowdisplaybreaks

\newcommand{\CAL}{Theoretical Astrophysics 350-17, California Institute of Technology, Pasadena, CA 91125, USA}
\newcommand{\UIUC}{Illinois  Center  for  Advanced  Studies  of  the  Universe \& Department of Physics, University of Illinois at Urbana-Champaign, Urbana, Illinois 61801, USA}
\newcommand{\UTexas}{Weinberg Institute, University of Texas at Austin, Austin, TX 78712, USA}

\begin{document}
    \title{Isospectrality breaking in the Teukolsky formalism}
    \author{Dongjun Li}
    \email{dlli@caltech.edu}
    \affiliation{\CAL}
	
    \author{Asad Hussain}
    \email{asadh@utexas.edu}
    \affiliation{\UTexas}
	
    \author{Pratik Wagle}
    \email{wagle2@illinois.edu}
    \affiliation{\UIUC}

    \author{Yanbei Chen}
    \affiliation{\CAL}

    \author{Nicol\'as Yunes}
    \affiliation{\UIUC}
    
    \author{Aaron Zimmerman}
    \affiliation{\UTexas}
	
    \date{\today}
	
    \begin{abstract}
    General relativity, though the most successful theory of gravity, has been continuously modified to resolve its incompatibility with quantum mechanics and explain the origin of dark energy or dark matter. One way to test these modified gravity theories is to study the gravitational waves emitted during the ringdown of binary mergers, which consist of quasinormal modes. In several modified gravity theories, the even- and odd-parity gravitational perturbations of non-rotating and slowly rotating black holes have different quasinormal mode frequencies, breaking the isospectrality of general relativity. For black holes with arbitrary spin in modified gravity, there were no avenues to compute quasinormal modes except numerical relativity, until recent extensions of the Teukolsky formalism. In this work, we describe how to use the modified Teukolsky formalism to study isospectrality breaking in modified gravity. We first introduce how definite-parity modes are defined through combinations of Weyl scalars in general relativity, and then, we extend this definition to modified gravity. We then use the eigenvalue perturbation method to show how the degeneracy in quasinormal mode frequencies of different parity is broken in modified gravity. To demonstrate our analysis, we also apply it to some specific modified gravity theories. Our work lays the foundation for studying isospectrality breaking of quasinormal modes in modified gravity for black holes with arbitrary spin.  
    \end{abstract}
    \maketitle
	
    \section{Introduction} \label{sec:introduction}
    
    The development of current and next-generation gravitational wave (GW) detectors allows us for the first time to study the extreme gravitational events that emit these waves and use them to test theories of gravity. 
    General relativity (GR), as one of the most successful gravity theories, has been widely tested \cite{Will2014}, but its incompatibility with quantum mechanics motivated the development of new theories of quantum gravity, such as string theory \cite{Damour:1994zq, Lozano:2021xxs, Nojiri:2010wj, Padmanabhan:2002ji} and loop quantum gravity \cite{Aharony:1999ti, Birrell:1982ix, Ashtekar:1997yu, Ashtekar:2004eh}. 
    Furthermore, to resolve observational anomalies, such as the asymmetry of matter and antimatter abundance in our universe \cite{Canetti:2012zc}, one can also modify the theory of gravity. Among these modified gravity theories, parity-violating ones have attracted great attention. One major subset of these can be understood as effective field theory (EFT) extensions of GR in Lorentzian geometry, such as dynamical Chern-Simons (dCS) gravity \cite{Jackiw:2003pm, Alexander:2009tp}, parity-violating ghost-free scalar-tensor gravity \cite{Crisostomi:2017ugk, Nishizawa:2018srh, Zhao:2019xmm}, certain versions of Horava-Lifshitz gravity \cite{Horava:2009uw, Zhu:2013fja}, and parity-violating corrections in higher-derivative gravity without extra fields \cite{Burgess_2004, Donoghue_2012, Endlich:2017tqa, Cano:2019ore}. Another subset is built instead on non-Riemannian geometry \cite{Qiao:2022mln}, such as parity-violating symmetric teleparallel gravity \cite{Conroy:2019ibo}.
	
    For most of these parity-breaking theories, the action itself is invariant under parity transformation, such as in dCS gravity. Parity is broken instead in these theories when astrophysical systems have a preferred axis of symmetry \cite{Cardoso:2009pk, Wagle_Yunes_Silva_2021, Loutrel:2022tbk} or cosmologically, when additional degrees of freedom acquire a non-zero vacuum expectation value \cite{Garcia-Bellido:2003wva, Alexander:2004xd, Alexander:2004us}. In the former case, axisymmetric geometries, such as those describing spinning black hole (BH) spacetimes, are modified. For example, unlike in GR, the odd-parity multiple moments (odd mass multipole moments and even current multipoles) of rotating BHs can be nonzero in some of these parity-breaking theories, so the equatorial symmetry of Kerr in GR is broken \cite{Endlich:2017tqa, Cardoso:2018ptl, Bena:2020uup, Bah:2021jno, Fransen:2022jtw, Cano:2022wwo}, which may be detectable via, for example, extreme mass-ratio inspirals (EMRIs) \cite{Fransen:2022jtw}. 
	
    Gravitational perturbations of these stationary geometries are also affected. Years ago, it was observed that the amplitude of left-circular or right-circular polarized GWs propagating in these parity-breaking theories decreases or increases with propagation, resulting in amplitude birefringence \cite{Alexander:2017jmt, Qiao:2019wsh, Zhao:2019xmm, Li:2022grj, Qiao:2022mln}. These left-circular and right-circular polarized modes can also propagate with different velocities, causing velocity birefringence \cite{Nishizawa:2018srh, Qiao:2019wsh, Zhao:2019xmm, Boudet:2022nub}. Both amplitude and velocity birefringences can be detected, in principle, with LIGO \cite{Yunes:2010yf, Alexander:2017jmt, Jenks:2023pmk}. These birefringence effects might also leave imprints at a larger scale, for example, generating chiral primordial GWs, which directly affect the cosmic wave background radiation \cite{Lue:1998mq, Alexander:2004wk, Contaldi:2008yz, Takahashi:2009wc, Yoshida:2017cjl, Bartolo:2018elp}, or circularly-polarized stochastic GW background, which can be detected by GW detectors \cite{Seto:2007tn, Orlando:2020oko, Martinovic:2021hzy}. 
	
    Besides propagation effects, gravitational perturbations of BHs in modified gravity can also have parity asymmetry during generation. One important feature of GWs emitted during the ringdown phase of binary BH mergers in GR, or quasinormal modes (QNMs), is that the modes with the same quantum number, but different parity, have the same frequency \cite{Chandrasekhar_1983, Berti:2009kk}, a result known as isospectrality. However, in parity-violating theories, isospectrality is generally broken, similar to the breaking of degeneracies in quantum mechanical perturbation theory. For example, in dCS gravity, it has been found that only odd-parity modes are modified for non-rotating BHs \cite{Cardoso:2009pk, Molina:2010fb, Pani_Cardoso_Gualtieri_2011}. For spinning BHs, both parities are modified but in different ways \cite{Wagle_Yunes_Silva_2021, Srivastava_Chen_Shankaranarayanan_2021}. Similar isospectrality breaking of QNMs has been observed in parity-violating corrections of higher-derivative gravity \cite{Cardoso_Kimura_Maselli_Senatore_2018, Cano:2021myl, Cano:2023tmv, Cano:2023jbk} and, more interestingly, in certain parity-preserving theories, such as parity-preserving corrections of higher-derivative gravity \cite{Cardoso_Kimura_Maselli_Senatore_2018, deRham_Francfort_Zhang_2020, Cano:2021myl, Cano:2023tmv, Cano:2023jbk} and EdGB theory\cite{Pani:2009wy, Blazquez-Salcedo:2016enn, Blazquez-Salcedo_Khoo_Kunz_2017, Pierini:2021jxd, Pierini:2022eim}. Such parity asymmetry in the generation of GWs may cause observable effects, depending on whether there is enough signal-to-noise ratio (SNR) to resolve the shifts to the ringdown frequencies (both real and imaginary parts) of the resultant modes \cite{Maximiliano:2021}.
	
    In this work, we focus on the isospectrality breaking of QNMs in these EFT extensions of GR. The study of QNMs has been an important topic in GR and modified gravity because their spectrum allows us to retrieve information on the exterior geometry of BHs and the dynamics of modified gravity theories, which is the idea of BH spectroscopy \cite{Dreyer:2003bv, Berti:2005ys, Berti:2018vdi}. 
    For these beyond GR (bGR) theories, extra non-metric fields (scalar, vector, or tensor) leave imprints on the QNMs. To study QNMs, one major approach is BH perturbation theory, where the gravitational perturbations of an isolated stationary BH are computed, given that the merger of binary BHs always settles down to a stationary geometry in GR. For non-rotating BHs, thanks to spherical symmetry, QNMs can be directly computed from metric perturbations in both GR \cite{Regge:PhysRev.108.1063, Zerilli:1971wd, Moncrief:1974am, Vishveshwara:1970cc, Vishveshwara:1970zz} and modified gravity \cite{Cardoso:2009pk, Molina:2010fb, Pani_Cardoso_Gualtieri_2011, Pani:2009wy, Blazquez-Salcedo:2016enn, Blazquez-Salcedo_Khoo_Kunz_2017, Cardoso_Kimura_Maselli_Senatore_2018, deRham_Francfort_Zhang_2020, Cardoso_Kimura_Maselli_Berti_Macedo_McManus_2019}. In this case, metric perturbations are separated into two pieces, one with even and one with odd parity. For each parity piece, one can find a single gauge-invariant function that characterizes all degrees of freedom, i.e., the Zerilli-Moncrief (ZM) function for even-parity perturbations and the Regge-Wheeler (RW) function for odd-parity ones, the governing equations of which are decoupled and separable. Since each of the metric perturbation functions has a definite parity, one can easily study isospectrality breaking in this approach.
	
    For rotating BHs, due to the lack of spherical symmetry, it is hard to decouple all the metric fields and find only two functions to represent all the metric components. For this reason, Teukolsky developed another approach for rotating BHs in GR \cite{Teukolsky:1973ha, Press:1973zz, Teukolsky:1974yv} within the framework of Newman and Penrose (NP) \cite{Newman-Penrose} and using spinor calculus. In the Teukolsky formalism, instead of solving for metric perturbations directly, curvature perturbations, characterized by the Weyl scalars $\Psi_0$ and $\Psi_4$, are solved for first, from which the metric can then be reconstructed \cite{Cohen_Kegeles_1975, Chrzanowski:1975wv, Kegeles_Cohen_1979, Lousto_Whiting_2002, Ori_2003, Whiting_Price_2005, Yunes_Gonzalez_2006, Keidl_Friedman_Wiseman_2007, Keidl_Shah_Friedman_Kim_Price_2010, Chandrasekhar_1983, Loutrel_Ripley_Giorgi_Pretorius_2020}. Both non-rotating and rotating BHs in GR can be mathematically classified as Petrov type D spacetimes \cite{Chandrasekhar_1983, Petrov:2000bs}, the leading-order gravitational perturbations of which are fully described by decoupled and separable Teukolsky equations. However, in the Teukolsky formalism, the modes are not naturally separated into definite parity. To study parity, one then needs to first find combinations of solutions to the Teukolsky equations that generate definite-parity metric perturbations. This work was first done in \cite{Chrzanowski:1975wv}, using metric reconstruction to map definite-parity metric perturbations to Teukolsky functions, and expressed in a simpler form in \cite{Nichols_Zimmerman_Chen_Lovelace_Matthews_Owen_Zhang_Thorne_2012}.
	
    In modified gravity, perturbations of spinning BHs were previously studied using metric perturbations in the slow-rotation expansion \cite{Wagle_Yunes_Silva_2021, Srivastava_Chen_Shankaranarayanan_2021, Pierini:2021jxd, Pierini:2022eim, Cano:2020cao, Cano:2021myl} and using numerical relativity for an arbitrary spin but with secularly-growing errors \cite{Okounkova:2019dfo, Okounkova_Stein_Moxon_Scheel_Teukolsky_2020}. However, most of the remnant BHs of binary BH mergers are rapidly rotating (at least 65\% of their maximum), as predicted theoretically \cite{Buonanno:2007sv} and confirmed observationally \cite{LIGOScientific:2021djp}. One can, in principle, extend the approach using metric perturbations in the slow-rotating expansion to higher orders in spin, but to produce reliable results for these fast-spinning BHs, one usually needs to go beyond fifth order in the slow-rotation expansion \cite{Pani:2011vy}. Extensions to such a high order will involve complicated couplings between different $l$ modes, so this approach might not be practically feasible. Although in EdGB, Ref.~\cite{Pierini:2022eim} recently found that by resuming the $\mathcal{O}(\chi^2)$, slow-rotation expansion of QNMs using Pad\'{e} approximants \cite{Damour:1997ub, Julie:2019sab, Julie:2022huo}, one might find accurate results for dimensionless spin up to $\chi=a/M\sim 0.7$, it is still worth developing a formalism without explicit reliance on a small spin expansion. An alternative approach, combining metric perturbations with spectral decomposition techniques, was recently developed for Schwarzschild BHs \cite{Chung:2023zdq} and Kerr BHs (valid up to $\chi\sim 0.95$) \cite{Chung:Kerr}. However, it is worth noting that, although promising, such spectral decomposition techniques have only been demonstrated for BHs in GR as of yet.
	
    Recently, Refs.~\cite{Li:2022pcy, Hussain:2022ins} showed that one can extend the Teukolsky formalism in GR to modified gravity for any deformed BHs that do not significantly deviate from their counterparts in GR so that they can be treated through an EFT approach. In this modified Teukolsky formalism, the Weyl scalars $\Psi_0$ and $\Psi_4$ are decoupled from other degrees of freedom of curvature perturbations, just like in GR. Their equations are also separable because the homogeneous part of the modified Teukolsky equation is of the same form as in GR, and the source terms can be separated by projection to spin-weighted spheroidal harmonics \cite{Li:2022pcy, Hussain:2022ins}. Later, Refs.~\cite{Cano:2023tmv, Cano:2023jbk} applied the approach of \cite{Li:2022pcy, Hussain:2022ins} to higher-derivative gravity up to $\mathcal{O}(\chi^{14})$. The authors successfully separated the equations into radial and angular parts and computed the QNM frequencies valid up to $\chi\sim0.7$. Their results also match well with previous calculations using metric perturbations in \cite{Cano:2020cao, Cano:2021myl}.
    
    Nonetheless, to study isospectrality breaking, one needs to first find out what the definite-parity modes are in these modified Teukolsky equations and derive their equations. In this work, we show that one can extend the definition in \cite{Nichols_Zimmerman_Chen_Lovelace_Matthews_Owen_Zhang_Thorne_2012} to the modified Teukolsky equations in \cite{Li:2022pcy, Hussain:2022ins}. Furthermore, we derive the equations that govern these definite-parity modes and prescribe how to evaluate the shifts of QNMs using the eigenvalue perturbation (EVP) method of \cite{Zimmerman:2014aha, Mark_Yang_Zimmerman_Chen_2015, Hussain:2022ins}. For simplicity, in this work, we only focus on spacetimes that are Petrov type D even in modified gravity, but our results can be easily extended to non-Petrov-type-D spacetimes, where the modified Teukolsky formalism of \cite{Li:2022pcy, Hussain:2022ins} still applies. We also assume that background spacetimes are parity invariant, which is true for non-rotating and slowly-rotating BHs in dCS \cite{Yunes:2009hc, Yagi:2012ya} and EdGB gravity \cite{Kanti:1995vq, Pani:2009wy, Maselli:2015tta}, so we can focus on the parity properties of dynamical perturbations.
	
    In the remainder of this paper, we present in more detail our formalism for studying isospectrality breaking of QNMs in modified gravity using the Teukolsky formalism. In Sec.~\ref{sec:modified_Teukolsky_equations}, we give a quick review of the modified Teukolsky formalism developed by \cite{Li:2022pcy, Hussain:2022ins}. In Sec.~\ref{sec:definite_parity_modes_GR}, we review the construction of definite-parity modes of Teukolsky equations in GR found by \cite{Nichols_Zimmerman_Chen_Lovelace_Matthews_Owen_Zhang_Thorne_2012}. In Sec.~\ref{sec:definite_parity_modes_bGR}, we show that the same definition of definite-parity modes in GR can be extended to Petrov type D spacetimes in modified gravity. For non-Petrov-type-D spacetimes, we discuss how one might extend our construction and leave details to future work. In Sec.~\ref{sec:definite_parity_equations}, we follow the discussion in \cite{Hussain:2022ins} to derive the shifts of QNM frequencies using the EVP method of \cite{Zimmerman:2014aha, Mark_Yang_Zimmerman_Chen_2015, Hussain:2022ins} and show how the degeneracy in QNM frequencies of even- and odd-parity modes is generally broken in modified gravity. We then derive the condition for the modified Teukolsky equation to have definite-parity solutions and present the shifts of their QNM frequencies. In Sec.~\ref{sec:application}, we apply our formalism to two specific bGR theories: dCS and EdGB gravity, and we show that our definite-parity equations agree qualitatively with the equations found by metric perturbations in \cite{Cardoso:2009pk, Molina:2010fb, Pani_Cardoso_Gualtieri_2011, Wagle_Yunes_Silva_2021, Srivastava_Chen_Shankaranarayanan_2021, Pani:2009wy, Blazquez-Salcedo:2016enn, Blazquez-Salcedo_Khoo_Kunz_2017, Pierini:2021jxd, Pierini:2022eim}. Finally, in Sec.~\ref{sec:discussion}, we discuss future avenues of this work and conclude.

    \section{Modified Teukolsky equations} \label{sec:modified_Teukolsky_equations}
	
    In this section, we review the modified Teukolsky formalism in bGR theories developed in \cite{Li:2022pcy, Hussain:2022ins}. Here, we focus on the equation of $\Psi_0$, and the equation of $\Psi_4$ can be found following the same procedure or via the GHP transformation \cite{Geroch_Held_Penrose_1973}.
	
    \subsection{bGR theories and expansion scheme}
    \label{sec:bGR_theory}
	
    As shown in \cite{Li:2022pcy}, for any modified gravity theory that admits an EFT description and allows perturbation theory, the gravitational perturbations of any non-Ricci-flat, Petrov type I BH can be studied via the curvature perturbation formalism. For this large subset of modified gravity theories, its Lagrangian can be schematically written as,
    \begin{equation} \label{eq:lagrangian}
        \mathcal{L} = \mathcal{L}_{\rm{\GR}} + \ell^p \mathcal{L}_{\rm{\bGR}} + \mathcal{L}_{\rm{\matter}} + \mathcal{L}_{\rm{\field}} \,,
    \end{equation}
    where $\mathcal{L}_{\GR}$ is the Einstein-Hilbert Lagrangian, and $\mathcal{L}_{\matter}$ is the Lagrangian of matter. In this work, we focus on vacuum backgrounds, so $\mathcal{L}_{\matter}=0$. $\mathcal{L}_{\field}$ is the Lagrangian of extra non-metric fields, including both kinetic and potential terms. The Lagrangian $\mathcal{L}_{\bGR}$ describes additional corrections to the Einstein-Hilbert Lagrangian, which may contain non-minimal couplings to the extra non-metric fields. The quantity $\ell$ with dimensions of length characterizes the strength of the GR correction, with $p$ introduced to ensure that the dimension of $\ell^p {\cal{L}}_{\rm{\bGR}}$ are correct. 
    
    Based on whether there are additional non-metric fields, we can divide the subset of modified gravity theories that our modified Teukolsky formalism applies to into the following two classes: 
    \begin{itemize}
    \centering
        \item $ \mathcal{L}_{\rm{\field}} \neq 0 \Longrightarrow$ Class A\,,
        \item $ \mathcal{L}_{\rm{\field}} = 0 \Longrightarrow$ Class B\,.
    \end{itemize}
    Some examples of class A bGR theories are dCS gravity \cite{Jackiw:2003pm,Alexander:2009tp}, EdGB gravity \cite{Gross_Sloan_1987, Kanti:1995vq, Moura:2006pz}, Horndeski theory \cite{Kobayashi:2019hrl}, scalar-tensor theories \cite{Sotiriou:2014yhm}, $f(R)$ gravity \cite{Sotiriou:2006hs, Sotiriou:2008rp}, Einstein-Aether theory \cite{Jacobson_2008}, and bi-gravity \cite{Schmidt-May:2015vnx}. There are also certain EFT extensions of GR that do not contain extra non-metric fields, so they can be classified as class B bGR theories, such as higher-derivative gravity \cite{Burgess_2004, Donoghue_2012, Endlich:2017tqa, Cano:2019ore}.
	
    To study gravitational perturbations in modified gravity in the formalism of~\cite{Li:2022pcy}, we need at least two expansion parameters. In this work, we follow the conventions in \cite{Li:2022pcy} and use $\zeta$ to denote the strength of bGR corrections and $\epsilon$ the size of GW perturbations, a parameter that also appears in the GR case. Both $\zeta$ and $\epsilon$ are dimensionless, so $\zeta$ is usually some power of the ratio of the scale $\ell$ to the BH mass. We also assume that the leading correction to the metric field due to modified gravity is at least of $\mathcal{O}(\zeta)$, so the correction to other non-metric fields enters at lower and non-integer order of $\zeta$ \cite{Li:2022pcy}. Reference~\cite{Li:2022pcy} additionally showed that if the background tetrad is carefully chosen, the bGR correction to all NP quantities also enters at $\mathcal{O}(\zeta)$. Then, all the NP quantities can be expanded in the following way:
    \begin{align} \label{eq:expansion_Weyl}
	    \Psi_i
	    &=\Psi_i^{(0)}+\epsilon\Psi_i^{(1)} \nonumber\\
	    &=\Psi_i^{(0,0)}+\zeta\Psi_{i}^{(1,0)}+\epsilon\Psi_{i}^{(0,1)}
	    +\zeta\epsilon\Psi_{i}^{(1,1)}\,,
    \end{align}
    where we have taken Weyl scalars as an example. In this work, we will hide the expansion in $\zeta$ from certain equations to minimize notational clutter, so the superscript will only stand for an expansion in $\epsilon$, as shown in the first line of Eq.~\eqref{eq:expansion_Weyl}.
	
    \subsection{Modified Teukolsky equation}
    \label{sec:modified_Teuk}
	
    Using the expansion scheme in Eq.~\eqref{eq:expansion_Weyl}, one can then derive the modified Teukolsky equation. For convenience, let us first define the following operators in terms of the NP spin coefficients and tetrad derivatives (see e.g.~\cite{Newman-Penrose,Stephani:2003tm}):
    \begin{subequations} \label{eq:np-der-op}
    \begin{align}
	    D_{[a,b,c,d]}
	    &=D+a\varepsilon+b\bar{\varepsilon}+c\rho+d\bar{\rho}\,, \\
	    \boldsymbol{\Delta}_{[a,b,c,d]}
	    &=\boldsymbol{\Delta}+a\mu+b\bar{\mu}+c\gamma+d\bar{\gamma}\,, \\
	    \delta_{[a,b,c,d]}
	    &=\delta+a\bar{\alpha}+b\beta+c\bar{\pi}+d\tau\,, \\
	    \bar{\delta}_{[a,b,c,d]}
	    &=\bar{\delta}+a\alpha+b\bar{\beta}+c\pi+d\bar{\tau}\,,
    \end{align}
    \end{subequations}
    where the overhead bar denotes complex conjugation.
    The equations we start from are two Ricci identities and one Bianchi identity, namely
    \begin{subequations} \label{eq:Bianchi_simplified}
    \begin{align} 
        & F_1\Psi_0-J_1\Psi_1-3\kappa\Psi_2=S_1\,, \label{eq:BianchiId_Psi0_1_simplify} \\
        & F_2\Psi_0-J_2\Psi_1-3\sigma\Psi_2=S_2\,, \label{eq:BianchiId_Psi0_2_simplify} \\
        & E_2\sigma-E_1\kappa-\Psi_0=0 \label{eq:RicciId_Psi0_simplify}\,,
    \end{align}
    \end{subequations}
    where the operators $F_{1,2}$, $J_{1,2}$, and $E_{1,2}$ are defined via
    \begin{equation} \label{eq:simplify_values}
    \begin{aligned}
        & F_1\equiv\bar{\delta}_{[-4,0,1,0]}\,,\quad
	    && F_2\equiv\boldsymbol{\Delta}_{[1,0,-4,0]}\,, \\
	    & J_1\equiv D_{[-2,0,-4,0]}\,,\quad
		&& J_2\equiv\delta_{[0,-2,0,-4]}\,, \\
        & E_1\equiv\delta_{[-1,-3,1,-1]}\,,\quad
		&& E_2\equiv D_{[-3,1,-1,-1]}\,, \\
    \end{aligned}
    \end{equation}
    and the source terms $S_{1,2}$ are
    \begin{subequations} \label{eq:source_bianchi}
    \begin{align}
        \label{eq:source_bianchi_1}
        \begin{split} 
		  S_1\equiv& \;\delta_{[-2,-2,1,0]}\Phi_{00}
		  -D_{[-2,0,0,-2]}\Phi_{01} \\
            & \;+2\sigma\Phi_{10}-2\kappa\Phi_{11}-\bar{\kappa}\Phi_{02}\,,
	    \end{split} \\
	    \label{eq:source_bianchi_2}
	    \begin{split} 
		  S_2\equiv& \;\delta_{[0,-2,2,0]}\Phi_{01}
		  -D_{[-2,2,0,-1]}\Phi_{02} \\
		  & \;-\bar{\lambda}\Phi_{00}+2\sigma\Phi_{11}-2\kappa\Phi_{12}\,.
	    \end{split}
    \end{align}
    \end{subequations}
    
    To derive the modified Teukolsky equation, Ref.~\cite{Li:2022pcy} made some convenient gauge choices for both the background spacetime and the dynamical perturbations, following Chandrasekhar \cite{Chandrasekhar_1983}. Since we care about BH spacetimes that are modifications of Petrov type D spacetimes in GR, one can use the Kinnersley tetrad to set 
    \begin{equation}
        \Psi_{0,1,3,4}^{(0,0)}=0\,.
    \end{equation}
    For dynamical perturbations, Ref.~\cite{Li:2022pcy} showed that one can rotate the $\mathcal{O}(\zeta^1,\epsilon^1)$ part of the tetrad, such that
    \begin{equation}
        \Psi_{1,3}^{(0,1)}=\Psi_{1,3}^{(1,1)}=0\,.
    \end{equation}
    In this gauge, one can then easily decouple $\Psi_{0}^{(1,1)}$ from other NP quantities and derive a single decoupled equation for $\Psi_{0}^{(1,1)}$, namely \cite{Li:2022pcy}
    \begin{align} \label{eq:master_eqn_non_typeD_Psi0}
        H_{0}^{\GR}\Psi_0^{(1,1)}
        =\mathcal{S}_{\geo}^{(1,1)}+\mathcal{S}^{(1,1)}\,,
    \end{align}
    where $H_{0}^{\GR}$ is the Teukolsky operator in GR, and the source terms $\mathcal{S}_{\geo}^{(1,1)}$ and $\mathcal{S}^{(1,1)}$ are given by
    \begin{align} \label{eq:source_def_geo}
        & \mathcal{S}_{\geo}^{(1,1)}=\mathcal{S}_{0,\typeD}^{(1,1)}
        +\mathcal{S}_{0,\nonD}^{(1,1)}+\mathcal{S}_{1,\nonD}^{(1,1)}\,, \nonumber\\
        & \mathcal{S}_{0,\typeD}^{(1,1)}=-H_0^{(1,0)}\Psi_0^{(0,1)}\,, \nonumber\\
        & \mathcal{S}_{0,\nonD}^{(1,1)}=-H_0^{(0,1)}\Psi_0^{(1,0)}\,, \nonumber\\
        & \mathcal{S}_{1,\nonD}^{(1,1)}=H_1^{(0,1)}\Psi_1^{(1,0)}\,,
    \end{align}
    and
    \begin{align} \label{eq:source_def_stress}
        \mathcal{S}^{(1,1)}
        =& \;\mathcal{E}_2^{(0,0)}S_2^{(1,1)}+
        \mathcal{E}_2^{(0,1)}S_2^{(1,0)}
        -\mathcal{E}_1^{(0,0)}S_1^{(1,1)} \nonumber\\
        & \;-\mathcal{E}_1^{(0,1)}S_1^{(1,0)}\,.
    \end{align}
    The operators $H_{0,1}$ and $\mathcal{E}_{1,2}$ are defined via
    \begin{equation} \label{eq:def_operators}
    \begin{aligned}
        & H_0 = \mathcal{E}_2F_2-\mathcal{E}_1 F_1-3\Psi_2\,,\quad
        H_1 = \mathcal{E}_2J_2-\mathcal{E}_1 J_1\,, \\
        & \mathcal{E}_1=E_1-\frac{1}{\Psi_2}\delta\Psi_2 \,,\quad
        \mathcal{E}_2=E_2-\frac{1}{\Psi_2}D\Psi_2\,.
    \end{aligned}
    \end{equation}
    The equation for $\Psi_4^{(1,1)}$ can be found in \cite{Li:2022pcy}. In Eq.~\eqref{eq:source_def_geo}, the source term $\mathcal{S}_{\geo}$ comes from the homogeneous part of the Bianchi identities and Ricci identities in Eq.~\eqref{eq:Bianchi_simplified}, so it is generated by modifications to the background spacetime and does not involve terms from the effective stress tensor.  Within $\mathcal{S}_{\geo}$, the terms $\mathcal{S}_{i,\nonD}$ only appear in non-Petrov-type-D spacetimes, while $\mathcal{S}_{i,\typeD}$ also appears in Petrov type D spacetimes. On the other hand, the source term $\mathcal{S}$ comes from the effective stress-energy tensor, so it depends on the details of the modified gravity theory and may contain extra non-metric fields. 
    
    Inspecting Eqs.~\eqref{eq:master_eqn_non_typeD_Psi0}--\eqref{eq:source_def_stress}, one notices that every NP quantity in $\mathcal{S}_{\geo}^{(1,1)}$ has lower order than $\mathcal{O}(\zeta^1,\epsilon^1)$, and the only terms at $\mathcal{O}(\zeta^1,\epsilon^1)$ are $\Psi_0^{(1,1)}$ and $\mathcal{S}^{(1,1)}$. In \cite{Li:2022pcy}, it was additionally shown that for class B bGR theories, $\mathcal{S}^{(1,1)}\sim h^{(1,0)}h^{(0,1)}$, while for class A bGR theories, $\mathcal{S}^{(1,1)}\sim\vartheta^{(1,0)}h^{(0,1)}+\vartheta^{(1,1)}g^{(0,0)}$, where $\vartheta$ represents extra (scalar, vector or tensor fields) non-metric fields. 
    For both cases, there are no factors of $h^{(1,1)}_{\mu\nu}$, so we have fully decoupled $\Psi_{0}^{(1,1)}$ from all metric fields. For class A bGR theories, we also have $\vartheta^{(1,1)}$, but as shown in \cite{Li:2022pcy, Hussain:2022ins}, these extra non-metric fields can be solved for first by following the order-reduction scheme in \cite{Okounkova_Stein_Scheel_Hemberger_2017}. The key idea is that for these non-minimal coupling class A bGR theories, the bGR corrections always drive non-metric fields first before driving GW perturbations \cite{Li:2022pcy, Hussain:2022ins}. Thus, when writing down $\vartheta^{(1,1)}$, we actually have absorbed the coupling constant into $\vartheta$, while it enters at a lower order. For details of decoupling non-metric fields from $\Psi_{0}$, one can refer to \cite{Hussain:2022ins}. 
    
    Besides using the gauge freedom of both the background spacetime and the dynamical perturbations, one can also derive the modified Teukolsky equation without making any explicit gauge choices, as done in the original derivation of the Teukolsky equation in GR \cite{Teukolsky:1973ha} and in modified gravity in \cite{Hussain:2022ins}. Reference~\cite{Hussain:2022ins} showed that instead of using the NP language from the beginning, one could follow the idea in \cite{Wald_1978} and work with the Einstein equations directly to then project to a modified Teukolsky equation at the end. In spite of the many differen approaches to derive the modified Teukolsky equation, the final master equation shares many similarities. One major feature is that the master equation always contains terms at $\mathcal{O}(\zeta^0,\epsilon^1)$, which requires metric reconstruction of GW perturbations in GR, as one can observe in Eqs.~\eqref{eq:source_def_geo} and \eqref{eq:source_def_stress}. In the next section, we will introduce one of these metric reconstruction procedures. For the terms at $\mathcal{O}(\zeta^0,\epsilon^0)$ and $\mathcal{O}(\zeta^1,\epsilon^0)$, one can directly compute them using the background metric. To transform the modified Teukolsky equation in Eq.~\eqref{eq:master_eqn_non_typeD_Psi0} to the one for definite-parity modes, the next step is to understand what definite-parity modes of the Teukolsky equation are in both GR and modified gravity.
    
    \section{Modes with Definite Parity in GR}  \label{sec:definite_parity_modes_GR}
	
    In this section, we review definite-parity solutions to the Teukolsky equation in GR, following \cite{Nichols_Zimmerman_Chen_Lovelace_Matthews_Owen_Zhang_Thorne_2012}. Our focus is on bGR, beyond-Kerr BH spacetimes, whereas we show below that isospectrality can be broken by dynamical effects. Since this is our primary goal, we further assume that, like in the Kerr solution, the stationary spacetime is invariant under the parity transformation. This assumption holds for known BH solutions in modified gravity theories, such as in dCS gravity \cite{Yunes:2009hc, Yagi:2012vf} and EdGB gravity\cite{Kanti:1995vq, Pani:2009wy, Maselli:2015tta}, and seems physically reasonable for a stationary BH. To make this more concrete, we first define what we mean by a parity transformation.
    
    Let the spacetime be a manifold $\mathcal{M}$, with an open set $U \subset \mathcal{M}$ inside it that contains Boyer-Lindquist-like coordinates, i.e., the metric $g$ on $\mathcal{M}$ has the functional form of a modified Kerr metric on $U$ in these coordinates. 
    Define the parity operator $\hat{P}$ as an operator that acts on functions in these Boyer-Lindquist coordinates. 
    The action of the operator is the following:
    \begin{equation} \label{eq:def_parity_trans}
        \hat{P}[f(t,r,\theta,\phi)] = f(t,r,\pi-\theta,\phi+\pi)\,.
    \end{equation}
    Then for metric perturbations \cite{Zerilli:1971wd,Regge:PhysRev.108.1063,Nichols_Zimmerman_Chen_Lovelace_Matthews_Owen_Zhang_Thorne_2012}, the modes with even and odd parity are defined to be
    \begin{equation} \label{eq:metric_definite_parity}
	    \hat{P}h^{\Even,\Odd}_{\mu\nu}=\pm(-1)^l h^{\Even,\Odd}_{\mu\nu}\,,
    \end{equation}
    where $l$ is the angular momentum number after decomposing metric perturbations into spheroidal harmonics.\footnote{We have followed the definition of definite-parity modes in Sec.~IC2 of \cite{Nichols_Zimmerman_Chen_Lovelace_Matthews_Owen_Zhang_Thorne_2012}.} To define parity for solutions to the Teukolsky equation, we then need to know the relation between metric perturbations and curvature perturbations. 
 
    \subsection{Metric reconstruction in GR} \label{sec:metric_reconstruct}
	
    To find curvature perturbations from metric perturbations, we can directly compute Weyl scalars from the perturbed metric. In contrast, reconstructing metric perturbations from Weyl scalars is a more complicated process. Fortunately, this procedure was developed for Kerr BHs or, more generally, Petrov type D spacetimes in GR either via an intermediate Hertz potential \cite{Cohen_Kegeles_1975, Chrzanowski:1975wv, Kegeles_Cohen_1979, Lousto_Whiting_2002, Ori_2003, Whiting_Price_2005, Yunes_Gonzalez_2006, Keidl_Friedman_Wiseman_2007, Keidl_Shah_Friedman_Kim_Price_2010} or by solving the Bianchi identities, Ricci identities, and commutation relations \cite{Chandrasekhar_1983, Loutrel_Ripley_Giorgi_Pretorius_2020}. In this work, we follow the approach using the Hertz potential, the so-called Chrzanowski-Cohen-Kegeles (CCK) procedure \cite{Keidl_Friedman_Wiseman_2007}, and the conventions in \cite{Keidl_Friedman_Wiseman_2007, Keidl_Shah_Friedman_Kim_Price_2010} due to more explicit algebraic relations between Weyl scalars and metric perturbations. In this section, we only provide a brief introduction, and more details can be found in Appendix \ref{appendix:metric_reconstruction_more}. Furthermore, since in this section we review metric reconstruction in GR, our expressions hold only to $\mathcal{O}(\zeta^0,\epsilon^1)$.
	
    As discussed in \cite{Keidl_Friedman_Wiseman_2007, Keidl_Shah_Friedman_Kim_Price_2010}, the Hertz potential $\Psi_{\Hertz}^{(0,1)}$ generates metric perturbations $h_{\mu\nu}^{(0,1)}$ that solve the linearized Einstein equations in GR. For simplicity, we will drop the superscripts of $h_{\mu\nu}^{(0,1)}$ and $\Psi_{\Hertz}^{(0,1)}$ for the rest of this section. In the outgoing radiation gauge (ORG) \cite{Keidl_Shah_Friedman_Kim_Price_2010}, where $n^{\mu}h_{\mu\nu}=0$ and $h\equiv g^{\mu\nu}h_{\mu\nu}=0$. The perturbed metric $h_{\mu\nu}$ is related to $\Psi_{\Hertz}$ via
    \begin{equation} \label{eq:metric_Hertz_ORG}
    \begin{aligned}
        h_{\mu\nu}
        =& \;-\rho^{-4}\left[n_{\mu}n_{\nu}\bar{\delta}_{[-3,-1,5,0]}
        \bar{\delta}_{[-4,0,1,0]}\right. \\
        & \;\left.+\bar{m}_{\mu}\bar{m}_{\nu}
        \boldsymbol{\Delta}_{[5,0,-3,1]}
        \boldsymbol{\Delta}_{[1,0,-4,0]}\right. \\
        & \;\left.-n_{(\mu}\bar{m}_{\nu)}
        \left(\bar{\delta}_{[-3,1,5,1]}
        \boldsymbol{\Delta}_{[1,0,-4,0]} \right.\right.\\
        & \;\left.\left.+\boldsymbol{\Delta}_{[5,-1,-3,-1]}
        \bar{\delta}_{[-4,0,1,0]}\right)\right]\Psi_{\Hertz}
        +\text {c.c}\,.
    \end{aligned}
    \end{equation}
    In the ingoing radiation gauge (IRG) \cite{Keidl_Friedman_Wiseman_2007}, where $l^{\mu}h_{\mu\nu}=0$ and $h=0$, we have instead that
    \begin{equation} \label{eq:metric_Hertz_IRG}
    \begin{aligned}
		h_{\mu\nu}
		=& \;\left[l_{\mu}l_{\nu}\bar{\delta}_{[1,3,0,-1]}
		\bar{\delta}_{[0,4,0,3]}\right.\\
		& \;+\bar{m}_{\mu}\bar{m}_{\nu}D_{[-1,3,0,-1]}D_{[0,4,0,3]} \\
		& \;-l_{(\mu}\bar{m}_{\nu)}\left(D_{[1,3,1,-1]}
		\bar{\delta}_{[0,4,0,3]}\right.\\
		& \;\left.\left.+\bar{\delta}_{[-1,3,-1,-1]}D_{[0,4,0,3]}
		\right)\right]\bar{\Psi}_{\Hertz}+\text{c.c}\,.
    \end{aligned}
    \end{equation}
    Notice that, since we have chosen the opposite signature from \cite{Keidl_Friedman_Wiseman_2007,Keidl_Shah_Friedman_Kim_Price_2010}, our $h_{\mu\nu}$ has a different sign.
	
    Using Eqs.~\eqref{eq:metric_Hertz_ORG} and \eqref{eq:metric_Hertz_IRG}, one can derive the relation between Weyl scalars and $\Psi_{\Hertz}$. For example, for perturbations of Kerr in ORG \cite{Keidl_Shah_Friedman_Kim_Price_2010},
    \begin{subequations} \label{eq:Hertz_ORG}
    \begin{align}
	    & \Psi_{4}
	    =-\frac{1}{32}\rho^{4}\Delta^{2}D^{\dagger4}
        \Delta^{2}\bar{\Psi}_{\Hertz}\,, \label{eq:Hertz_ORG_Psi4} \\
	    & \Psi_{0}=-\frac{1}{8}\left[\mathcal{L}^{4}\bar{\Psi}_{\Hertz}
	    +12M\partial_{t}\Psi_{\Hertz}\right]\,, \label{eq:Hertz_ORG_Psi0}
    \end{align}
    \end{subequations} 
    while in IRG
    \begin{subequations} \label{eq:Hertz_IRG}
    \begin{align}
	    & \Psi_{0}=-\frac{1}{2}D^{4}\bar{\Psi}_{\Hertz}\,, \label{eq:Hertz_IRG_Psi0} \\
	    & \Psi_{4}
	=-\frac{1}{8}\rho^{4}\left[\mathcal{L}^{\dagger4}
        \bar{\Psi}_{\Hertz}-12M\partial_{t}\Psi_{\Hertz}\right]\,,
        \label{eq:Hertz_IRG_Psi4}
    \end{align}
    \end{subequations}
    where
    \begin{equation}
    \begin{aligned}
	    & \Delta=r^2-2Mr+a^2\,,\quad
	    \rho=-\frac{1}{1-ia\cos{\theta}}\,, \\
	    & D=l^{\mu}\partial_{\mu}
	    =\frac{r^2+a^2}{\Delta}\partial_t+\partial_r
	    +\frac{a}{\Delta}\partial_{\phi}\,, \\
	    & D^{\dagger}=-\frac{r^2+a^2}{\Delta}\partial_t
	    +\partial_r-\frac{a}{\Delta}\partial_{\phi}\,, \\
	    & \mathcal{L}_{s}=-ia\sin{\theta}\partial_t
	    -\left[\partial_{\theta}-s\cot{\theta}
	    +i\csc{\theta}\partial_{\phi}\right]\,, \\
	    & \mathcal{L}_{s}^{\dagger}=ia\sin{\theta}\partial_t-
	    \left[\partial_{\theta}-s\cot{\theta}
	    -i\csc{\theta}\partial_{\phi}\right]\,, \\
    \end{aligned}
    \end{equation}
    and $\mathcal{L}^4=\mathcal{L}_1\mathcal{L}_0
    \mathcal{L}_{-1}\mathcal{L}_{-2}$, $\mathcal{L}^{\dagger 4}=\mathcal{L}_1^{\dagger}\mathcal{L}_0^{\dagger}
    \mathcal{L}_{-1}^{\dagger}\mathcal{L}_{-2}^{\dagger}$. Here, we have also dropped the superscript $(0,1)$ of $\Psi_{0,4}$ for simplicity. All the Weyl scalars in this subsection are assumed to be at $\mathcal{O}(\zeta^0,\epsilon^1)$.
	
    In this work, we focus on the modified Teukolsky equation for $\Psi_0$, so it is convenient to work with IRG, where $\Psi_{\Hertz}$ can be reconstructed from $\Psi_0$ by inverting Eq.~\eqref{eq:Hertz_IRG_Psi0} using the Teukolsky-Starobinsky identities \cite{Teukolsky:1974yv, Starobinsky:1973aij,Starobinskil:1974nkd}, e.g.,
    \begin{equation} \label{eq:reconstruct_Hertz}
        \bar{\Psi}_{\Hertz}
        =-2\mathfrak{C}^{-1}\Delta^{2}\left(D^{\dagger}\right)^{4}
        \left[\Delta^{2}\Psi_{0}\right]\,.
    \end{equation}
    Here, $\mathfrak{C}$ is a real constant \cite{Chandrasekhar_1983, Ori_2003},\footnote{$\mathfrak{C}$ is the squared modulus of the Teukolsky-Starobinsky constant $|\mathscr{C}|^2$ in \cite{Chandrasekhar_1983} and the constant $p$ in \cite{Ori_2003}.} 
    \begin{equation} \label{eq:ST_coef}
    \begin{aligned}
        \mathfrak{C}=& \;\lambda^2(\lambda+2)^2-8\omega^2\lambda
        \left[\alpha^2(5\lambda+6)-12a^2\right] \\
        & \;+144\omega^4\alpha^4+144\omega^2M^2\,,
    \end{aligned}
    \end{equation}
    where $\omega$ is the QNM frequency associated with a specific $(l,m,\omega)$ mode of $\Psi_0$, $\lambda$ is the separation constant used by Chandrasekhar \cite{Chandrasekhar_1983}, and $\alpha^2 \equiv a^2-am/\omega$. The relation between $\Psi_{\Hertz}$ and other Weyl scalars in IRG can be found in Appendix~\ref{appendix:metric_reconstruction_more}. For the Schwarzschild background, these relations greatly simplify. Since we use the relation between $\Psi_2$ and $\Psi_{\Hertz}$ in the Schwarzschild limit frequently in Sec.~\ref{sec:application}, we present it here for convenience,
    \begin{equation} \label{eq:Hertz_IRG_Psi2}
	    \Psi_{2}=-\frac{1}{2}D^{2}(\bar{\delta}+2\beta)
	    (\bar{\delta}+4\beta)\bar{\Psi}_{\Hertz}\,.
    \end{equation}
	
    \subsection{Definition of even- and odd-parity modes} \label{sec:define_parity_modes}
	
    With the relation in Eq.~\eqref{eq:metric_Hertz_IRG}, we can now define the modes with definite parity in GR. For convenience, let us define an operator $\hat{\mathcal{P}} \equiv\hat{\mathcal{C}}\hat{P}$, where $\hat{P}$ is the parity transformation, and $\hat{\mathcal{C}}$ is the complex conjugation,
    \begin{equation} \label{eq:def_P}
		\hat{\mathcal{P}}f=\hat{\mathcal{C}}\hat{P}f=\hat{\mathcal{C}}f(\pi-\theta,\phi+\pi)=\bar{f}(\pi-\theta,\phi+\pi)\,.
    \end{equation}
    In \cite{Miguel:2023rzp}, the same $\hat{\mathcal{P}}$ operator was also constructed and used for studying scalar and vector QNMs and their isospectrality in EFT extensions of GR. With the definition above and $\hat{P}^2=\hat{\mathcal{C}}^2=\hat{\mathcal{I}}$, where $\hat{\mathcal{I}}$ is the identity operator, one can easily show that 
    \begin{equation} \label{eq:pCP_relations}
	    \hat{\mathcal{P}}^2=\hat{\mathcal{I}}\,,\quad
	    \hat{\mathcal{C}}\hat{\mathcal{P}}=\hat{P}\,,\quad
	    \hat{P}\hat{\mathcal{P}}=\hat{\mathcal{C}}\,,
    \end{equation}
    because $\hat{P}$, $\hat{\mathcal{C}}$, and $\hat{\mathcal{P}}$ commute with each other. Using Eq.~\eqref{eq:pCP_relations}, we will replace $\hat{P}$ with $\hat{\mathcal{C}}$ and $\hat{\mathcal{P}}$ in most places. When $\hat{\mathcal{P}}$ acts on another operator $\hat{X}$, we have
    \begin{equation}
        \hat{\mathcal{P}}[\hat{X}]=\hat{\mathcal{P}}\hat{X}\hat{\mathcal{P}},
    \end{equation}
    and similarly for $\hat{\mathcal{C}}$. Other useful properties of $\hat{P}$, $\hat{\mathcal{C}}$, and $\hat{\mathcal{P}}$ are listed in Appendix~\ref{appendix:parity_operators}.
    
    As discussed in \cite{Nichols_Zimmerman_Chen_Lovelace_Matthews_Owen_Zhang_Thorne_2012}, at $\mathcal{O}(\zeta^0,\epsilon^0)$ in the Kinnersley tetrad of Kerr, 
    \begin{equation} \label{eq:NP_parity_Kerr}
    \begin{aligned}
	    & \hat{\mathcal{P}}\left\{D,\boldsymbol{\Delta}\right\}
	    =\left\{D,\boldsymbol{\Delta}\right\}\,,\;
	    && \hat{\mathcal{P}}\left\{\delta,\bar{\delta}\right\}
	    =-\left\{\delta,\bar{\delta}\right\}\,, \\
	    & \hat{\mathcal{P}}\left\{\rho,\mu,\gamma\right\}
	    =\left\{\rho,\mu,\gamma\right\}\,,\;
	    && \hat{\mathcal{P}}\left\{\alpha,\beta,\pi,\tau\right\}
	    =-\left\{\alpha,\beta,\pi,\tau\right\}\,,
    \end{aligned}
    \end{equation}
    and other spin coefficients are zero.  For convenience, let us rewrite Eqs.~\eqref{eq:metric_Hertz_ORG} and \eqref{eq:metric_Hertz_IRG} as
    \begin{equation} \label{eq:metric_Hertz_scheme0}
        h_{\mu\nu}=\mathcal{O}_{\mu\nu}\bar{\Psi}_{\Hertz}
        +\bar{\mathcal{O}}_{\mu\nu}\Psi_{\Hertz}\,,
    \end{equation}
    where $\mathcal{O}_{\mu\nu}$ denotes the operator converting $\bar{\Psi}_{\Hertz}$ to $h_{\mu\nu}$. Using Eq.~\eqref{eq:NP_parity_Kerr}, one can show that 
    \begin{equation} \label{eq:O_parity}
        \hat{\mathcal{P}}\mathcal{O}_{\mu\nu}=\mathcal{O}_{\mu\nu}\,.
    \end{equation}
    Thus,
    \begin{equation} \label{eq:Pmetric_Hertz_scheme0}
        \hat{\mathcal{P}}h_{\mu\nu}
        =\mathcal{O}_{\mu\nu}\hat{\mathcal{P}}\bar{\Psi}_{\Hertz}
        +\bar{\mathcal{O}}_{\mu\nu}\hat{\mathcal{P}}\Psi_{\Hertz}\,.
    \end{equation}
    For even- and odd-parity metric perturbations, $\hat{\mathcal{P}}h_{\mu\nu}^{\Even,\Odd}=\pm(-1)^lh_{\mu\nu}^{\Even,\Odd}$. Comparing Eq.~\eqref{eq:Pmetric_Hertz_scheme0} to Eq.~\eqref{eq:metric_Hertz_scheme0}, we find that the Hertz potentials $\Psi_{\Hertz}^{\Even,\Odd}$ generated from $h_{\mu\nu}^{\Even,\Odd}$ must transform as
    \begin{equation} \label{eq:transform_Hertz_parity}
	    \hat{\mathcal{P}}\Psi_{\Hertz}^{\Even,\Odd}
	    =\pm(-1)^{l}\Psi_{\Hertz}^{\Even,\Odd}\,.
    \end{equation}
    Since the operators converting $\bar{\Psi}_{\Hertz}$ to $\Psi_{0,4}$ in Eqs.~\eqref{eq:Hertz_ORG_Psi4} and \eqref{eq:Hertz_IRG_Psi0} are invariant under $\hat{\mathcal{P}}$, the even- and odd-parity modes of $\Psi_{0,4}$ must transform in the same way as Eq.~\eqref{eq:transform_Hertz_parity}. 

    We can then define the definite-parity modes $\Psi^{\Even,\Odd}_{lm\omega}$ of $\Psi_{0,4}$ to be
    \begin{equation} \label{eq:def_parity_modes}
		\Psi^{\Even,\Odd}_{lm\omega}
	    \coloneqq \Psi_{lm\omega}\pm(-1)^{l}\hat{\mathcal{P}}\Psi_{lm\omega}
    \end{equation}
    because then we have that
    \begin{equation}
    \begin{aligned}
        \hat{\mathcal{P}}\Psi_{lm\omega}^{\Even,\Odd}
        =& \;\hat{\mathcal{P}}\Psi_{lm\omega}\pm(-1)^l\Psi_{lm\omega} \\
        =& \;\pm(-1)^l\left[\Psi_{lm\omega}
        \pm(-1)^{l}\hat{\mathcal{P}}\Psi_{lm\omega}\right] \\
        =& \;\pm(-1)^{l}\Psi_{lm\omega}^{\Even,\Odd}\,,
    \end{aligned}
    \end{equation}
    where $\Psi_{lm\omega}$ is a single $(l,m,\omega)$ mode solving the Teukolsky equation of either $\Psi_0$ or $\Psi_4$. For simplicity, let us define
    \begin{align} \label{eq:def_parity_modes_simple}
        \Psi_{\Even,\Odd}\coloneqq\Psi\pm(-1)^{l}\hat{\mathcal{P}}\Psi\,,\quad
        \Psi\coloneqq\Psi_{lm\omega}\,,
    \end{align}
    where we have dropped the mode label $lm\omega$ for all the fields in Eq.~\eqref{eq:def_parity_modes}. Henceforth, we always assume that $\Psi$ is a single $(l,m,\omega)$ mode. Using Eq.~\eqref{eq:def_parity_modes_simple}, we can also express $\Psi$ and $\hat{\mathcal{P}}\Psi$ in terms of $\Psi_{\Even,\Odd}$,
    \begin{equation} \label{Weyl_scalar_even_odd}
		\Psi=\frac{1}{2}\left(\Psi_{\Even}+\Psi_{\Odd}\right)\,,\quad 
		\hat{\mathcal{P}}\Psi=\frac{(-1)^l}{2}\left(\Psi_{\Even}-\Psi_{\Odd}\right)\,, 
    \end{equation}
    which provides the inverse map from definite-parity modes back to the full solutions to the Teukolsky equation.
    
    \subsection{Relation between definite-parity modes and solutions to the Teukolsky equation}
    \label{sec:relation_to_Teuk_sol}
    
    To generate metric perturbations with definite parity, besides the transformation property in Eq.~\eqref{eq:transform_Hertz_parity}, we also need $\Psi_{\Even,\Odd}$ to solve the Teukolsky equation; otherwise, the metric perturbations generated will not solve the Einstein equations. The modes in Eq.~\eqref{eq:def_parity_modes} are not necessarily solutions to the modified Teukolsky equation since $\hat{\mathcal{P}}\Psi$ is not guaranteed to be a solution except in some special cases.
	
    For Kerr, one can show that Eq.~\eqref{eq:def_parity_modes} are solutions to the Teukolsky equation by using the transformation properties of Teukolsky functions under $\hat{P}$ and $\hat{\mathcal{C}}$. According to \cite{Nichols_Zimmerman_Chen_Lovelace_Matthews_Owen_Zhang_Thorne_2012}, the radial Teukolsky functions ${}_{s}R_{lm\omega}(r)$ and the angular Teukolsky functions ${}_{s}S_{lm\omega}(\theta)$ for Kerr in GR satisfy
    \begin{equation} \label{eq:parity_Teuk_func}
    \begin{aligned}
		& {}_{s}\bar{R}_{lm\omega}=(-1)^{m}{}_{s}R_{l-m-\bar{\omega}}\,, \\
		& {}_{s}S_{lm\omega}(\pi-\theta)
		=(-1)^{m+l}{}_{-s}S_{lm\omega}(\theta)\,, \\
		& {}_{s}\bar{S}_{lm\omega}(\theta)
		=(-1)^{m+s}{ }_{-s}S_{l-m-\bar{\omega}}(\theta)\,.
    \end{aligned}
    \end{equation}
    Thus, by applying the above relations for $s=\pm2$, we can rewrite $\hat{\mathcal{P}}\Psi$ as
    \begin{equation} \label{eq:PPsi_Kerr}
    \begin{aligned}
		\hat{\mathcal{P}}\Psi
		=& \;\bar{\Psi}_{lm\omega}(\pi-\theta,\phi+\pi) \\
		=& \;{}_{\pm2}\bar{R}_{lm\omega}
		e^{-i\left(m(\phi+\pi)-\bar{\omega}t\right)}
		{}_{\pm2}\bar{S}_{lm\omega}(\pi-\theta) \\
		=& \;(-1)^{l}{}_{\pm2}R_{l-m-\bar{\omega}}e^{-i\left(m\phi-\bar{\omega}t\right)}
		{}_{\pm2}S_{l-m-\bar{\omega}}\,, \\
    \end{aligned}	
    \end{equation}
    where in the third line we have used the relations in Eq.~\eqref{eq:parity_Teuk_func}. Thus, for Kerr, we also have
    \begin{equation} \label{eq:parity_modes_Kerr}
	    \Psi^{\Even,\Odd}_{lm\omega}=\Psi_{lm\omega}\pm\Psi_{l-m-\bar{\omega}}\,,
	\end{equation}
    which is the definition used in \cite{Nichols_Zimmerman_Chen_Lovelace_Matthews_Owen_Zhang_Thorne_2012}.
	
    In general, the modes in Eq.~\eqref{eq:parity_modes_Kerr} do not necessarily satisfy the transformation rule in Eq.~\eqref{eq:transform_Hertz_parity} as one can explicitly check. For Kerr, due to the relations in Eq.~\eqref{eq:parity_Teuk_func}, we can transform $\Psi_{l-m-\bar{\omega}}$ to $\hat{\mathcal{P}}\Psi_{lm\omega}$, so Eq.~\eqref{eq:parity_modes_Kerr} becomes modes of definite parity. Generally, we need to define the even and odd modes for $\Psi_{lm\omega}$ and $\Psi_{l-m\bar{\omega}}$ separately using Eq.~\eqref{eq:def_parity_modes_simple}. For Kerr, since $\hat{\mathcal{P}}\Psi_{lm\omega}=(-1)^l\Psi_{l-m\bar{\omega}}$, the even and odd modes for $\Psi_{lm\omega}$ and $\Psi_{l-m\bar{\omega}}$ are degenerate, $\Psi^{\Even,\Odd}_{lm\omega}=\pm\Psi^{\Even,\Odd}_{l-m\bar{\omega}}$.
 
    Another feature of the definition in Eq.~\eqref{eq:def_parity_modes_simple} is that the modes with definite parity are linear combinations of modes with frequency $\omega$ and the (negative of its) conjugate $-\bar{\omega}$. One may wonder whether we can define modes with definite parity without mixing modes with different frequencies. The answer is no. For a generic Hertz potential, we can always decompose it into modes with different frequencies, e.g., $\Psi=\sum_{\omega}A_{\omega}(r,\theta,\phi)e^{-i\omega t}$. Since modes with definite parity need to transform as in Eq.~\eqref{eq:transform_Hertz_parity}, we must have 
    \begin{equation}
		\sum_{\omega}A_{\omega}(r,\theta,\phi)e^{-i\omega t} \\
		=\pm(-1)^l\sum_{\omega}\bar{A}_{\omega}(r,\pi-\theta,\phi+\pi)
		e^{i\bar{\omega}t}\,,
    \end{equation}
    so $A_{-\bar{\omega}}(r,\theta,\phi)=\pm(-1)^l\bar{A}_{\omega}(r,\pi-\theta,\phi+\pi)$. Thus, any mode with frequency $\omega$ must be accompanied by a mode with frequency $-\bar{\omega}$. This additionally shows that any solution of the Teukolsky equation in GR has the decomposition in Eq.~\eqref{eq:def_parity_modes_simple} if it is a definite-parity mode.
	
    To summarize, the definition in Eq.~\eqref{eq:def_parity_modes} is a more fundamental definition than the one in Eq.~\eqref{eq:parity_modes_Kerr} since it does not rely on the specific properties of the solutions to the Teukolsky equation in Kerr. As such, we use Eqs.~\eqref{eq:def_parity_modes} and \eqref{eq:def_parity_modes_simple} for the rest of the work. The major goal of this work is to study the correction to the definite-parity Teukolsky solutions in GR defined in Eq.~\eqref{eq:def_parity_modes_simple}. In many cases, we do not expect that the modified Teukolsky equation admits solutions with definite parity for generic systems that generate GWs. We also do not expect the modes defined in Eq.~\eqref{eq:def_parity_modes_simple} to generate metric perturbations with definite parity in modified gravity. Nonetheless, in the next section, we show that the definition in Eqs.~\eqref{eq:def_parity_modes} and \eqref{eq:def_parity_modes_simple} still work for Petrov type D spacetimes that are perturbations of Kerr in modified gravity.

    \section{Modes with Definite Parity in Modified Gravity}
    \label{sec:definite_parity_modes_bGR}
	
    In the previous section, we have shown that the Teukolsky functions defined in Eq.~\eqref{eq:def_parity_modes_simple} generate metric perturbations with definite parity in GR using metric reconstruction. However, for general modified gravity theories, the procedure to reconstruct metric perturbations from the modified Teukolsky functions is not known. On the other hand, one may also wonder whether the definition in Eq.~\eqref{eq:def_parity_modes_simple} still works for modified gravity. Although one can find bGR corrections to the GR, definite-parity QNMs without knowing whether the corrected modes have definite parity, it is still important to understand what definite-parity QNMs are in modified gravity. Along with the results of Sec.~\ref{sec:definite_parity_equations}, one can then directly check whether a modified gravity theory admits modes with definite parity as solutions to the modified Teukolsky equations. In this section, we show that the definition in Eq.~\eqref{eq:def_parity_modes_simple} also works for Petrov type D spacetimes in modified gravity.
	
    Since metric reconstruction in modified gravity is generally unknown, we first start from metric perturbations with definite parity and then compute the parity transformation of the Weyl scalars generated from these metric perturbations. In our derivation, we make certain gauge choices such that all the NP quantities have simple transformation properties under $\hat{\mathcal{P}}$. In Sec.~\ref{sec:parity_gauge_invariant}, we further show that although our derivation is not manifestly gauge-invariant, the parity properties of Weyl scalars $\Psi_{0,4}$ are gauge-invariant. In this work, we only aim to define definite-parity modes for Petrov type D spacetimes in modified gravity. More specifically, we consider Petrov type D BHs that are modifications of Kerr. For non-Petrov-type-D spacetimes in modified gravity, there are additional complexities, so we discuss a potential strategy and leave further investigations for future work. 
	
    \subsection{$\hat{\mathcal{P}}$-transformation of stationary NP quantities}
    \label{sec:parity_property_stationary}
	
    In this section, we compute the parity transformation of NP quantities at $\mathcal{O}(\epsilon^0)$. For Kerr BHs in GR, we can use the Kinnersley tetrad such that at $\mathcal{O}(\zeta^0,\epsilon^0)$,
    \begin{equation} \label{eq:tetrad_00_parity}
    \begin{aligned}
	    & \hat{\mathcal{P}}D^{(0,0)}= D^{(0,0)}\,,\quad
	&& \hat{\mathcal{P}}\boldsymbol{\Delta}^{(0,0)}=      \boldsymbol{\Delta}^{(0,0)}\,, \\
	    & \hat{\mathcal{P}}\delta^{(0,0)}= -\delta^{(0,0)}\,,\quad
	    && \hat{\mathcal{P}}\bar{\delta}^{(0,0)}= -\bar{\delta}^{(0,0)}\,.
    \end{aligned}
    \end{equation}
    For tetrads corrected at $\mathcal{O}(\zeta^1,\epsilon^0)$, we can follow \cite{Li:2022pcy} to choose
    \begin{equation}
	    \delta e_{a\mu}^{(1,0)}
	    =-\frac{1}{2}e_{a\nu}^{(0,0)}h^{\;\nu(1,0)}_{\mu}\,,
    \end{equation}
    such that all the orthogonality conditions of NP tetrads are satisfied. Recall that we assume the BH spacetime invariant under parity, so $\hat{\mathcal{P}}h_{\mu\nu}^{(1,0)}=\hat{P}h_{\mu\nu}^{(1,0)}=h_{\mu\nu}^{(1,0)}$, where we have used that $h_{\mu\nu}^{(1,0)}$ is real. Then, using Eq.~\eqref{eq:tetrad_00_parity}, we find
    \begin{equation} \label{eq:tetrad_10_parity}
    \begin{aligned}
	    & \hat{\mathcal{P}}D^{(1,0)}= D^{(1,0)}\,,\quad
	    && \hat{\mathcal{P}}\boldsymbol{\Delta}^{(1,0)}=\boldsymbol{\Delta}^{(1,0)}\,, \\
	    & \hat{\mathcal{P}}\delta^{(1,0)}=-\delta^{(1,0)}\,,\quad
	    && \hat{\mathcal{P}}\bar{\delta}^{(1,0)}=-\bar{\delta}^{(1,0)}\,.
    \end{aligned}
    \end{equation}
    Since the background tetrad in GR and modified gravity have the same transformation properties under $\hat{\mathcal{P}}$, we do not distinguish them below by suppressing the index orders in $\zeta$ so that the superscripts give the order in the expansion of $\epsilon$ only.
	
    With Eqs.~\eqref{eq:tetrad_00_parity} and \eqref{eq:tetrad_10_parity}, one can then compute the $\hat{\mathcal{P}}$-transformation of any spin coefficient or Ricci rotation coefficient using the definition, 
    \begin{equation} \label{eq:ricci_rot_def}
	    \gamma_{cab}=e_{a\mu;\nu}e_{c}^{\mu}e_{b}^{\nu}\,.
    \end{equation}
    From Eqs.~\eqref{eq:tetrad_00_parity}, \eqref{eq:tetrad_10_parity}, and \eqref{eq:ricci_rot_def}, we know that the spin coefficients with even number of $m^{\mu}$, $\bar{m}^{\mu}$, and their corresponding directional derivatives are invariant under $\hat{\mathcal{P}}$, while the ones with odd numbers of $m^{\mu}$, $\bar{m}^{\mu}$, and their corresponding directional derivatives pick up a minus sign under $\hat{\mathcal{P}}$. 
    For example, $\kappa=\gamma_{131}$ only contains one $m^{\mu}$, so $\hat{\mathcal{P}}\kappa^{(0)}=-\kappa^{(0)}$. Although $\kappa^{(0)}$ vanishes in Petrov type D spacetimes, we still list its parity transformation property since it only depends on the transformation rules in Eqs.~\eqref{eq:tetrad_00_parity} and \eqref{eq:tetrad_10_parity}. Similarly, we find
    \begin{equation} \label{eq:spin_coeff_00_10_parity}
    \begin{aligned}
	    & \;\hat{\mathcal{P}}\left\{\sigma^{(0)},\lambda^{(0)},
	    \varepsilon^{(0)},\rho^{(0)},\mu^{(0)},\gamma^{(0)}\right\} \\
	    & \;=\left\{\sigma^{(0)},\lambda^{(0)},
	    \varepsilon^{(0)},\rho^{(0)},\mu^{(0)},\gamma^{(0)}\right\}\,, \\
	    & \;\hat{\mathcal{P}}\left\{\kappa^{(0)},\nu^{(0)},\alpha^{(0)},
	    \beta^{(0)},\tau^{(0)},\pi^{(0)}\right\} \\
	    & \;=-\left\{\kappa^{(0)},\nu^{(0)},\alpha^{(0)},
	\beta^{(0)},\tau^{(0)},\pi^{(0)}\right\}\,,
    \end{aligned}
    \end{equation}
    which are consistent with the results in \cite{Nichols_Zimmerman_Chen_Lovelace_Matthews_Owen_Zhang_Thorne_2012} for GR.
	
    \subsection{$\hat{\mathcal{P}}$-transformation of dynamical NP quantities}
    \label{sec:parity_property_dynamical}
	
    At $\mathcal{O}(\epsilon^{1})$, the tetrad can be expressed in terms of the tetrad at $\mathcal{O}(\epsilon^{0})$. As found in \cite{Campanelli_Lousto_1999,Loutrel_Ripley_Giorgi_Pretorius_2020}, one can use the tetrad freedom to choose
    \begin{equation} \label{eq:tetrad_01_11}
    \begin{aligned}
        D^{(1)}=&\;-\frac{1}{2}h_{ll}^{(1)}\boldsymbol{\Delta}^{(0)}\,, \\
        \boldsymbol{\Delta}^{(1)}=&\;-\frac{1}{2}h_{nn}^{(1)}D^{(0)}
        -h_{ln}^{(1)}\boldsymbol{\Delta}^{(0)}\,, \\
        \delta^{(1)}=&\;-h_{nm}^{(1)}D^{(0)}
        -h_{lm}^{(1)}\boldsymbol{\Delta}^{(0)}
        +\frac{1}{2}h_{m\bar{m}}^{(1)}\delta^{(0)}
        +\frac{1}{2}h_{mm}^{(1)}\bar{\delta}^{(0)}\,.
    \end{aligned}
    \end{equation}
    Since this tetrad choice is possible at both $\mathcal{O}(\zeta^{0},\epsilon^{1})$ and $\mathcal{O}(\zeta^{1},\epsilon^{1})$, we also suppress the expansion in $\zeta$ here for simplicity. Taking $h_{\mu\nu}^{(1)}$ to be the metric perturbations with definite parity of Eq.~\eqref{eq:metric_definite_parity}, we find that the dynamical tetrad transforms under $\hat{\mathcal{P}}$ as
    \begin{equation} \label{eq:tetrad_01_11_parity}
    \begin{aligned}
	    & \hat{\mathcal{P}}D^{(1)}=\pm(-1)^lD^{(1)}\,, \;
	    && \hat{\mathcal{P}}\boldsymbol{\Delta}^{(1)}= 
	    \pm(-1)^l\boldsymbol{\Delta}^{(1)}\,, \\
	    & \hat{\mathcal{P}}\delta^{(1)}=\mp(-1)^l\delta^{(1)}\,,\quad
	    && \hat{\mathcal{P}}\bar{\delta}^{(1)}=
	    \mp(-1)^l\bar{\delta}^{(1)}\,,
    \end{aligned}
    \end{equation}
    where the factor $\pm$ of $D^{(1)}$ and $\boldsymbol{\Delta}^{(1)}$ depends on the parity of $h_{\mu\nu}^{(1)}$ generating these tetrad perturbations, e.g., $+$ for even parity and $-$ for odd parity, and similarly for the factor $\mp$ of $\delta^{(1)}$ and $\bar{\delta}^{(1)}$. The eigenvalues of $\hat{\mathcal{P}}$ in Eq.~\eqref{eq:tetrad_01_11_parity} have contributions from both the background tetrad [Eqs.~\eqref{eq:tetrad_00_parity} and \eqref{eq:tetrad_10_parity}] and the perturbed metric $h_{\mu\nu}^{(1)}$ [$\pm(-1)^l$ for even and odd parity, respectively]. For example, since $\hat{\mathcal{P}}h_{ll}^{(1)}=\pm(-1)^lh_{ll}^{(1)}$ for even- or odd-parity $h_{\mu\nu}^{(1)}$, respectively, and $\hat{\mathcal{P}}\boldsymbol{\Delta}^{(0)}=\boldsymbol{\Delta}^{(0)}$, we find $\hat{\mathcal{P}}D^{(1)}=\pm(-1)^lD^{(1)}$. In total, $D^{(1)}$ and $\boldsymbol{\Delta}^{(1)}$ preserve the parity of $h_{\mu\nu}^{(1)}$, while $\delta^{(1)}$ and $\bar{\delta}^{(1)}$ flip its parity.
	
    To find the $\hat{\mathcal{P}}$-transformation of the spin coefficients at $\mathcal{O}(\epsilon^1)$, one can first express the spin coefficients in terms of metric perturbations. This can be done by linearizing the commutation relation, e.g.~following \cite{Chandrasekhar_1983}. We have listed the results in Appendix~\ref{appendix:metric_reconstruction_more}, which are consistent with the results in \cite{Loutrel_Ripley_Giorgi_Pretorius_2020}. Then, using Eqs.~\eqref{eq:tetrad_00_parity}, \eqref{eq:tetrad_10_parity}, \eqref{eq:spin_coeff_00_10_parity}, and \eqref{eq:tetrad_01_11_parity}, one can find that
    \begin{equation} \label{eq:spin_coeff_01_11_parity}
    \begin{aligned}
	    & \;\hat{\mathcal{P}}\left\{\sigma^{(1)},\lambda^{(1)},\varepsilon^{(1)},
	    \rho^{(1)},\mu^{(1)},\gamma^{(1)}\right\} \\
	    & \;=\pm(-1)^l\left\{\sigma^{(1)},\lambda^{(1)},\varepsilon^{(1)},
	    \rho^{(1)},\mu^{(1)},\gamma^{(1)}\right\}\,, \\
	    & \;\hat{\mathcal{P}}\left\{\kappa^{(1)},\nu^{(1)},\alpha^{(1)},
	    \beta^{(1)},\tau^{(1)},\pi^{(1)}\right\} \\
	    & \;=\mp(-1)^l\left\{\kappa^{(1)},\nu^{(1)},\alpha^{(1)},
	    \beta^{(1)},\tau^{(1)},\pi^{(1)}\right\}\,,
    \end{aligned}
    \end{equation}
    which have very similar transformation properties as the spin coefficients at $\mathcal{O}(\epsilon^{0})$ in Eq.~\eqref{eq:spin_coeff_00_10_parity} up to the overall factor $\pm(-1)^l$ of $h_{\mu\nu}^{(1)}$ under $\hat{\mathcal{P}}$.
    
    With this in hand, let us now study the transformation properties of the Weyl scalars. Using the Ricci identity, one has that
    \begin{equation}
	    \Psi_0=D_{[-3,1,-1,-1]} \sigma-\delta_{[-1,-3,1,-1]}\kappa\,.
    \end{equation}
    Using Eqs.~\eqref{eq:tetrad_00_parity}, \eqref{eq:tetrad_10_parity}, \eqref{eq:spin_coeff_00_10_parity}, \eqref{eq:tetrad_01_11_parity}, and \eqref{eq:spin_coeff_01_11_parity}, we then find that
    \begin{equation} \label{eq:transform_Psi0_perturbed}
	    \hat{\mathcal{P}}\Psi_0^{(1)}=\pm(-1)^l\Psi_0^{(1)}\,,
    \end{equation}
    which includes both $\mathcal{O}(\zeta^0,\epsilon^1)$ and $\mathcal{O}(\zeta^1,\epsilon^1)$ contributions and is consistent with Eq.~\eqref{eq:transform_Hertz_parity}. The same, of course, is also true for $\Psi_4^{(1)}$. In Sec.~\ref{sec:relation_to_Teuk_sol}, we have shown that for any mode to have the transformation properties of Eq.~\eqref{eq:transform_Psi0_perturbed}, the mode has to have the decomposition of Eq.~\eqref{eq:def_parity_modes_simple}. This confirms that the modes defined in Eq.~\eqref{eq:def_parity_modes_simple} are definite-parity Teukolsky solutions for Petrov type D spacetimes in modified gravity. Reference~\cite{Cano:2023tmv} has found the same definite-parity modes of $\Psi_{0,4}$ via metric reconstruction, but they only intended to construct these definite-parity modes in GR for evaluating QNM shifts. Nonetheless, our work extends the definition in GR to Petrov type D spacetimes in modified gravity.	
    \subsection{Gauge invariance}
    \label{sec:parity_gauge_invariant}
	
    In the analysis above, we have chosen a specific tetrad in Eqs.~\eqref{eq:tetrad_00_parity}, \eqref{eq:tetrad_10_parity}, and \eqref{eq:tetrad_01_11_parity}, so one may wonder whether our definition in Eq.~\eqref{eq:def_parity_modes_simple} still works if we choose a different tetrad. Moreover, besides the freedom of rotating the tetrad, one can also perform local coordinate transformations, which one may worry also affect our analysis. In this section, we show that, although the proof above selected a specific tetrad, its argument about definite-parity modes is both tetrad- and coordinate-invariant.
	
    Generally, for perturbations at $\mathcal{O}(\zeta^1,\epsilon^1)$, one should consider both coordinate and tetrad transformations at $\mathcal{O}(\zeta^1,\epsilon^0)$, $\mathcal{O}(\zeta^0,\epsilon^1)$, and $\mathcal{O}(\zeta^1,\epsilon^1)$. For simplicity, let us focus on $\Psi_0$, and an analogous argument can be made for $\Psi_4$. For a general combination of type I, II, and type III rotations with rotation parameters $a$, $b$, $A$, and $\vartheta$, where $a$ and $b$ are complex functions, and $A$ and $\vartheta$ are real functions, the Weyl scalar $\Psi_{0}$ transforms as \cite{Chandrasekhar_1983},
    \begin{equation} \label{eq:rotated_Psi0_general}
	    \Psi_0\rightarrow A^{-2}e^{2i\vartheta}\Psi_0
	    +4b\Psi_1+6b^2\Psi_2+4b^3\Psi_3+b^4\Psi_4\,,
    \end{equation}
    where we have kept all orders of the rotation parameters. For tetrad rotations at $\mathcal{O}(\zeta^1,\epsilon^1)$, Eq.~\eqref{eq:rotated_Psi0_general} reduces to
    \begin{equation}
	    \Psi_0^{(1,1)}\rightarrow \Psi_0^{(1,1)}-2[\delta A^{(1,1)}-i\vartheta^{(1,1)}]\Psi_0^{(0,0)}
	    +4b^{(1,1)}\Psi_1^{(0,0)}\,,
    \end{equation}
    where we have define $\delta A=A-1$. Since we are interested in background spacetimes that are modifications of Petrov type D spacetimes in GR, $\Psi_0^{(0,0)}=\Psi_1^{(0,0)}=0$, so $\Psi_0^{(1,1)}$ is invariant under tetrad rotations at $\mathcal{O}(\zeta^1,\epsilon^1)$, which is consistent with the result in \cite{Li:2022pcy}. 
    Similarly, for tetrad rotations at $\mathcal{O}(\zeta^0,\epsilon^1)$,
    \begin{equation}
	    \Psi_0^{(1,1)}\rightarrow \Psi_0^{(1,1)}-2[\delta A^{(0,1)}-i\vartheta^{(0,1)}]\Psi_0^{(1,0)}
	    +4b^{(0,1)}\Psi_1^{(1,0)}\,.
    \end{equation}
    For Petrov type D spacetimes in modified gravity, $\Psi_0^{(1,0)}=\Psi_1^{(1,0)}=0$, so $\Psi_0^{(1,1)}$ is also invariant under tetrad rotations at $\mathcal{O}(\zeta^0,\epsilon^1)$. 
    
    However, for Petrov type I spacetimes, since $\Psi_{0,1}^{(1,0)}$ are nonzero, $\Psi_0^{(1,1)}$ is not invariant under tetrad rotations at $\mathcal{O}(\zeta^0,\epsilon^1)$. 
    Then to justify our arguments in Secs.~\ref{sec:parity_property_stationary} and \ref{sec:parity_property_dynamical}, we need to first construct a tetrad- and coordinate-invariant dynamical curvature perturbation from $\Psi_0^{(1,1)}$. Although such a quantity has not been found in modified gravity yet, there have been similar efforts for second-order GW perturbations in GR. Reference~\cite{Campanelli_Lousto_1999} found that $\Psi_{0,4}^{(0,2)}$ are also not invariant under tetrad rotations and coordinate transformations at $\mathcal{O}(\zeta^0,\epsilon^1)$. Solutions to this issue include adding correction terms to $\Psi_{0,4}^{(0,2)}$ to construct a tetrad- and coordinate-invariant quantity \cite{Campanelli_Lousto_1999} or studying GW perturbations in an asymptotically flat representation of Kerr \cite{Ripley_Loutrel_Giorgi_Pretorius_2021}. As shown in \cite{Li:2022pcy}, our modified Teukolsky formalism can be directly mapped to the second-order Teukolsky formalism in GR in \cite{Campanelli_Lousto_1999}, so this issue in modified gravity can probably be solved using similar techniques. We leave the construction of a tetrad- and coordinate-invariant dynamical curvature perturbation and the definition of definite-parity modes in Petrov type I spacetimes in modified gravity to our future work. 
	
    Besides tetrad rotations at $\mathcal{O}(\zeta^1,\epsilon^1)$ and $\mathcal{O}(\zeta^0,\epsilon^1)$, we can also rotate the tetrad at $\mathcal{O}(\zeta^1,\epsilon^0)$. In this case, 
    \begin{equation}
	    \Psi_0^{(1,1)}\rightarrow \Psi_0^{(1,1)}-2[\delta A^{(1,0)}-i\vartheta^{(1,0)}]\Psi_0^{(0,1)}
	    +4b^{(1,0)}\Psi_1^{(0,1)}\,.
    \end{equation}
    Since $\Psi_{0,1}^{(0,1)}$ are GW perturbations in GR, which are nonzero in general, $\Psi_{0}^{(1,1)}$ is not invariant under tetrad rotations at $\mathcal{O}(\zeta^1,\epsilon^0)$. However, this behavior is very similar to tetrad rotations at $\mathcal{O}(\zeta^0,\epsilon^0)$ in GR, which also shift $\Psi_{0,4}^{(0,1)}$ and correspond to large Lorentz transformations of the background spacetime. In contrast, gauge transformations enter at the same order as GW perturbations, or $\mathcal{O}(\epsilon^1)$ in our notation. For Kerr BHs in GR, the Teukolsky equation is decoupled and separable in Boyer-Lindquist coordinates and in the Kinnersley tetrad, where definite-parity modes can also be easily defined. 
    However, we do not expect the same situation in other coordinates and tetrads in general. Similarly, there are convenient coordinates and tetrads in modified gravity, where parity can be easily studied, such as the tetrad in Eq.~\eqref{eq:tetrad_10_parity}, which transforms in the same way as the Kinnersley tetrad under $\hat{\mathcal{P}}$. 
    We can always perform tetrad rotations at $\mathcal{O}(\zeta^1,\epsilon^0)$, such that the modes defined in Eq.~\eqref{eq:def_parity_modes_simple} are not even of definite-parity, but this only indicates that we are in some frame where some degrees of freedom of these originally definite-parity modes of $\Psi_{0,4}^{(1,1)}$ are shifted to other Weyl scalars such that parity cannot be easily defined for $\Psi_{0,4}^{(1,1)}$. 
    Thus, it is not an issue that our discussion of parity is not invariant under tetrad rotations at $\mathcal{O}(\zeta^1,\epsilon^0)$, and we can stick to the background tetrad in Eqs.~\eqref{eq:tetrad_00_parity} and \eqref{eq:tetrad_10_parity}.
	
    Finally, we can also perform coordinate transformations at $\mathcal{O}(\zeta^1,\epsilon^1)$ and $\mathcal{O}(\zeta^0,\epsilon^1)$. For the same reasons as discussed above, we do not care about coordinate transformations at $\mathcal{O}(\zeta^0,\epsilon^0)$ and $\mathcal{O}(\zeta^1,\epsilon^0)$. Under coordinate transformations $x^{\mu}\rightarrow x^{\mu}+ \epsilon\xi^{\mu(0,1)}+\zeta\epsilon\xi^{\mu(1,1)}$, $\Psi_0^{(1,1)}$ transforms as \cite{Campanelli_Lousto_1999},
    \begin{equation}
	    \Psi_0^{(1,1)}\rightarrow
	    \Psi_0^{(1,1)}+\xi^{\mu(0,1)}\partial_{\mu}\Psi_0^{(1,0)}
	    +\xi^{\mu(1,1)}\partial_{\mu}\Psi_0^{(0,0)}\,. 
    \end{equation}
    For Petrov type D spacetimes in modified gravity, $\Psi_0^{(0,0)}=\Psi_0^{(1,0)} = 0$ , so $\Psi_0^{(1,1)}$ is also coordinate-invariant. Thus, our arguments in Secs.~\ref{sec:parity_property_stationary} and \ref{sec:parity_property_dynamical} to find the transformation rule in Eq.~\eqref{eq:transform_Psi0_perturbed} are both tetrad- and coordinate-invariant. For Petrov type I spacetimes, although one can set $\Psi_0^{(1,0)}\neq0$ by rotating the background tetrad \cite{Chandrasekhar_1983}, we may have to work with the tetrad where $\Psi_0^{(1,0)} = 0$ to preserve our perturbation scheme in Sec.~\ref{sec:bGR_theory}, as argued in \cite{Li:2022pcy}. Thus, we leave the construction of a tetrad- and coordinate-invariant dynamical curvature perturbation to our future work.

    \section{Isospectrality breaking in modified gravity} \label{sec:definite_parity_equations}
	
    In Secs.~\ref{sec:definite_parity_modes_GR} and \ref{sec:definite_parity_modes_bGR}, we have found the QNMs of Weyl scalars that generate definite-parity metric perturbations of Petrov type D spacetimes in GR and modified gravity. In this section, we compute the shift of QNM frequencies of these definite-parity modes using the modified Teukolsky equation found in \cite{Li:2022pcy, Hussain:2022ins}. We first need to extract the source terms having overlaps with the QNMs in GR, which shift the QNM frequencies.
    
    As shown in Sec.~\ref{sec:metric_reconstruct}, metric reconstruction mixes the modes with frequency $\omega$ and $-\bar{\omega}$, so we need to disentangle the source terms with frequency $\omega$ from the terms with frequency $-\bar{\omega}$ within the modified Teukolsky equation. After finding the equations with definite-frequency source terms, we then apply the EVP approach of \cite{Zimmerman:2014aha, Mark_Yang_Zimmerman_Chen_2015, Hussain:2022ins} to compute the shifts in QNMs. In the rest of this section, we show that the solutions form a two-dimensional subspace, the eigenvectors of which are two linear combinations of $(l,m)$ and $(l,-m)$ modes determined by the source terms. The frequencies of these two linear combinations are generally different, so the degeneracy of each $(l,m,\omega)$ mode in GR is broken, like in quantum mechanics, as observed, for example, in dCS gravity \cite{Cardoso:2009pk, Molina:2010fb, Pani_Cardoso_Gualtieri_2011, Wagle_Yunes_Silva_2021, Srivastava_Chen_Shankaranarayanan_2021}, EdGB gravity \cite{Pani:2009wy, Blazquez-Salcedo:2016enn, Blazquez-Salcedo_Khoo_Kunz_2017, Pierini:2021jxd, Pierini:2022eim}, and higher-derivative gravity \cite{Cano:2021myl, Cano:2023tmv, Cano:2023jbk}. Nonetheless, in the special case that the source terms are invariant under the $\hat{\mathcal{P}}$-transformation, these eigenvectors become even- and odd-parity modes. One can thus see this section as a non-trivial extension of Sec.~IVC in \cite{Hussain:2022ins}.
	
    \subsection{Identification of the source terms that shift QNM frequencies}
    \label{sec:identify_QNM_shift}
	
    In this subsection, we extract the source terms within the modified Teukolsky equations having overlaps with the QNMs in GR. In other words, we are interested in the terms of Eqs.~\eqref{eq:source_def_geo} and \eqref{eq:source_def_stress} that are driven by $h_{\mu\nu}^{(0,1)}$ or $\Psi_{0,4}^{(0,1)}$. 
	
    In Sec.~\ref{sec:modified_Teukolsky_equations}, we have discussed that there are two types of source terms. First, the source term $\mathcal{S}_{\geo}^{(1,1)}$ [Eq.~\eqref{eq:source_def_geo}] comes from the correction to the homogeneous part of the Bianchi and Ricci identities, so it is purely geometrical and only depends on corrections to the background geometry. The terms in $\mathcal{S}_{\geo}^{(1,1)}$ take the form of either $H_{i}^{(0,1)}\Psi_{i}^{(1,0)}$ or $H_{i}^{(1,0)}\Psi_{i}^{(0,1)}$, where $H_{i}$ are operators that involve only the metric. Both $H_{i}^{(0,1)}$ and $\Psi_{i}^{(0,1)}$ are driven by $h_{\mu\nu}^{(0,1)}$ and can be reconstructed from the Hertz potential $\Psi_{\Hertz}^{(0,1)}$.

    The next set of source terms is encoded in $\mathcal{S}^{(1,1)}$ [Eq.~\eqref{eq:source_def_stress}], which comes from the effective stress-energy tensor and depends on the details of the modified gravity theory. As discussed in Sec.~\ref{sec:modified_Teukolsky_equations}, this type of source term contains two classes. For bGR theories of class B, there are no extra non-metric fields, so $\mathcal{S}^{(1,1)}$ is driven by $h_{\mu\nu}^{(0,1)}$ directly as shown in detail in \cite{Li:2022pcy}. For bGR theories of class A, there are extra non-metric fields, so we need to be more careful. Using the order-reduction scheme in \cite{Okounkova_Stein_Scheel_Hemberger_2017}, one can argue that all these non-metric fields are driven first by the metric fields in GR \cite{Li:2022pcy}. Thus, for these dynamical extra fields to shift the QNM frequencies, they must be driven by $h_{\mu\nu}^{(0,1)}$. Nonetheless, the homogeneous part of these extra non-metric fields can oscillate at other frequencies, for example, as observed with the scalar perturbations in dCS gravity \cite{Wagle_Yunes_Silva_2021, Okounkova:2019dfo, Okounkova_Stein_Moxon_Scheel_Teukolsky_2020}.
	
    Incorporating only the source terms driven by $h_{\mu\nu}^{(0,1)}$, we can then write the modified Teukolsky equation in Eqs.~\eqref{eq:master_eqn_non_typeD_Psi0}--\eqref{eq:source_def_stress} as
    \begin{equation} \label{eq:modified_Teuk_scheme_1}
		H\Psi^{(1,1)}
		=\mathcal{S}^{\mu\nu}h_{\mu\nu}^{(0,1)}\,,
    \end{equation}
    where $H=H_0^{\GR}$ and $\Psi^{(1,1)}=\Psi_{0}^{(1,1)}$ for $\Psi_0$. The equation that $\Psi_4^{(1,1)}$ satisfies is of the same form, but with the replacements $H_0^{\GR} \to H_4^{\GR}$ and $\Psi_{0}^{(1,1)} \to \Psi_{4}^{(1,1)}$. In Sec.~\ref{sec:metric_reconstruct}, we have shown how to reconstruct $h_{\mu\nu}^{(0,1)}$ from the Hertz potential $\Psi_{\Hertz}^{(0,1)}$, i.e.,
    \begin{equation} \label{eq:metric_Hertz_scheme}
		h_{\mu\nu}^{(0,1)}
		=\left(\mathcal{O}_{\mu\nu}
        +\bar{\mathcal{O}}_{\mu\nu}\hat{\mathcal{C}}\right)\bar{\Psi}_{\Hertz}^{(0,1)}\,.
    \end{equation}
    Here, we have pulled out the complex conjugation operator $\hat{\mathcal{C}}$ acting on $\bar{\Psi}_{\Hertz}^{(0,1)}$ in Eq.~\eqref{eq:metric_Hertz_scheme0} since it transforms any mode with frequency $\omega$ to $-\bar{\omega}$. 
    Furthermore, if one reconstructs $\bar{\Psi}_{\Hertz}^{(0,1)}$ from $\Psi_{0}^{(0,1)}$ in IRG or $\Psi_{4}^{(0,1)}$ in ORG, the operators acting on $\bar{\Psi}_{\Hertz}^{(0,1)}$ do not mix modes with different frequencies [i.e., Eq.~\eqref{eq:reconstruct_Hertz}], as shown in \cite{Ori_2003}. Thus, we can absorb these operators into $\mathcal{O}_{\mu\nu}$ and $\bar{\mathcal{O}}_{\mu\nu}$, so Eq.~\eqref{eq:modified_Teuk_scheme_1} can be further written as
    \begin{equation} \label{eq:modified_Teuk_scheme_2}
		H\Psi^{(1,1)}=\mathcal{S}^{\mu\nu}
	    \left(\mathcal{O}_{\mu\nu}+\bar{\mathcal{O}}_{\mu\nu}\hat{\mathcal{C}}\right)\Psi^{(0,1)}\,.
    \end{equation}
    In principle, one can also have additional operators $\hat{\mathcal{C}}$ within $\mathcal{S}^{\mu\nu}$, but since $h_{\mu\nu}^{(0,1)}$ is real, we do not need to pull them out. 
    In the rest of this section, we take Eq.~\eqref{eq:modified_Teuk_scheme_2} as our modified Teukolsky equation.
	
    \subsection{Degeneracy breaking}
    \label{sec:degeneray_breaking}
    
    In Eq.~\eqref{eq:modified_Teuk_scheme_2}, we notice that the modes with frequency $\omega$ and $-\bar{\omega}$ are mixed due to the operator $\hat{\mathcal{C}}$ acting on $\Psi^{(0,1)}$. Thus, to solve Eq.~\eqref{eq:modified_Teuk_scheme_2}, one generally needs to consider the linear combinations
    \begin{equation} \label{eq:ansatz}
    \begin{aligned}
        \Psi_{\eta}^{(0,1)}
        =& \;\Psi^{(0,1)}+\eta\Psi_{\hat{\mathcal{P}}}^{(0,1)}\,, \\
        \Psi_{\eta}^{(1,1)}
        =& \;\Psi^{(1,1)}+\eta\Psi_{\hat{\mathcal{P}}}^{(1,1)}\,.
    \end{aligned}
    \end{equation}
    Here, $\Psi_{\hat{\mathcal{P}}}^{(0,1)}$ and $\Psi_{\hat{\mathcal{P}}}^{(1,1)}$ are the modes with GR QNM frequency that is the negative complex conjugate of the frequency of $\Psi^{(0,1)}$ and $\Psi^{(1,1)}$ so that we can solve Eq.~\eqref{eq:modified_Teuk_scheme_2} consistently. The constant $\eta$ is some complex number, though it is not completely defined at this moment since one can in principle absorb it into the definition of $\Psi_{\hat{\mathcal{P}}}^{(0,1)}$ and $\Psi_{\hat{\mathcal{P}}}^{(1,1)}$. In Eq.~\eqref{eq:PsiP_ansatz}, we will fix the relative normalization between $\Psi^{(0,1)}$ and $\Psi_{\hat{\mathcal{P}}}^{(0,1)}$, so $\eta$ will become well defined. 
    
    In Eq.~\eqref{eq:ansatz}, the modes $\Psi^{(0,1)}$ and $\Psi^{(1,1)}$ refer to a specific $(l,m,\omega)$ mode of $\mathcal{O}(\zeta^0,\epsilon^1)$ and $\mathcal{O}(\zeta^1,\epsilon^1)$ perturbations of $\Psi_{0,4}$, respectively, i.e.,
    \begin{equation} \label{eq:Psi_expansion}
    \begin{aligned}
        & \Psi^{(0,1)}=\psi_{lm}^{(0,1)}(r,\theta)
        \exp{\left[-i\left(\omega_{lm}^{(0)}
        +\zeta\omega_{lm}^{(1)}\right)t+im\phi\right]}\,, \\
        & \Psi^{(1,1)}=\psi_{lm}^{(1,1)}(r,\theta)
        \exp{\left[-i\left(\omega_{lm}^{(0)}
        +\zeta\omega_{lm}^{(1)}\right)t+im\phi\right]}\,,
    \end{aligned}
    \end{equation}
    where we have suppressed indices corresponding to the spin weight $s$ and the overtone number $n$ in these modes for simplicity. Notice that, in Eq.~\eqref{eq:Psi_expansion}, we have perturbed the frequencies of both the GR QNM $\Psi^{(0,1)}$ and the bGR QNM $\Psi^{(1,1)}$, following the EVP method in \cite{Zimmerman:2014aha, Mark_Yang_Zimmerman_Chen_2015, Hussain:2022ins}. Moreover, the QNM frequency shifts of $\Psi^{(0,1)}$ and $\Psi^{(1,1)}$ are the same; otherwise, the two sides of Eq.~\eqref{eq:modified_Teuk_scheme_2} cannot balance. This approach is the same, in essence, as the Poincar\'{e}-Lindstedt method of solving secular perturbation problems, introducing shifts of the QNM frequency to cancel off secularly growing terms due to the GR QNMs resonantly driving the modified Teukolsky equation. The shift in the QNM frequency plays a similar role to the slow timescale of multiple-scale analysis \cite{bender2013advanced}, which has been applied to spin-precessing systems and post-Newtonian dynamics in GR \cite{Klein:2013qda, Gerosa:2015tea, Chatziioannou:2017tdw}. 
    
    Similarly, the modes $\Psi_{\hat{\mathcal{P}}}^{(0,1)}$ and $\Psi_{\hat{\mathcal{P}}}^{(1,1)}$ correspond to the $(l,-m,-\bar{\omega}_{lm}^{(0)})$ mode of $\Psi^{(0,1)}$ and its perturbations, respectively, i.e.,
    \begin{equation} \label{eq:PsiP_expansion}
    \begin{aligned}
        & \Psi_{\hat{\mathcal{P}}}^{(0,1)}
        =\psi_{\hat{\mathcal{P}}\,l-m}^{(0,1)}(r,\theta)
        \exp{\left[-i\left(-\bar{\omega}_{lm}^{(0)}
        +\zeta\omega_{l-m}^{(1)}\right)t-im\phi\right]}\,, \\
        & \Psi_{\hat{\mathcal{P}}}^{(1,1)}
        =\psi_{\hat{\mathcal{P}}\;l-m}^{(1,1)}(r,\theta)
        \exp{\left[-i\left(-\bar{\omega}_{lm}^{(0)}
        +\zeta\omega_{l-m}^{(1)}\right)t-im\phi\right]}\,,
    \end{aligned}
    \end{equation}
    where we have used that in GR, for any $\omega_{lm}^{(0)}$, there exists a $\omega_{l-m}^{(0)}$ such that $\omega_{l-m}^{(0)}=-\bar{\omega}_{lm}^{(0)}$. The modes $\Psi^{(1,1)}$ and $\Psi_{\hat{\mathcal{P}}}^{(1,1)}$ can be directly mapped to the modes $\psi_{s}^{\pm(2)}$ in \cite{Hussain:2022ins}. In Eq.~\eqref{eq:ansatz}, we have distinguished the mode $\Psi_{\hat{\mathcal{P}}}^{(0,1)}$ and $\Psi_{\hat{\mathcal{P}}}^{(1,1)}$ from $\hat{\mathcal{P}}\Psi^{(0,1)}$ and $\hat{\mathcal{P}}\Psi^{(1,1)}$, the $\hat{\mathcal{P}}$-transformation of $\Psi^{(0,1)}$ and $\Psi^{(1,1)}$, since we do not know the relation between $\omega_{l-m}^{(1)}$ and $-\bar{\omega}_{lm}^{(1)}$ at this stage of the calculation in modified gravity. In the case that $\Psi_{\mathcal{P}}^{(0,1)}=\hat{\mathcal{P}}\Psi^{(0,1)}$ and $\Psi_{\mathcal{P}}^{(1,1)}=\hat{\mathcal{P}}\Psi^{(1,1)}$, the modes $\Psi_{\eta}^{(0,1)}+\zeta\Psi_{\eta}^{(1,1)}$ with $\eta=\pm1$ have definite parity in Petrov type D spacetimes in modified gravity, as we have shown in Sec.~\ref{sec:definite_parity_modes_bGR}.
    
    Inserting the ansatz in Eq.~\eqref{eq:ansatz} into Eq.~\eqref{eq:modified_Teuk_scheme_2}, we find 
    \begin{widetext}
    \begin{equation} \label{eq:general_ansatz}
		H\left(\Psi^{(1,1)}+\eta\Psi_{\hat{\mathcal{P}}}^{(1,1)}\right)
		=\mathcal{S}^{\mu\nu}\left[\left(\mathcal{O}_{\mu\nu}\Psi^{(0,1)}
		+\bar{\eta}\bar{\mathcal{O}}_{\mu\nu}
        \hat{\mathcal{C}}\Psi^{(0,1)}_{\hat{\mathcal{P}}}\right)_{A}
		+\left(\bar{\mathcal{O}}_{\mu\nu}
        \hat{\mathcal{C}}\Psi^{(0,1)}
		+\eta\mathcal{O}_{\mu\nu}\Psi_{\hat{\mathcal{P}}}^{(0,1)}\right)_{B}
		\right]\,,
    \end{equation}
    \end{widetext}
    where the first and last term on the right-hand side of Eq.~\eqref{eq:general_ansatz} come from acting $\mathcal{O}_{\mu\nu}$ in Eq.~\eqref{eq:modified_Teuk_scheme_2} on $\Psi^{(0,1)}$ and  $\Psi_{\hat{\mathcal{P}}}^{(0,1)}$, respectively. The second and third term come from acting $\bar{\mathcal{O}}_{\mu\nu}\hat{\mathcal{C}}$ in Eq.~\eqref{eq:modified_Teuk_scheme_2} on $\Psi^{(0,1)}_{\hat{\mathcal{P}}}$ and $\Psi^{(0,1)}$, respectively.
    We have grouped the first two terms on the right-hand side together (i.e., group $A$) since they have the same GR QNM frequency $\omega_{\ell m}^{(0)}$. Similarly, the last two terms in group $B$ have the same GR QNM frequency $-\bar{\omega}_{\ell m}^{(0)}$. In addition, we also need to match the bGR phase within group $A$ or group $B$. Since the bGR QNM frequency of the first and second term in the group $A$ is $\omega_{\ell m}^{(1)}$ and $-\bar{\omega}_{l-m}^{(1)}$, respectively, we have to impose
    \begin{equation} \label{eq:freq_constraint}
        \omega_{l-m}^{(1)}=-\bar{\omega}_{l m}^{(1)}\,. 
    \end{equation}
    The same constraint can also be obtained by requiring the terms in group $B$ to have the same bGR phase. After imposing Eq.~\eqref{eq:freq_constraint}, the phase of $H\Psi^{(1,1)}$ and $H\Psi_{\hat{\mathcal{P}}}^{(1,1,)}$ also match the phase of group $A$ and $B$, respectively, so Eq.~\eqref{eq:general_ansatz} is completely balanced and solvable. The frequency of $\Psi_{\hat{\mathcal{P}}}^{(0,1)}$ is now the complex conjugate of the frequency of $\Psi^{(0,1)}$. Since $\Psi_{\hat{\mathcal{P}}}^{(0,1)}$ is a $(l,-m)$ mode of the solution to the Teukolsky equation in GR, and $\hat{\mathcal{P}}\Psi_{lm\omega}^{(0,1)}=(-1)^l\Psi_{l-m-\bar{\omega}}$ [i.e., Eq.~\eqref{eq:PPsi_Kerr}], we can conveniently choose
    \begin{equation} \label{eq:PsiP_ansatz}
        \Psi_{\hat{\mathcal{P}}}^{(0,1)}=\hat{\mathcal{P}}\Psi^{(0,1)}
    \end{equation}
    such that Eq.~\eqref{eq:general_ansatz} becomes
    \begin{widetext}
    \begin{equation} \label{eq:general_ansatz2}
        H\left(\Psi^{(1,1)}+\eta\Psi_{\hat{\mathcal{P}}}^{(1,1)}\right)
		=\mathcal{S}^{\mu\nu}\left[\left(\mathcal{O}_{\mu\nu}
		+\bar{\eta}\bar{\mathcal{O}}_{\mu\nu}
        \hat{\mathcal{C}}\hat{\mathcal{P}}\right)\Psi^{(0,1)}
		+\left(\bar{\mathcal{O}}_{\mu\nu}
        \hat{\mathcal{C}}\hat{\mathcal{P}}
		+\eta\mathcal{O}_{\mu\nu}\right)
		\hat{\mathcal{P}}\Psi^{(0,1)}\right]\,,
    \end{equation}
    \end{widetext}
    where we have pulled out a factor of $\hat{\mathcal{P}}$ in the second and third term using $\hat{\mathcal{P}}^2=1$. Notice, the operator $\hat{\mathcal{C}}\hat{\mathcal{P}}=\hat{P}$ does not change the frequency of the mode it acts on.

    Separating the equation into two parts for $\Psi^{(1,1)}$ and $\Psi_{\hat{\mathcal{P}}}^{(1,1)}$, we find
    \begin{subequations} \label{eq:general_eqns}
    \begin{align}
		& H\Psi^{(1,1)}
		=\mathcal{S}^{\mu\nu}\left(\mathcal{O}_{\mu\nu}
		+\bar{\eta}\bar{\mathcal{O}}_{\mu\nu}\hat{\mathcal{C}}
        \hat{\mathcal{P}}\right)\Psi^{(0,1)}\,, \label{eq:general_Psi} \\
		& \eta H\Psi_{\hat{\mathcal{P}}}^{(1,1)}
		=\mathcal{S}^{\mu\nu}\left(\eta\mathcal{O}_{\mu\nu}
		+\bar{\mathcal{O}}_{\mu\nu}
        \hat{\mathcal{C}}\hat{\mathcal{P}}\right)
        \hat{\mathcal{P}}\Psi^{(0,1)}\,.
		\label{eq:general_PsiP}
    \end{align}
    \end{subequations}
    Acting $\hat{\mathcal{P}}$ on Eq.~\eqref{eq:general_eqns}, we also find
    \begin{subequations} \label{eq:general_eqns_P}
    \begin{align}
		& H\left(\hat{\mathcal{P}}\Psi^{(1,1)}\right)
		=(\hat{\mathcal{P}}\mathcal{S}^{\mu\nu})\left(\mathcal{O}_{\mu\nu}
        +\eta\bar{\mathcal{O}}_{\mu\nu}
        \hat{\mathcal{C}}\hat{\mathcal{P}}\right)
        \hat{\mathcal{P}}\Psi^{(0,1)}\,, \label{eq:general_PPsi} \\
		& \bar{\eta}H\left(\hat{\mathcal{P}}\Psi_{\hat{\mathcal{P}}}^{(1,1)}\right)
		=(\hat{\mathcal{P}}\mathcal{S}^{\mu\nu})
        \left(\bar{\eta}\mathcal{O}_{\mu\nu}
		+\bar{\mathcal{O}}_{\mu\nu}
        \hat{\mathcal{C}}\hat{\mathcal{P}}\right)\Psi^{(0,1)}\,, \label{eq:general_PPsiP}
    \end{align}
    \end{subequations}
    where we have used that
    \begin{equation} \label{eq:PO}
	    \hat{\mathcal{P}}H=H\,,\quad
	    \hat{\mathcal{P}}\mathcal{O}_{\mu\nu}=\mathcal{O}_{\mu\nu}\,.
    \end{equation}
     
    One can notice that Eq.~\eqref{eq:general_eqns} is redundant with respect to Eq.~\eqref{eq:general_eqns_P}, since the latter is just a $\hat{\mathcal{P}}$-transformation of the former, both of which have to be satisfied simultaneously. Therefore, one can either solve the pair of Eqs.~\eqref{eq:general_Psi} and \eqref{eq:general_PPsiP} or the pair of Eqs.~\eqref{eq:general_PsiP} and \eqref{eq:general_PPsi}. Let us focus on the first pair and apply the EVP method in \cite{Zimmerman:2014aha, Mark_Yang_Zimmerman_Chen_2015, Hussain:2022ins} to remove the wave functions at $\mathcal{O}(\zeta^1,\epsilon^1)$. References~\cite{Zimmerman:2014aha, Mark_Yang_Zimmerman_Chen_2015, Hussain:2022ins} have defined an inner product 
    \begin{equation} \label{eq:self_adjoin_product}
	    \langle \tilde{H}\psi(r,\theta)|\varphi(r,\theta)\rangle
        =\langle \psi(r,\theta)|\tilde{H}\varphi(r,\theta)\rangle\,,
    \end{equation}
    that makes the Teukolsky operator $H$ in GR self-adjoint,
    where $\tilde{H}$ is the harmonic decomposition of $H$ into modes $e^{-i\omega t+im\phi}$. The functions $\psi(r,\theta)$ and $\varphi(r,\theta)$ are the $(r,\theta)$ parts of any modes satisfying the QNM boundary conditions, such as $\psi_{lm}^{(0,1)}(r,\theta)$ and $\psi_{lm}^{(1,1)}(r,\theta)$ in Eq.~\eqref{eq:Psi_expansion}. Expanding $\omega_{\ell m}$ about $\zeta$ on the left-hand side of Eqs.~\eqref{eq:general_Psi} and \eqref{eq:general_PPsiP} and taking the inner product [i.e., Eq.~\eqref{eq:self_adjoin_product}] of these two equations with $\Psi^{(0,1)}$, we find
    \begin{widetext}
    \begin{equation} \label{eq:matrix_general}
        \frac{1}{\langle \partial_{\omega}\tilde{H}\rangle_{lm}}
	    \begin{pmatrix}
            \langle \mathcal{S}^{\mu\nu}
            \mathcal{O}_{\mu\nu}\rangle_{lm}
            & \langle \mathcal{S}^{\mu\nu}
            \mathcal{\bar{O}}_{\mu\nu}
            \hat{\mathcal{C}}\hat{\mathcal{P}}\rangle_{lm} \\
            \langle (\hat{\mathcal{P}}\mathcal{S}^{\mu\nu})
            \mathcal{\bar{O}}_{\mu\nu}
            \hat{\mathcal{C}}\hat{\mathcal{P}}\rangle_{lm}
            & \langle (\hat{\mathcal{P}}\mathcal{S}^{\mu\nu})
            \mathcal{O}_{\mu\nu} \rangle_{lm}
        \end{pmatrix}
        \begin{pmatrix}
            1 \\ \bar{\eta}
        \end{pmatrix}
        =\omega_{lm}^{(1)}\begin{pmatrix}
            1 \\ \bar{\eta}
        \end{pmatrix}\,,
    \end{equation}    
    \end{widetext}
    where we have defined the shorthand notation
    \begin{equation} \label{eq:inner_product_shorthand}
	    \langle \mathcal{O}\rangle_{lm}
	    =\langle\psi_{lm}^{(0,1)}|\tilde{\mathcal{O}}
        \psi^{(0,1)}_{lm}\rangle\,,
    \end{equation}
    where $\tilde{\mathcal{O}}$ is the harmonic decomposition of the operator $\mathcal{O}$ into modes $e^{-i\omega t+im\phi}$. In Eq.~\eqref{eq:matrix_general}, to remove $\psi_{lm}^{(1,1)}$, we have used that $\psi_{lm}^{(0,1)}$ solves the Teukolsky equation in GR, i.e., $\tilde{H}\psi_{lm}^{(0,1)}=0$, such that $\langle\psi_{lm}^{(0,1)}|\tilde{H}\psi_{lm}^{(1,1)}\rangle=0$.

    Equation~\eqref{eq:matrix_general} is a standard eigenvalue problem in degenerate perturbation theory. One can either calculate the eigenvalues of Eq.~\eqref{eq:matrix_general} first or follow \cite{Hussain:2022ins} to solve for $\bar{\eta}$ first. Multiplying the first equation in Eq.~\eqref{eq:matrix_general} by $\bar{\eta}$ and equating the left-hand side of it to the left-hand side  of the second equation, we find a quadratic equation in $\bar{\eta}$,
    \begin{widetext}
    \begin{equation} \label{eq:eta_eqn}
	    \bar{\eta}^2\left\langle\mathcal{S}^{\mu\nu}
	    \bar{\mathcal{O}}_{\mu\nu}
        \hat{\mathcal{C}}\hat{\mathcal{P}}\right\rangle_{lm}
	    +\bar{\eta}\left\langle\left[\mathcal{S}^{\mu\nu}
	    -(\hat{\mathcal{P}}\mathcal{S}^{\mu\nu})\right]
	    \mathcal{O}_{\mu\nu}\right\rangle_{lm}
	    -\left\langle(\hat{\mathcal{P}}\mathcal{S}^{\mu\nu})
	    \bar{\mathcal{O}}_{\mu\nu}\hat{\mathcal{C}}
        \hat{\mathcal{P}}\right\rangle_{lm}=0\,.
    \end{equation} 
    \end{widetext}
    Since Eq.~\eqref{eq:eta_eqn} is quadratic, $\bar{\eta}$ has two solutions $\bar{\eta}_1$ and $\bar{\eta}_2$ that can be computed in terms of these inner products. For each solution to $\bar{\eta}$, one finds a correction $\omega_{lm}^{(1)}$ to the frequency of the mode $(l,m,\omega_{lm}^{(0)})$ in GR,
    \begin{equation} \label{eq:QNM_freq}
	    \omega_{i,lm}^{(1)}
	    =\frac{\left\langle\mathcal{S}^{\mu\nu}\left(\mathcal{O}_{\mu\nu}
		+\bar{\eta}_i\bar{\mathcal{O}}_{\mu\nu}\hat{\mathcal{C}}
        \hat{\mathcal{P}}\right)\right\rangle_{lm}}
		{\langle\partial_{\omega}\tilde{H}\rangle_{lm}}\,,\quad
        i\in\{1,2\}\,.
    \end{equation}
    
    One can take the difference of $\omega_{1,lm}^{(1)}$ and $\omega_{2,lm}^{(1)}$ to characterize the degree of isospectrality breaking, i.e.,
    \begin{equation}
        \delta \omega_{lm}^{(1)}
        =\omega_{1,lm}^{(1)}-\omega_{2,lm}^{(1)}
	    =\left(\bar{\eta}_1-\bar{\eta}_2\right)
        \frac{\left\langle\mathcal{S}^{\mu\nu}
        \bar{\mathcal{O}}_{\mu\nu}\hat{\mathcal{C}}
        \hat{\mathcal{P}}\right\rangle_{lm}}
		{\langle\partial_{\omega}\tilde{H}\rangle_{lm}}\,.
    \end{equation}
    For isospectrality to be preserved beyond GR [i.e., $\delta\omega_{lm}^{(1)}=0$], there are two general possibilities: $\left\langle\mathcal{S}^{\mu\nu}
    \bar{\mathcal{O}}_{\mu\nu}\hat{\mathcal{C}}
    \hat{\mathcal{P}}\right\rangle_{lm}=0$ or $\bar{\eta}_1=\bar{\eta}_2\neq 0$. For the first possibility, Eq.~\eqref{eq:eta_eqn} becomes linear. In addition, $\left\langle\mathcal{S}^{\mu\nu}
    \bar{\mathcal{O}}_{\mu\nu}\hat{\mathcal{C}}
    \hat{\mathcal{P}}\right\rangle_{lm}$ and $\left\langle(\hat{\mathcal{P}}\mathcal{S}^{\mu\nu})
    \bar{\mathcal{O}}_{\mu\nu}\hat{\mathcal{C}}
    \hat{\mathcal{P}}\right\rangle_{lm}$ have to vanish simultaneously for the following reason. If $\bar{\eta}=0$, this condition is satisfied automatically from Eq.~\eqref{eq:eta_eqn}. If $\bar{\eta}\neq 0$, for any $(l,m)$, Eq.~\eqref{eq:eta_eqn} needs to be satisfied for $(l,m)$ and $(l,-m)$ together, since we jointly solve for these two modes. More specifically, one can set $\Psi^{(0,1)}$ and $\Psi^{(1,1)}$ to be the $(l,-m)$ mode in Eq.~\eqref{eq:ansatz}, repeat the same argument above, and replace $\langle\cdots\rangle_{lm}$ in Eq.~\eqref{eq:eta_eqn} with $\langle \cdots\rangle_{l-m}$ in the end. Due to $\hat{\mathcal{P}}\Psi^{(0,1)}_{lm}=(-1)^l\Psi^{(0,1)}_{l-m}$ [Eq.~\eqref{eq:PPsi_Kerr}] and Eq.~\eqref{eq:PsiP_ansatz}, the $(l,-m)$ mode of $\Psi_{\eta}^{(0,1)}$ in Eq.~\eqref{eq:ansatz} only differs from the $(l,m)$ mode by an overall constant, so $\omega_{lm}^{(1)}=\omega_{l-m}^{(1)}$, and any constraint on isospectrality must be redundant for $(l,\pm m)$. Thus, we must simultaneously have $\left\langle\mathcal{S}^{\mu\nu}
    \bar{\mathcal{O}}_{\mu\nu}\hat{\mathcal{C}}\hat{\mathcal{P}}
    \right\rangle_{l\pm m}=0$. Furthermore, one can derive an equation for $\eta$ using Eqs.~\eqref{eq:general_PsiP} and \eqref{eq:general_PPsi} instead, which results in a $\hat{\mathcal{P}}$-transformation of Eqs.~\eqref{eq:eta_eqn}, leading to $\left\langle(\hat{\mathcal{P}}\mathcal{S}^{\mu\nu})
    \bar{\mathcal{O}}_{\mu\nu}\hat{\mathcal{C}}
    \hat{\mathcal{P}}\right\rangle_{l\mp m}=0$, where we use again $\hat{\mathcal{P}}\Psi^{(0,1)}_{lm}=(-1)^l\Psi^{(0,1)}_{l-m}$. In this case, the matrix in Eq.~\eqref{eq:matrix_general} becomes diagonal, so its eigenvectors are $(1,0)$ and $(0,1)$. The first eigenvector can be directly found from Eq.~\eqref{eq:eta_eqn}. The second eigenvector is not directly captured by Eq.~\eqref{eq:eta_eqn}, since we have fixed the normalization of $\Psi^{(0,1)}$ and $\Psi^{(1,1)}$ to be unity in Eq.~\eqref{eq:ansatz}. Nonetheless, Eqs.~\eqref{eq:matrix_general} and \eqref{eq:eta_eqn} need to be satisfied for $(l,-m)$, the solution to which corresponds to the second eigenvector. This indicates that the $(l,m)$ and $(l,-m)$ modes decouple in Eq.~\eqref{eq:general_ansatz2}. 
    
    For $\left\langle\mathcal{S}^{\mu\nu}
    \bar{\mathcal{O}}_{\mu\nu}\hat{\mathcal{C}}
    \hat{\mathcal{P}}\right\rangle_{lm}=0$, there are several sub-cases. First, $\mathcal{S}^{\mu\nu}$ annihilates $\bar{\mathcal{O}}_{\mu\nu}$, indicating that the source terms do not contain any complex conjugation $\hat{\mathcal{C}}$ acting on the GR QNMs according to Eq.~\eqref{eq:modified_Teuk_scheme_2}. This condition usually cannot happen since most operators in the source terms $\mathcal{S}^{(1,1)}$ and $\mathcal{S}_{\geo}^{(1,1)}$ [i.e., Eqs.~\eqref{eq:Bianchi_simplified}, \eqref{eq:simplify_values}, and \eqref{eq:def_operators}] contain both the NP quantities and their complex conjugates. Furthermore, as we will see in Sec.~\ref{sec:application} and Appendix~\ref{appendix:source_terms}, the Ricci tensor in the NP basis in many bGR theories also contains both types of terms. At $\mathcal{O}(\zeta^0,\epsilon^1)$, this mixing of NP quantities and their complex conjugates results in a mixing of terms proportional to $\Psi_{0}^{(0,1)}$ and $\bar{\Psi}_{0}^{(0,1)}$, where $\mathcal{S}^{\mu\nu}$ cannot annihilate $\bar{\mathcal{O}}_{\mu\nu}$, as one can observe in the reconstructed quantities in Sec.~\ref{sec:metric_reconstruct} and Appendix~\ref{appendix:metric_reconstruction_more}. 
    
    One exception is when the source terms in Eq.~\eqref{eq:modified_Teuk_scheme_2} only contain $\mathcal{S}_{\geo}^{(1,1)}$, and the bGR background spacetime is Petrov type D, so $\mathcal{S}_{\geo}^{(1,1)}$ takes the form $\mathcal{S}_{\geo}^{(1,1)}=-H_{0}^{(1,0)}\Psi_{0}^{(0,1)}$ [Eq.~\eqref{eq:source_def_geo}]. In this case, no metric reconstruction is needed, and no complex conjugation $\hat{\mathcal{C}}$ acts on $\Psi_0$, so isospectrality is preserved for this correction, as we will discuss in more detail in Sec.~\ref{sec:application}.

    Second, $\mathcal{S}^{\mu\nu}\bar{\mathcal{O}}_{\mu\nu}$ annihilates the mode $\psi_{lm}^{(0,1)}$ or $\hat{\mathcal{C}}\hat{\mathcal{P}}\psi_{lm}^{(0,1)}$. This can happen, for example, when $\mathcal{S}^{\mu\nu}\bar{\mathcal{O}}_{\mu\nu}$ is proportional to the Teukolsky operator (or with additional operators acting on it). Third, the mode $\mathcal{S}^{\mu\nu}\bar{\mathcal{O}}_{\mu\nu}\hat{\mathcal{C}}\hat{\mathcal{P}}\psi_{lm}^{(0,1)}$ and the mode $\psi_{lm}^{(0,1)}$ are orthogonal, so their inner product vanishes. For example, if $\mathcal{S}^{\mu\nu}\bar{\mathcal{O}}_{\mu\nu}\hat{\mathcal{C}}\hat{\mathcal{P}}$ shifts the $l$ of $\psi_{lm}^{(0,1)}$, the two modes are orthogonal due to the orthogonality of spin-weighted spheroidal harmonics.

    For the second possibility that $\bar{\eta}_1=\bar{\eta}_2\neq0$, we get a constraint on $\mathcal{S}^{\mu\nu}$ using Eq.~\eqref{eq:eta_eqn}, i.e,
    \begin{widetext}
    \begin{equation} \label{eq:eta_constraint}
        \left\langle\left[\mathcal{S}^{\mu\nu}
	    -(\hat{\mathcal{P}}\mathcal{S}^{\mu\nu})\right]
	    \mathcal{O}_{\mu\nu}\right\rangle_{lm}^2
        +4\left\langle\mathcal{S}^{\mu\nu}
	    \bar{\mathcal{O}}_{\mu\nu}
        \hat{\mathcal{C}}\hat{\mathcal{P}}\right\rangle_{lm}
        \left\langle(\hat{\mathcal{P}}\mathcal{S}^{\mu\nu})
	    \bar{\mathcal{O}}_{\mu\nu}\hat{\mathcal{C}}
        \hat{\mathcal{P}}\right\rangle_{lm}=0\,,\quad
        \left\langle\mathcal{S}^{\mu\nu}
	    \bar{\mathcal{O}}_{\mu\nu}
        \hat{\mathcal{C}}\hat{\mathcal{P}}\right\rangle_{lm}\neq0\,.
    \end{equation}
    \end{widetext}
    Since $\left\langle\mathcal{S}^{\mu\nu}
    \bar{\mathcal{O}}_{\mu\nu}\hat{\mathcal{C}}
    \hat{\mathcal{P}}\right\rangle_{lm}$ and $\left\langle(\hat{\mathcal{P}}\mathcal{S}^{\mu\nu})
    \bar{\mathcal{O}}_{\mu\nu}\hat{\mathcal{C}}
    \hat{\mathcal{P}}\right\rangle_{lm}$ have to vanish simultaneously, the last term in Eq.~\eqref{eq:eta_constraint} never vanishes. This indicates that even when the source term is preserved under parity, i.e., $\hat{\mathcal{P}}\mathcal{S}^{\mu\nu}=\mathcal{S}^{\mu\nu}$, isospectrality can still break, as we will discuss in more detail in Sec.~\ref{sec:solns_definite_parity}. Therefore, the only way to have Eq.~\eqref{eq:eta_constraint} satisfied is for the second term to cancel with the first term. Since $\mathcal{O}_{\mu\nu}$ and $\bar{\mathcal{O}}_{\mu\nu}$ are fixed by the metric reconstruction procedures, we can only tweak $\mathcal{S}^{\mu\nu}$. However, it is almost impossible to construct such a $\mathcal{S}^{\mu\nu}$ consistently for all $(l,m)$, since $\mathcal{S}^{\mu\nu}$ is symmetric with only $10$ components. Moreover, when $\bar{\eta}_1=\bar{\eta}_2\neq0$, the matrix in Eq.~\eqref{eq:matrix_general} becomes rank one, which indicates that the full solution to Eq.~\eqref{eq:general_eqns} may also contain modes that are not harmonic in time, similar to the critically damping case of a simple harmonic oscillator.
     
    Except for these special cases, the isospectrality of even- and odd-parity modes in the QNM frequencies at each $(l,m)$ is broken by modified gravity corrections. For instance, in Sec.~\ref{sec:application}, we will see that neither $\bar{\eta}_1=\bar{\eta}_2$ nor $\mathcal{S}^{\mu\nu}$ annihilates $\bar{\mathcal{O}}_{\mu\nu}$ (after summing up the contribution from both $\mathcal{S}^{(1,1)}$ and $\mathcal{S}_{\geo}^{(1,1)}$) in all the examples.
    
    A similar analysis specifically for higher-derivative gravity was done by \cite{Cano:2021myl} using metric perturbations, and by \cite{Cano:2023tmv, Cano:2023jbk} using the modified Teukolsky equation. Here, by following \cite{Hussain:2022ins}, we provide a more general equation of QNM frequency shifts [e.g., Eq.~\eqref{eq:QNM_freq}], valid for a broad class of modified gravity theories. Our results are consistent with \cite{Hussain:2022ins}, but simplified using the parity properties of $H$ and $\mathcal{O}_{\mu\nu}$. This allows for a systematic study of the relation between modified gravity corrections and the structure of isospectrality breaking.

    \subsection{Solutions with definite parity}
    \label{sec:solns_definite_parity}

    One may also want to know when the modified Teukolsky equation still admits definite-parity solutions, i.e., solutions for which $\hat{\mathcal{P}}\Psi_{\Even,\Odd}=\pm(-1)^l\Psi_{\Even,\Odd}$ [i.e., Eq.~\eqref{eq:transform_Hertz_parity}]. For this reason, let us consider the special case that $\eta=\pm(-1)^l$ and $\Psi_{\hat{\mathcal{P}}}^{(1,1)}=\hat{\mathcal{P}}\Psi^{(1,1)}$, which corresponds to even- and odd-parity modes for Petrov type D spacetimes in modified gravity, as shown in Sec.~\ref{sec:definite_parity_modes_bGR}. Inserting $\eta=\pm(-1)^l$ into Eq.~\eqref{eq:general_Psi}, we find the definite-parity modified Teukolsky equations to be
    \begin{equation} \label{eq:definite_parity_eqn}
	    H\Psi_{\Even,\Odd}^{(1,1)}
		=\mathcal{S}^{\mu\nu}\left(\mathcal{O}_{\mu\nu}
		\pm(-1)^l\bar{\mathcal{O}}_{\mu\nu}
        \hat{\mathcal{C}}\hat{\mathcal{P}}\right)\Psi^{(0,1)}\,.
    \end{equation}
    The solutions to Eq.~\eqref{eq:definite_parity_eqn} can then be obtained from Eq.~\eqref{eq:QNM_freq}, i.e.,
    \begin{equation} \label{eq:QNM_freq_parity}
	    \omega_{lm}^{\Even,\Odd(1)}
	    =\frac{\left\langle\mathcal{S}^{\mu\nu}
        \left(\mathcal{O}_{\mu\nu}
		\pm(-1)^l\bar{\mathcal{O}}_{\mu\nu}
        \hat{\mathcal{C}}\hat{\mathcal{P}}\right)\right\rangle_{lm}}
		{\langle\partial_{\omega}\tilde{H}\rangle_{lm}}\,.
    \end{equation}
    On the other hand, Eqs.~\eqref{eq:general_Psi} and \eqref{eq:general_PPsiP} need to be satisfied simultaneously, and we have only used Eq.~\eqref{eq:general_Psi} to get Eq.~\eqref{eq:QNM_freq_parity}. From Eq.~\eqref{eq:general_PPsiP}, we similarly find
    \begin{equation} \label{eq:QNM_freq_parity_P}
        \omega_{lm}^{\Even,\Odd(1)}
	    =\frac{\left\langle\hat{\mathcal{P}}
        \mathcal{S}^{\mu\nu}\left(\mathcal{O}_{\mu\nu}
		\pm(-1)^l\bar{\mathcal{O}}_{\mu\nu}
        \hat{\mathcal{C}}\hat{\mathcal{P}}\right)\right\rangle_{lm}}
		{\langle\partial_{\omega}\tilde{H}\rangle_{lm}}\,.
    \end{equation}
    Comparing Eqs.~\eqref{eq:QNM_freq_parity_P} to \eqref{eq:QNM_freq_parity}, we find a constraint on $\mathcal{S}^{\mu\nu}$, i.e.,
    \begin{equation} \label{eq:constraint_on_S}
	    \hat{\mathcal{P}}\mathcal{S}^{\mu\nu}=\mathcal{S}^{\mu\nu}\,,
    \end{equation}
    which implies that the source term $\mathcal{S}^{\mu\nu}$ needs to transform in the same way as the Teukolsky operator $H$ in GR under $\hat{\mathcal{P}}$. On the other hand, if one assumes $\hat{\mathcal{P}}\mathcal{S}^{\mu\nu}=\mathcal{S}^{\mu\nu}$, one finds that $\eta=\pm 1$ using Eq.~\eqref{eq:eta_eqn} and $\Psi_{\hat{\mathcal{P}}}^{(1,1)}=\hat{\mathcal{P}}\Psi^{(1,1)}$ using Eqs.~\eqref{eq:general_Psi} and \eqref{eq:general_PPsiP}. Thus, for Petrov type D spacetimes in modified gravity, the solutions to the modified Teukolsky equation generate definite-parity perturbations if and only if $\hat{\mathcal{P}}\mathcal{S}^{\mu\nu}=\mathcal{S}^{\mu\nu}$.
	
    The constraint in Eq.~\eqref{eq:constraint_on_S} is closely related to how one diagonalizes the correction to the Hamiltonian for degenerate systems in quantum mechanics. For degenerate perturbation theory in quantum mechanics, the modes that naturally diagonalize the perturbed Hamiltonian are the eigenvectors of a certain Hermitian operator that commutes with both the original Hamiltonian and the perturbation to the Hamiltonian. In our case, the operator $\hat{\mathcal{P}}$ commutes with both the Teukolsky equation in GR and the modified Teukolsky equation, since according to Eqs.~\eqref{eq:PO} and \eqref{eq:constraint_on_S},
    \begin{subequations}
    \begin{align}
	    & [\hat{\mathcal{P}},H]f
	    =(\hat{\mathcal{P}}H)(\hat{\mathcal{P}}f)-H(\hat{\mathcal{P}}f)=0\,, \\
	    & [\hat{\mathcal{P}},\mathcal{S}^{\mu\nu}]f
	    =(\hat{\mathcal{P}}\mathcal{S}^{\mu\nu})(\hat{\mathcal{P}}f)
	    -\mathcal{S}^{\mu\nu}(\hat{\mathcal{P}}f)=0\,.
    \end{align}
    \end{subequations}
    Thus, the even- and odd-parity modes naturally ``diagonalize" the modified Teukolsky equation when $\hat{\mathcal{P}}\mathcal{S}^{\mu\nu}=\mathcal{S}^{\mu\nu}$. 
    In more general cases, when $\hat{\mathcal{P}}$ does not commute with the source terms, one must diagonalize manually as in Sec.~\ref{sec:degeneray_breaking}. In the next section, we apply the analysis developed in this section to two specific modified gravity theories: dCS and EdGB gravity.

    \section{Application} \label{sec:application}

    In this section, we apply the formalism above to specific corrections to the Teukolsky equation for two relatively simple examples. In particular, we consider two widely studied modified gravity theories: dCS and EdGB gravity. We will not present the details of these two theories here, since one can find them in \cite{Jackiw:2003pm, Alexander:2009tp, Yunes:2009hc, Yagi:2012ya, Cardoso:2009pk, Molina:2010fb, Pani_Cardoso_Gualtieri_2011, Wagle_Yunes_Silva_2021, Srivastava_Chen_Shankaranarayanan_2021} for dCS and in \cite{Kanti:1995vq, Pani:2009wy, Yunes:2011we, Blazquez-Salcedo:2016enn, Blazquez-Salcedo_Khoo_Kunz_2017, Witek:2018dmd, Pierini:2021jxd, Pierini:2022eim} for EdGB gravity. We choose to follow the convention of the action in \cite{Yunes:2009hc} for dCS and \cite{Pierini:2021jxd} for EdGB theory. As discussed in Secs.~\ref{sec:modified_Teukolsky_equations} and \ref{sec:definite_parity_equations}, modifications to the Teukolsky equation generally originate from two different places:
    \begin{enumerate}
	    \item The modification to the background geometry, e.g., $\mathcal{S}_{\geo}^{(1,1)}$.
	    \item The effective stress-energy tensor specific to each modified gravity theory, e.g., $\mathcal{S}^{(1,1)}$.
    \end{enumerate}
    For all these examples of modified gravity theories, we discuss the leading contribution to both types of source terms. In this work, we also choose to focus on the Petrov type D backgrounds and leave the generalization to the non-Petrov-type D case for future work. This implies we must focus only on slowly rotating BHs in dCS and EdGB gravity to linear order in rotation. Thus, we must carry out an additional expansion in the dimensionless spin $\chi=a/M$, such that all the quantities are expanded as
    \begin{equation}
        \Psi=\Psi^{(0,0,0)}+\zeta\Psi^{(1,0,0)}+\chi\Psi^{(0,1,0)}
        +\epsilon\Psi^{(0,0,1)}+\cdots\,.
    \end{equation}
    For simplicity, we also focus on the equation that governs the evolution of $\Psi_0$, while the equation for $\Psi_4$ can be easily obtained by a GHP transformation \cite{Li:2022pcy}.
	
    \subsection{dCS gravity} \label{sec:dCS}
	
    In this theory, there is no correction to the background geometry for non-rotating BHs, since the Pontryagin density vanishes for spherically-symmetric spacetimes \cite{Yunes:2009hc}. In this case, the leading order correction to the background geometry enters at $\mathcal{O}(\zeta^1,\chi^1,\epsilon^0)$ with
    \begin{equation} \label{eq:dCS_coupling}
	    \zeta=\zeta_{\dCS}\equiv\frac{\alpha_\dCS^2}{\kappa_g M^4}\,,\quad
        \kappa_g\equiv\frac{1}{16\pi G}\,,
    \end{equation}
    where $\alpha_{\dCS}$ is the coupling constant of dCS gravity, and we have chosen the coupling constant of the pseudoscalar field action $\beta=1$. For simplicity, we will drop the subscript labeling the modified gravity theory. Then, the leading-order contribution to $\mathcal{S}_{\geo}^{(1,1)}$ enters at $\mathcal{O}(\zeta^1,\chi^1,\epsilon^1)$. Since slowly rotating BHs at $\mathcal{O}(\zeta^1,\chi^1,\epsilon^0)$ in dCS gravity are of Petrov type D \cite{Yunes:2009hc}, $\mathcal{S}_{\geo}^{(1,1,1)}$ only depends on the correction to the Teukolsky operator $H_{0}^{(1,1,0)}$. On the other hand, since the pseudoscalar field $\vartheta$ is driven by the GW perturbations, there is a nonzero effective stress tensor at $\mathcal{O}(\zeta^1,\chi^0,\epsilon^1)$ \cite{Cardoso:2009pk, Molina:2010fb, Pani_Cardoso_Gualtieri_2011}. Thus, the leading contribution to $\mathcal{S}^{(1,1)}$ is $\mathcal{S}^{(1,0,1)}$.
	
    \subsubsection{Correction due to $\mathcal{S}_{\geo}^{(1,1)}$ in dCS}
    \label{sec:dCS_S_geo}
	
    Expanding the Teukolsky equation in GR to $\mathcal{O}(\chi^1)$, we first find  
    \begin{align}
	    & H_0^{(0,0)}=H_0^{(0,0,0)}+\chi H_0^{(0,1,0)}+\cdots\,, \nonumber\\
	    & \begin{aligned}
		H_{0}^{(0,0,0)}
		=& \;\frac{1}{r-r_s}\left[-6(r-r_s)+4r(r-3M)\partial_{t}
		+r^3\partial_{t}^2\right] \\
		& \;-6(r-M)\partial_{r}-r(r-r_s)\partial_{r}^2
		-\cot{\theta}\partial_{\theta}-\partial_{\theta}^{2} \\
		& \;+\csc{\theta}^2\left(4-4i\cos{\theta}\partial_{\phi}
		-\partial_{\phi}^{2}\right)\,, \nonumber\\
        \end{aligned} \\
        & \begin{aligned}
	H_{0}^{(0,1,0)}
		=& \;8Mr^2\left\{\frac{1}{r(r-r_s)}
		\left[(r-M)\partial_{\phi}+Mr\partial_t\partial_{\phi}\right]\right. \\
		& \;\left.-i\cos{\theta}\partial_t\right\}\,,
		\end{aligned}
    \end{align}
    where $r_s=2M$ is the Schwarzschild radius, $H_{0}^{(0,0,0)}$ is the Teukolsky operator on the Schwarzschild background in GR, and $H_{0}^{(0,1,0)}$ is the leading slow-rotation correction to $H_{0}^{(0,0,0)}$. We have also restored the full coordinate dependence of these operators here. Under the $\hat{\mathcal{P}}$ transformation, we find that $\hat{\mathcal{P}}H_{0}^{(0,0,0)}=H_{0}^{(0,0,0)}$, so the Teukolsky equation on the Schwarzschild background in GR admits definite-parity solutions. In addition, $\hat{\mathcal{P}}H_{0}^{(0,1,0)}=H_{0}^{(0,1,0)}$, which is expected since the Kerr background admits GW perturbations of definite parity. There is also no isospectrality breaking due to $\mathcal{S}_{\geo}^{(0,1,1)}=H_0^{(0,1,0)}\Psi_{0}^{(0,0,1)}$, since there is no mixing of modes with frequency $\omega$ and $-\bar{\omega}$. 
	
    Next, let us compute $H_0^{(1,1,0)}$. 
    In this work, we only sketch the key steps, and more details can be found in \cite{dcstyped1}, where the complete modified Teukolsky equation (before separation into definite-parity parts) is found up to $\mathcal{O}(\zeta^1,\chi^1,\epsilon^1)$. 
    To compute $H_0^{(1,1,0)}$, one needs to first find a corrected tetrad that satisfies all the orthogonality conditions at $\mathcal{O}(\zeta^1,\chi^1,\epsilon^0)$. The background spacetime at $\mathcal{O}(\zeta^1,\chi^1,\epsilon^0)$ was found in \cite{Yunes:2009hc}, where all the components of $h_{\mu\nu}^{(1,1,0)}$ vanish except
    \begin{equation} \label{eq:metric_dCS}
		h_{t\phi}^{(1,1,0)}=
		\frac{5M^5}{8r^4}\left(1+\frac{12M}{7r} +\frac{27M^2}{10r^2}\right)\sin^2\theta\,.
    \end{equation}
    One can explicitly check that $\hat{\mathcal{P}}h_{\mu\nu}^{(1,1,0)}=h_{\mu\nu}^{(1,1,0)}$ in dCS gravity, consistent with our assumption. Since the background is still Petrov type D \cite{Yunes:2009hc}, by tetrad rotations, one can find a frame where $\Psi_{i}^{(1,1,0)}=0$ for $i = \{0,1,3,4\}$. Notice that in this frame, one does not necessarily have $\kappa^{(1,1,0)}=\sigma^{(1,1,0)}=\lambda^{(1,1,0)}=\nu^{(1,1,0)}=0$, as implied by the Goldberg-Sachs theorem \cite{Chandrasekhar_1983}, since we are in a non-Ricci-flat spacetime \cite{dcstyped1}. One can now use this modified tetrad and Eq.~\eqref{eq:def_operators} to compute $H_0^{(1,1,0)}$,
    \begin{equation} \label{eq:H0110dCS}
    \begin{aligned}
		H_{0,\dCS}^{(1,1,0)}
		=& \;\frac{M^4}{448r^9(r-r_s)}\left(C_1(r)\partial_{\phi}
		-4r^2C_2(r)\partial_{\phi}\partial_t\right) \\
        & \;-\frac{iM^4}{16r^9}\cos{\theta}
		\left(C_3(r)+\frac{r^2D_4(r)}{2}\partial_t\right) \\
		& \;\frac{iM^4}{32r^8}\left[(r-r_s)C_4(r)\cos{\theta}\partial_r
		-\frac{C_5(r)}{2r}\sin{\theta}\partial_{\theta}\right]\,,
    \end{aligned}
    \end{equation}
    where $C_{i}(r)$ are functions of $r$ found in \cite{dcstyped1} and listed in Appendix~\ref{appendix:radial_functions} for convenience. 
    We can check that $\hat{\mathcal{P}}H_{0,\dCS}^{(1,1,0)}=H_{0,\dCS}^{(1,1,0)}$, so the modified Teukolsky equation up to $\mathcal{O}(\zeta^1, \chi^1, \epsilon^1)$ admits definite-parity solutions if we ignore the source term $\mathcal{S}^{(1,1,1)}$ associated with $\vartheta$. Similar to the $\mathcal{O}(\zeta^0,\chi^1,\epsilon^1)$ correction, since there is no mixing of modes with different frequencies, $\mathcal{S}_{\geo}^{(1,1,1)}$ preserves isospectrality. 
	
    \subsubsection{Correction due to $\mathcal{S}^{(1,1)}$ in dCS}
    \label{sec:dCS_S}

    In this subsection, we compute the correction due to $\mathcal{S}^{(1,1)}$. As discussed above, the leading contribution to $\mathcal{S}^{(1,1)}$ is $\mathcal{S}^{(1,0,1)}$ in dCS gravity. 
    In these previous works \cite{Cardoso:2009pk, Molina:2010fb, Pani_Cardoso_Gualtieri_2011}, they found that only the odd-parity modes are modified for non-rotating BHs in dCS gravity. We now verify this result using our formalism based on the Teukolsky framework.
    
    As found in \cite{Yunes:2009hc}, the trace-reverse Einstein equations takes the form
    \begin{equation} \label{eq:Ricci_dCS}
    \begin{aligned}
		R_{\mu\nu}=
		& \;-\left(\frac{1}{\kappa_g}\right)^{1/2}M^2
		\Bigg[\left(\nabla^{\sigma}\vartheta\right)
		\epsilon_{\sigma\delta\alpha(\mu}
		\nabla^{\alpha}R_{\nu)}{}^{\delta} \\
		& \;+\left(\nabla^{\sigma}\nabla^{\delta}
		\vartheta\right)~^*\!R_{\delta(\mu\nu)\sigma}\Bigg]
		+\frac{1}{2\kappa_g\zeta}\left(\nabla_{\mu}\vartheta\right)
		\left(\nabla_{\nu}\vartheta\right)\,.
    \end{aligned}
    \end{equation}
    To be consistent with \cite{Li:2022pcy, dcstyped1}, we have absorbed an additional factor of $\zeta^{1/2}$ into the expansion of $\vartheta$, so its expansion also follows Eq.~\eqref{eq:expansion_Weyl}. In other words, we have multiplied the first and second terms in Eq.~\eqref{eq:Ricci_dCS} by $\zeta^{-1/2}$ while the third term by $\zeta^{-1}$. The equation of motion of $\vartheta$ at $\mathcal{O}(\zeta^1,\epsilon^1)$ is then \cite{Li:2022pcy}
    \begin{equation} \label{eq:EOM_theta_11_dCS}
		\square^{(0,0)}\vartheta^{(1,1)}
	    =-\pi^{-1/2}M^2\left[R~^*\!R\right]^{(0,1)}
	    -\square^{(0,1)}\vartheta^{(1,0)}\,.
    \end{equation}
	
    In this work, we are interested in modified BH spacetimes that are vacuum in GR, so $R_{\mu\nu}=0$ at $\mathcal{O}(\zeta^0)$. As argued in \cite{Li:2022pcy}, all the metric fields in $R_{\mu\nu}$ have to enter at $\mathcal{O}(\zeta^0)$ in Eq.~\eqref{eq:Ricci_dCS}, so the first term in this equation vanishes. In addition, since there is no correction to the background metric at $\mathcal{O}(\zeta^1,\chi^0,\epsilon^0)$, $\vartheta^{(1,0,0)}=0$ \cite{Yunes:2009hc}. 
    At $\mathcal{O}(\zeta^1,\chi^0,\epsilon^1)$, the last term in Eq.~\eqref{eq:Ricci_dCS} is proportional to $\left(\nabla_{\mu}\vartheta^{(1,0,1)}\right)\left(\nabla_{\nu}\vartheta^{(1,0,0)}\right)$, but $\vartheta^{(1,0,0)}=0$, so this term vanishes. In the end, only the second term in Eq.~\eqref{eq:Ricci_dCS} contributes at $\mathcal{O}(\zeta^1,\chi^0,\epsilon^1)$. 
    Since only $\vartheta^{(1,0,1)}$ is nonzero, the term $~^*\!R_{\delta(\mu\nu)\sigma}$ coupled to it has to be stationary, so we do not need metric reconstruction here. 
    Now the only term that can mix modes with different frequencies, and thus break isospectrality, is $\vartheta^{(1,0,1)}$, so we need to focus on Eq.~\eqref{eq:EOM_theta_11_dCS}.
	
    Since $\vartheta^{(1,0,0)}=0$, the last term in Eq.~\eqref{eq:EOM_theta_11_dCS} vanishes, and only the first term in Eq.~\eqref{eq:EOM_theta_11_dCS} is important. Projecting this term into the NP basis, one can find that
    \begin{equation} \label{eq:R*R}
	    R~^*\!R=8i(3\Psi_2^2-4\Psi_1\Psi_3+\Psi_0\Psi_4-c.c.)\,,
    \end{equation}
    which is made up of quadratic terms in Weyl scalars. Since we are interested in $\mathcal{O}(\epsilon^1)$ corrections, one of the Weyl scalars in each pair has to be stationary. For Petrov type D spacetimes, all the Weyl scalars vanish except $\Psi_2$, so
    \begin{equation} \label{eq:R*R_01}
	    [R~^*\!R]^{(0,1)}=48i\left(\Psi_2^{(0,0)}\Psi_2^{(0,1)}
	    -\bar{\Psi}_2^{(0,0)}\bar{\Psi}_2^{(0,1)}\right)\,.
    \end{equation}
    In Schwarzschild, $\Psi_2^{(0,0,0)}=-M/r^3$ is real, so
    \begin{align} \label{eq:R*R_001}
	    [R~^*\!R]^{(0,0,1)}
	    =& \;-\frac{48iM}{r^3}\left(\Psi_2^{(0,0,1)}-\bar{\Psi}_2^{(0,0,1)}\right) \nonumber\\
	    =& \;\frac{96M}{r^3}\mathcal{I}\left[\Psi_2^{(0,0,1)}\right]\,,
    \end{align}
    where $\mathcal{I}[f]$ refers to the imaginary part of $f$.
	
    One can now naturally ask whether we can remove $[R~^*\!R]^{(0,0,1)}$ via a tetrad rotation or a coordinate transformation at $\mathcal{O}(\zeta^0,\chi^0,\epsilon^1)$. The answer is no. First, one can explicitly check that all the tetrad rotations at $\mathcal{O}(\zeta^0,\chi^0,\epsilon^1)$ leave $\Psi_2^{(0,0,1)}$ unchanged since $\Psi_1^{(0,0,0)}=\Psi_3^{(0,0,0)}=0$. Second, under the coordinate transformation $x^{\mu}\rightarrow x^{\mu}+\xi^{\mu}$, where $\xi$ is at $\mathcal{O}(\zeta^0,\chi^0,\epsilon^1)$, $\Psi_2^{(0,0,1)}$ transforms as \cite{Campanelli_Lousto_1999}
    \begin{equation}
	    \Psi_2^{(0,0,1)}\rightarrow\Psi_2^{(0,0,1)}
	    +\xi^{\mu(0,0,1)}\partial_\mu\Psi_2^{(0,0,0)}\,,
    \end{equation}
    which implies that
    \begin{equation}
	    \mathcal{I}\left[\Psi_2^{(0,0,1)}\right]
	    \rightarrow\mathcal{I}\left[\Psi_2^{(0,0,1)}\right]
	    +\xi^{\mu(0,0,1)}\partial_\mu\mathcal{I}
        \left[\Psi_2^{(0,0,0)}\right]\,.	
     \end{equation}
    Since $\Psi_2^{(0,0,0)}$ is real, $\mathcal{I}\left[\Psi_2^{(0,0,0)}\right]=0$, so $[R~^*\!R]^{(0,0,1)}$ is invariant under both tetrad and coordinate transformations at $\mathcal{O}(\zeta^0,\chi^0,\epsilon^1)$. 
    
    More generally, for an arbitrary $\chi$, the source term in Eq.~\eqref{eq:EOM_theta_11_dCS} is still tetrad- and coordinate-invariant. The tetrad invariance is easy to confirm since $\Psi_2^{(0,1)}$ is invariant under tetrad rotations at $\mathcal{O}(\zeta^0,\epsilon^1)$ if the background spacetime is Petrov type D at $\mathcal{O}(\zeta^0,\epsilon^0)$. Furthermore, $\square^{(0,1)}$ is tetrad-invariant. On the other hand, unlike at $\mathcal{O}(\chi^0)$, $\Psi_2^{(0,0)}$ is complex, so the coordinate invariance needs to be shown in another way. For convenience, let us denote the source term in Eq.~\eqref{eq:EOM_theta_11_dCS} as $\mathcal{S}_{\vartheta}^{(1,1)}$. 
    Then under the coordinate transformation $x^{\mu}\rightarrow x^{\mu}+\xi^{\mu}$ at $\mathcal{O}(\zeta^0,\epsilon^1)$, we find
    \begin{equation}
        \mathcal{S}_{\vartheta}^{(1,1)}\rightarrow
        \mathcal{S}_{\vartheta}^{(1,1)}
        +\xi^{\mu(0,1)}\partial_{\mu}\mathcal{S}_{\vartheta}^{(0,0)}\,,
    \end{equation}
    but notice that
    \begin{equation}
        \mathcal{S}_{\vartheta}^{(0,0)}
        =-\pi^{-1/2}M^2[R~^*\!R]^{(0,0)}
        -\square^{(0,0)}\vartheta^{(1,0)}=0
    \end{equation}
    due to the same equation of Eq.~\eqref{eq:EOM_theta_11_dCS} at $\mathcal{O}(\zeta^1,\epsilon^0)$, i.e.,
    \begin{equation}
        \square^{(0,0)}\vartheta^{(1,0)}
	    =-\pi^{-1/2}M^2\left[R~^*\!R\right]^{(0,0)}\,.
    \end{equation}
    Thus, $\mathcal{S}_{\vartheta}^{(1,1)}$ is both tetrad- and coordinate-invariant.

    By pure order counting, one can write the source term $\mathcal{S}$ in Eq.~\eqref{eq:source_def_stress} as 
    \begin{equation} \label{eq:source_dCS}
	    \mathcal{S}_{\dCS}^{(1,0,1)}=\mathcal{F}^{\dCS}\left(\Psi_2^{(0,0,1)}-\bar{\Psi}_2^{(0,0,1)}\right)\,,
    \end{equation}
    where $\mathcal{F}^{\dCS}$ is some complicated operator converting the source term driving $\vartheta$ in Eq.~\eqref{eq:EOM_theta_11_dCS} to the source term in Eq.~\eqref{eq:source_def_stress}. 
    The operator $\mathcal{F^{\dCS}}$ contains three parts: 
    \begin{enumerate}
	    \item The inversion of $\square$ in Eq.~\eqref{eq:EOM_theta_11_dCS} to solve for $\vartheta$.
	    \item The NP quantities and derivatives acting on $\vartheta$ in Eq.~\eqref{eq:Ricci_dCS}.
	    \item The tetrad acting on $R_{\mu\nu}$ to convert it to NP Ricci scalars and the operators in Sec.~\ref{sec:modified_Teukolsky_equations} to convert NP Ricci scalars to $\mathcal{S}_{\dCS}$.
    \end{enumerate}
    Despite being a complicated operator, $\mathcal{F}^{\dCS}$ only contains stationary terms, so it will not mix modes with different frequencies. Thus, to study isospectrality-breaking properties without computing QNM shifts, we do not need to know the exact form of $\mathcal{F}^{\dCS}$.
	
    To show that only the odd-parity modes are modified, we essentially need to show that Eq.~\eqref{eq:QNM_freq} vanishes for $\eta=(-1)^l$ but not for $\eta=(-1)^{(l+1)}$. 
    Then, we need to use metric reconstruction to rewrite $\Psi_2^{(0,0,1)}$ in terms of $\Psi_0^{(0,0,1)}$ or the Hertz potential, e.g., Eq.~\eqref{eq:Hertz_IRG_Psi2}. 
    For simplicity, we can absorb the operator $-D^2/2$ into $\mathcal{F}^{\dCS}$ since $D$ is a real operator. We then rewrite Eq.~\eqref{eq:source_dCS} as
    \begin{equation} \label{eq:source_dCS_2}
	    \mathcal{S}_{\dCS}=\mathcal{F}^{\dCS}\left(\mathcal{O}-\bar{\mathcal{O}}\hat{\mathcal{C}}\right)\bar{\Psi}_{\Hertz}^{(0,0,1)}\,,\;
	    \mathcal{O}=(\bar{\delta}+2\beta)(\bar{\delta}+4\beta)\,.
    \end{equation}
    Comparing Eq.~\eqref{eq:source_dCS_2} to Eq.~\eqref{eq:modified_Teuk_scheme_2}, one can extract that
    \begin{equation} \label{eq:SO_to_FO}
	    \mathcal{S}^{\mu\nu}\mathcal{O}_{\mu\nu}
	    =\mathcal{F}^{\dCS}\mathcal{O}\,,\quad
	    \mathcal{S}^{\mu\nu}\bar{\mathcal{O}}_{\mu\nu}
	    =-\mathcal{F}^{\dCS}\bar{\mathcal{O}}\,,
    \end{equation}
    so Eq.~\eqref{eq:QNM_freq} becomes
    \begin{equation} \label{eq:QNM_freq_dCS}
	    \omega_{lm}^{\dCS(1)}
	    =\frac{\left\langle\mathcal{F}^{\dCS}\left(\mathcal{O}
        \mp(-1)^l\bar{\mathcal{O}}
        \hat{\mathcal{C}}\hat{\mathcal{P}}\right)\mathscr{D}\right\rangle_{lm}}
		{\langle\partial_{\omega}\tilde{H}_0\rangle_{lm}}\,,
    \end{equation}
    where $\mathscr{D}$ denotes the operator converting $\Psi_0^{(0,0,1)}$ to $\bar{\Psi}_{\Hertz}^{(0,0,1)}$ in Eq.~\eqref{eq:reconstruct_Hertz}.
 
    With this expression at hand, we can now show that $\mathcal{O}\mathscr{D}\bar{\Psi}_{0}^{(0,0,1)}=(-1)^l\bar{\mathcal{O}}\hat{\mathcal{C}}\hat{\mathcal{P}}\mathscr{D}\bar{\Psi}_{0}^{(0,0,1)}$ or $\mathcal{O}\bar{\Psi}_{\Hertz}^{(0,0,1)}=(-1)^l\bar{\mathcal{O}}\hat{P}\bar{\Psi}_{\Hertz}^{(0,0,1)}$, and thus, only the odd modes are modified. Using Eqs.~\eqref{eq:NP_parity_Kerr} and \eqref{eq:source_dCS_2}, one can find that
    \begin{equation}
        \hat{\mathcal{P}}\mathcal{O}=\mathcal{O}\,,
    \end{equation}
    so $\bar{\mathcal{O}}=\hat{P}\mathcal{O}$. Then, it is equivalent to show that $\hat{P}\left(\mathcal{O}\bar{\Psi}_{\Hertz}^{(0,0,1)}\right)=(-1)^l\mathcal{O}\bar{\Psi}_{\Hertz}^{(0,0,1)}$. 
    In other words, $\mathcal{O}\bar{\Psi}_{\Hertz}^{(0,0,1)}$ transforms in the same way as $Y_{lm}(\theta,\phi)$ under the standard parity transformation $\hat{P}$. The easiest way to show this is to recognize that
    \begin{align}
	    & \left(\bar{\delta}+2s\beta\right)f
	    =-\frac{1}{\sqrt{2}r}\bar{\eth}f\,,\nonumber\\
	    & \bar{\eth} f=-\left(\partial_{\theta}-i\csc{\theta}\partial_\phi
	    +s\cot{\theta}\right)f\,,
    \end{align}
    where $f$ has spin weight $s$, and $\bar{\eth}$ is the operator lowering spin weight by $1$ \cite{Goldberg:1966uu}, e.g.,
    \begin{equation}
        \bar{\eth}~_{s}\!Y_{lm}=-[(l+s)(l-s+1)]^{1/2}~_{s-1}\!Y_{lm}\,.
    \end{equation}
    From Eq.~\eqref{eq:reconstruct_Hertz}, one can notice that $\bar{\Psi}_{\Hertz}$ has the same spin weight as $\Psi_0$ in IRG, e.g., $\bar{\Psi}_{\Hertz}^{(0,0,1)}\propto{}_{2}\!Y_{lm}(\theta,\phi)$. Then,
    \begin{equation}
	    \mathcal{O}\bar{\Psi}_{\Hertz}^{(0,0,1)}
	    =\frac{1}{2r^2}\bar{\eth}^{2}\bar{\Psi}_{\Hertz}^{(0,0,1)}
	    \propto Y_{lm}(\theta,\phi)\,,
    \end{equation}
    so $\mathcal{O}\bar{\Psi}_{\Hertz}^{(0,0,1)}$ transforms like $Y_{lm}(\theta,\phi)$ under parity transformations, i.e., $\hat{P}\left(\mathcal{O}\bar{\Psi}_{\Hertz}^{(0,0,1)}\right)=(-1)^l\mathcal{O}\bar{\Psi}_{\Hertz}^{(0,0,1)}$. 
	
    To check whether the modified Teukolsky equation has definite-parity solutions, we need to check whether $\hat{\mathcal{P}}\mathcal{S}^{\mu\nu(1,0,1)}=\mathcal{S}^{\mu\nu(1,0,1)}$. Since $\hat{\mathcal{P}}\mathcal{O}_{\mu\nu}=\mathcal{O}_{\mu\nu}$ and $\hat{\mathcal{P}}\mathcal{O}=\mathcal{O}$, using Eq.~\eqref{eq:SO_to_FO}, one can alternatively check whether $\hat{\mathcal{P}}\mathcal{F}^{\dCS}=\mathcal{F}^{\dCS}$. In this case, one has to know the exact expression of $\mathcal{S}_{\dCS}$. Generally, this is a non-trivial calculation, but for non-rotating BHs in dCS gravity, since only $\vartheta$ is dynamical in $\mathcal{S}_{\dCS}$, one can easily evaluate $\mathcal{S}_{\dCS}$ in terms of $\vartheta$ and the background metric without doing metric reconstruction. 
    In Appendix~\ref{appendix:source_terms_dCS}, we show how to evaluate $\mathcal{S}_{\dCS}^{(1,0,1)}$ in dCS gravity in detail and provide the result of $\mathcal{F}_{\dCS}$ in Eq.~\eqref{eq:F_dCS}. One can now easily verify that $\hat{\mathcal{P}}\mathcal{F}^{\dCS}=\mathcal{F}^{\dCS}$, so the modified Teukolsky equation at $\mathcal{O}(\zeta^1,\chi^0,\epsilon^1)$ in dCS gravity still admits definite-parity solutions.
	
    In this subsection, we have so far shown successfully that only the odd modes are modified for non-rotating BHs in dCS gravity using the NP language developed in this work, which is consistent with \cite{Cardoso:2009pk, Molina:2010fb, Pani_Cardoso_Gualtieri_2011}. 
    One can also carry out the same calculation at $\mathcal{O}(\zeta^1,\chi^1,\epsilon^0)$ and compare to the results using metric perturbations in \cite{Wagle_Yunes_Silva_2021, Srivastava_Chen_Shankaranarayanan_2021}. In Sec.~\ref{sec:dCS_S_geo}, we have found the correction due to $\mathcal{S}_{\geo}^{(1,1,1)}$. 
    The correction due to $\mathcal{S}_{\dCS}^{(1,1,1)}$ is much more complicated, and can be found in \cite{dcstyped1}. In \cite{dcstyped2}, we will apply the formalism developed in this work and the expression found in \cite{dcstyped1} to compute the correction to QNMs directly and compare to \cite{Wagle_Yunes_Silva_2021, Srivastava_Chen_Shankaranarayanan_2021}. 
    Another interesting avenue for future work is to find a direct mapping between the modified RW (ZM) equations in \cite{Cardoso:2009pk, Molina:2010fb, Pani_Cardoso_Gualtieri_2011, Wagle_Yunes_Silva_2021, Srivastava_Chen_Shankaranarayanan_2021} and the odd (even) modified Teukolsky equations in this work, as we discuss in Sec.~\ref{sec:discussion}.
	
    \subsection{EdGB gravity} \label{sec:EdGB}
    
    The structure of the metric shifts to the EdGB BH solution is qualitatively different from that of the dCS case. In EdGB, we follow \cite{Ayzenburg:2014} to define the expansion parameter $\zeta$ to be
    \begin{equation}
        \zeta=\zeta_{\EdGB}\equiv\frac{\alpha_{\EdGB}^2}{M^4}\,,
    \end{equation}
    where $\alpha_{\EdGB}$ is the coupling constant of EdGB gravity in \cite{Pierini:2021jxd}, and we have chosen the coupling constant of the scalar field action $\beta=1$.\footnote{Note that we choose to follow the convention of the EdGB action in \cite{Pierini:2021jxd}, whereas the expressions in Eq.~\eqref{eq:AyzenburgEquations} are taken from \cite{Ayzenburg:2014}. In this case, the coupling constant $\alpha_{\EdGB}$ in \cite{Ayzenburg:2014} is $\frac{1}{64\pi}$ of the one in \cite{Pierini:2021jxd}. We adjust the expressions in Eqs.~\eqref{eq:AyzenburgEquations} and \eqref{eq:Dfunctions} to account for the change in convention.} One finds that the shifts in the background metric at $\mathcal{O}(\zeta)$ actually contain a component that is independent of spin. In other words, EdGB gravity perturbs the non-rotating BH solution. Since non-rotating BHs in EdGB gravity are Petrov Type D \cite{Ayzenburg:2014}, the leading correction to $\mathcal{S}_{\geo}^{(1,1)}$ is $\mathcal{S}_{\geo}^{(1,0,1)}$, completely determined by $H_{0}^{(1,0,0)}$. Similarly, the leading contribution to $\mathcal{S}^{(1,1)}$ is $\mathcal{S}^{(1,0,1)}$.
	
    \subsubsection{Correction due to $\mathcal{S}_{\geo}^{(1,1)}$ in EdGB}
    \label{sec:EdGB_S_geo}
	
    The corrections due to $\mathcal{S}^{(1,1)}_{\geo}$ are made of several parts as noted in Eq.~\eqref{eq:source_def_geo}. 
    However, note that at the leading order $(\zeta^1, \chi^0, \epsilon^0)$ in EdGB, we know that $\Psi_{0,1,3,4}^{(1,0)} = 0$.  Hence, the only contribution to $\mathcal{S}^{(1,1)}_{\geo}$ comes from $\mathcal{S}_{0,D}^{(1,1)}=H_{0}^{(1,0)}\Psi^{(0,1)}_0$. By following similar procedures to what we presented in Sec.~\ref{sec:dCS_S_geo} for dCS, and using the metric from \cite{Ayzenburg:2014}, 
    \begin{equation}
    \begin{aligned}
    \label{eq:AyzenburgEquations}
        h_{tt}^{(1,0,0)}
        =& \;-\frac{1}{(64\pi)^2}\frac{M^3}{3r^3}
        \left[1+\frac{26M}{r}+\frac{66M^2}{5r^2} \right.\\
        & \;\left.+\frac{96M^3}{5r^3}-\frac{80M^4}{r^4}\right]\,, \\
        h_{rr}^{(1,0,0)}
        =& \;-\frac{1}{(64\pi)^2}\frac{M^2}{(r-2M)^2}
        \left[1+\frac{M}{r}+\frac{52M^2}{3r^2}
        +\frac{2M^3}{r^3}\right.\\
        & \;\left.+\frac{16M^4}{5r^4}-\frac{368M^5}{3r^5}\right],
    \end{aligned}
    \end{equation}
    where he other components of $h_{\mu\nu}^{(1,0,0)}$ vanish, we find
    \begin{equation} \label{eq:H0100EdGB}
        H_{0,\EdGB}^{(1,0,0)}
        =D_1(r)+D_2(r)\partial_t+D_3(r)\partial_r + \\ 
        D_4(r)\partial^2_{t}+D_5(r)\partial^2_{r}\,.
    \end{equation}
    Note that the factor of ${1}/{(64\pi)^2}$ comes from the use of the conventions of $\alpha_{\EdGB}$ in \cite{Pierini:2021jxd}. The full form of the radial functions $D_i(r)$ can be found in Appendix~\ref{appendix:radial_functions}. One can also easily check that $\hat{\mathcal{P}}h_{\mu\nu}^{(1,0,0)}=h_{\mu\nu}^{(1,0,0)}$ for EdGB, consistent with our assumption. Since $H_{0,\EdGB}^{(1,0,0)}$ does not depend on the $\theta$ or $\phi$ coordinates, we can use Eq.~\eqref{eq:P_properties} to easily show that
    \begin{equation}
        \hat{\mathcal{P}}H_{0,\EdGB}^{(1,0,0)}=H_{0,\EdGB}^{(1,0,0)}\,.
    \end{equation}
    In this case, the modified Teukolsky equation still admits definite-parity solutions up to $\mathcal{O}(\zeta^1,\chi^0,\epsilon^1)$ if we ignore the source terms driven by the nonminimally coupled scalar field $\varphi$. Furthermore, in EdGB, $\mathcal{S}^{(1,0,1)}_{\geo}$ does not break isospectrality because it does not mix the modes with frequencies $\omega$ and $-\bar{\omega}$.
    
    \subsubsection{Correction due to $\mathcal{S}^{(1,1)}$ in EdGB}
    \label{sec:EdGB_S}
	
    In this subsection, we compute $\mathcal{S}^{(1,0,1)}$ in EdGB gravity, where the  calculation is similar to the dCS case. First, the trace-reversed Einstein equations in EdGB gravity take the form \cite{Witek:2018dmd, Pierini:2021jxd},
	\begin{align}
        \label{eq:tracericci-sgb}
        R_{\mu \nu} 
        &= -\kappa_g^{1/2}M^2\left(\mathcal{K}_{\mu\nu}
        -\frac{1}{2}g_{\mu\nu}\mathcal{K}\right)
        +\frac{1}{2\zeta}(\nabla_\mu\varphi)(\nabla_\nu\varphi)\,, \nonumber \\
        \mathcal{K}_{\mu\nu}
        &=\frac{1}{8}\left(g_{\mu\rho}g_{\nu\sigma}
        +g_{\mu\sigma}g_{\nu\rho}\right)\epsilon^{\delta\sigma\gamma\alpha} \nabla_\beta\left(^*\!R^{\rho\beta}{}_{\gamma\alpha}
        e^{\varphi}\nabla_\delta\varphi\right)\,, \nonumber\\
        \mathcal{K}&=g_{\mu\nu}\mathcal{K}^{\mu\nu}\,, 
    \end{align}
    with the equation of the scalar field $\varphi$ at $\mathcal{O}(\zeta^1,\epsilon^1)$ being \cite{Li:2022pcy}
    \begin{equation} \label{eq:EdGB_EOM_scalar_11}
        \square^{(0,0)}\varphi^{(1,1)}
        =-\pi^{-\frac{1}{2}}M^2\mathcal{G}^{(0,1)}-\square^{(0,1)}\varphi^{(1,0)}\,,
    \end{equation}
    where the Gauss-Bonnet invariant $\mathcal{G}$ is defined to be
    \begin{equation}
        \mathcal{G}
        =R^{\mu\nu\rho\sigma}R_{\mu\nu\rho\sigma}
        -4R^{\mu\nu}R_{\mu\nu}+R^2\,.
    \end{equation}
    Similar to the dCS case, we have absorbed a factor of $\zeta$ into the expansion of $\varphi$ by multiplying the first term of $R_{\mu\nu}$ in Eq.~\eqref{eq:tracericci-sgb} by $\zeta^{-1/2}$ and the second term by $\zeta^{-1}$. 

    Unlike in dCS gravity, since $\varphi^{(1,0,0)}\neq0$ for non-rotating BHs in EdGB, we need metric reconstruction to evaluate $h_{\mu\nu}^{(0,0,1)}$ coupled to  $\varphi^{(1,0,0)}$ for $\mathcal{O}(\zeta^1,\chi^0,\epsilon^1)$ corrections. For simplicity, in this work, we only consider the terms in Eq.~\eqref{eq:tracericci-sgb} driven by $\varphi^{(1,0,1)}$ or its derivatives. The evaluation of the terms proportional to $h_{\mu\nu}^{(0,0,1)}$ or its derivatives in Eq.~\eqref{eq:tracericci-sgb} is more complicated, while a similar calculation in dCS gravity has been done in \cite{dcstyped1}. Under this simplification, all the metric fields in Eq.~\eqref{eq:tracericci-sgb} can be evaluated at the Schwarzschild background, and we can focus on Eq.~\eqref{eq:EdGB_EOM_scalar_11}. To evaluate the source terms in Eq.~\eqref{eq:EdGB_EOM_scalar_11}, we need to compute $\square$ and $\mathcal{G}$ in the NP basis. At $\mathcal{O}(\zeta^1,\chi^0,\epsilon^1)$, since $\varphi^{(1,0,0)}\neq0$, unlike the pseudoscalar field in dCS, both terms in Eq.~\eqref{eq:EdGB_EOM_scalar_11} will contribute. 
    
    For the term $\mathcal{G}^{(0,1)}$, we find in the NP basis,
    \begin{equation} \label{eq:GB_NP}
	    \mathcal{G}=8(3\Psi_2^2-4\Psi_1\Psi_3+\Psi_0\Psi_4+c.c.)\,.
    \end{equation}
    We can notice that, in the dCS case, $R~^*\!R$ is proportional to the imaginary part of $3\Psi_2^2-4\Psi_1\Psi_3+\Psi_0\Psi_4$, while $\mathcal{G}$ is proportional to the real part of the same quantity. Expanding Eq.~\eqref{eq:GB_NP} to $\mathcal{O}(\zeta^0,\epsilon^1)$, we find
    \begin{equation} \label{eq:G_01}
        \mathcal{G}^{(0,1)}=48\left(\Psi_2^{(0,0)}\Psi_2^{(0,1)}
	    +\bar{\Psi}_2^{(0,0)}\bar{\Psi}_2^{(0,1)}\right)\,,
    \end{equation}
    which in Schwarzschild becomes 
    \begin{align} \label{eq:G_001}
	    \mathcal{G}^{(0,0,1)}
	    =& \;-\frac{48M}{r^3}\left(\Psi_2^{(0,0,1)}+\bar{\Psi}_2^{(0,0,1)}\right) \nonumber\\
	    =& \;-\frac{96M}{r^3}\mathcal{R}\left[\Psi_2^{(0,0,1)}\right]\,,
    \end{align}
    where $\mathcal{R}[f]$ refers to the real part of $f$. Following the same reasoning as in Sec.~\ref{sec:dCS_S}, one can argue that the part of $\mathcal{S}^{(1,0,1)}$ generated by $\mathcal{G}^{(0,0,1)}$ takes the form
    \begin{equation} \label{eq:source_EdGB}
        \mathcal{F}^{\EdGB}\left(\Psi_2^{(0,0,1)}+\bar{\Psi}_2^{(0,0,1)}\right)\,,
    \end{equation}
    where $\mathcal{F}^{\EdGB}$ contains pieces similar to $\mathcal{F}^{\dCS}$ but with the effective Ricci tensor given by Eq.~\eqref{eq:tracericci-sgb}. If one only considers the shift of QNM frequencies due to this term, then $\omega_{lm}^{(1)}$ is given by Eq.~\eqref{eq:QNM_freq_dCS}, with $\mathcal{F}_{\dCS}$ replaced by $\mathcal{F}_{\EdGB}$ and the sign $\mp$ between the terms proportional to $\mathcal{O}$ and $\mathcal{O}\hat{\mathcal{C}}\hat{\mathcal{P}}$ replaced by $\pm$, so the QNMs of odd-parity modes are not modified by these terms. 
    
    For the contribution from $\square^{(0,0,1)}\varphi^{(1,0,0)}$, we have shown in detail how to reconstruct $\square^{(0,0,1)}$ in \cite{dcstyped1}, so here we just present the results we found,
    \begin{equation}
        \square^{(0,0,1)}\varphi^{(1,0,0)}
        =-\frac{1}{2r^3}(r\partial_r^2+2\partial_r)\Phi(r)\bar{\eth}^{2}
        \left(\Psi_{\Hertz}^{(0,0,1)}+\bar{\Psi}_{\Hertz}^{(0,0,1)}\right)\,,
    \end{equation}
    where we have used that $\varphi^{(0,0,1)}=\Phi(r)$ is a pure radial function in EdGB gravity. Similar to the case of dCS gravity, the source term in Eq.~\eqref{eq:EdGB_EOM_scalar_11} is tetrad- and coordinate-invariant, following the same argument in Sec.~\ref{sec:dCS_S}.
    
    In total, following the same procedures in Sec.~\ref{sec:dCS_S}, we find
    \begin{equation}
        \omega_{lm}^{\EdGB(1)}=
        \frac{\left\langle
	    \mathcal{F}^{\EdGB}\left(\mathcal{O}'
        \pm(-1)^l\bar{\mathcal{O}}'\hat{\mathcal{C}}\hat{\mathcal{P}}\right)
		\mathscr{D}\right\rangle_{lm}}
		{\langle\partial_{\omega}\tilde{H}_0\rangle_{lm}}\,,
    \end{equation}
    where
    \begin{align} \label{eq:O'}
	    & \mathcal{O}'\bar{\Psi}_{\Hertz}^{(0,0,1)} \nonumber\\
	    &=\left[\frac{1}{2r}\left(r\partial_r^2+2\partial_r\right)\Phi(r)
	    -12\pi^{-1/2}\left(\frac{M}{r}\right)^3D^2\right]
	    \frac{1}{r^2}\bar{\eth}^{2}\bar{\Psi}_{\Hertz}^{(0,0,1)} \nonumber\\
	    & \propto Y_{lm}(\theta,\phi)\,,
	\end{align}
    and the last term of Eq.~\eqref{eq:O'} comes from the $\mathcal{O}$ operator defined in Eq.~\eqref{eq:source_dCS_2}. 
    Unlike in Sec.~\ref{sec:dCS_S}, we have not absorbed the factor of $-D^2/2$ in $\mathcal{O}$ into the definition of $\mathcal{F}^{\EdGB}$. Following the same argument in Sec.~\ref{sec:dCS_S}, one can easily see that for this type of contribution, only the QNM frequencies of even parity are shifted. This result is consistent with \cite{Pani:2009wy, Blazquez-Salcedo:2016enn, Blazquez-Salcedo_Khoo_Kunz_2017} since what we have essentially shown is that the scalar field $\varphi$ is only driven by even-parity gravitational perturbations at $\mathcal{O}(\zeta^1,\chi^0,\epsilon^1)$. In Appendix~\ref{appendix:source_terms_EdGB}, we further show that the modified Teukolsky equation driven by this contribution still admits solutions of definite parity by showing that $\hat{\mathcal{P}}\mathcal{F}^{\EdGB}=\mathcal{F}^{\EdGB}$.

    \section{Discussion} \label{sec:discussion}
	
    In this work, we developed a framework to study the isospectrality breaking of QNMs in modified gravity using the modified Teukolsky formalism developed in \cite{Li:2022pcy, Hussain:2022ins}. To analyze isospectrality breaking using the Teukolsky formalism, one has to first know how definite-parity modes are defined in terms of Weyl scalars $\Psi_{0,4}$. In GR, we followed \cite{Nichols_Zimmerman_Chen_Lovelace_Matthews_Owen_Zhang_Thorne_2012} to construct definite-parity modes of $\Psi_{0,4}^{(0,1)}$ by using the relation between metric perturbations and the Hertz potential. We found that at each $(l,m)$, the Weyl scalars generating definite-parity metric perturbations are linear combinations of the mode $(l,m,\omega)$ and its $\hat{\mathcal{P}}$-transformation, where $\hat{\mathcal{P}}$ is the parity transformation, but with an additional complex conjugation. Due to the transformation properties of Teukolsky functions under $\hat{\mathcal{P}}$, these modes are equal to the sum (difference) of the modes $(l,m,\omega)$ and $(l,-m,-\bar{\omega})$ for even (odd) parity, consistent with the definition in \cite{Nichols_Zimmerman_Chen_Lovelace_Matthews_Owen_Zhang_Thorne_2012}. 
	
    In modified gravity, we showed that the same definition in GR still applies for Petrov type D spacetimes. Since the relation between metric perturbations and the Hertz potential is not known in modified gravity in general, we instead started from definite-parity metric perturbations and derived the parity properties of $\Psi_{0,4}^{(1,1)}$ directly. The entire procedure is closely related to reconstructing NP quantities from metric perturbations. Using tetrad rotations, we first found some convenient gauges where the transformation property of both the background and the dynamical tetrad is simple under $\hat{\mathcal{P}}$. The transformation property of spin coefficients was then determined from commutation relations. Using Ricci identities, we finally obtained the $\hat{\mathcal{P}}$-transformation of $\Psi_{0,4}^{(1,1)}$ generated by definite-parity metric perturbations, which is the same as in GR. 
	
    After defining definite-parity modes of $\Psi_{0,4}^{(0,1)}$ and $\Psi_{0,4}^{(1,1)}$, we then proceeded to derive the equations that govern them from the modified Teukolsky equation. Since the source terms that shift QNM frequencies are those having overlaps with QNMs in GR, we first extracted these source terms. To evaluate the latter, one needs to perform metric reconstruction, which mixes the modes with frequency $\omega$ and $-\bar{\omega}$. Thus, the solutions to the modified Teukolsky equation are also linear combinations of these two modes in general. 
    Using the EVP method developed in \cite{Zimmerman:2014aha, Mark_Yang_Zimmerman_Chen_2015, Hussain:2022ins}, we then found that the solutions form a two-dimensional subspace, so the degeneracy in QNM frequencies of even- and odd-parity modes is broken in modified gravity in general, consistent with the finding in \cite{Hussain:2022ins}.
	
    In the special case that the solutions become even- and odd-parity modes, the source terms of the modified Teukolsky equation are constrained to transform in the same way as the Teukolsky operator in GR under $\hat{\mathcal{P}}$. 
    This constraint is closely related to how one solves the degenerate perturbation problem in quantum mechanics. We showed that the invariance of the source operator $\mathcal{S}^{\mu\nu}$ and the Teukolsky operator $H$ under $\hat{\mathcal{P}}$ implies that they commute with $\hat{\mathcal{P}}$. Similarly, in quantum mechanics, one can diagonalize the perturbed Hamiltonian by using the eigenstates of a Hermitian operator commuting with both the original and the perturbed Hamiltonian.
	
    To demonstrate our framework, we then applied this analysis of isospectrality breaking to two specific cases: dCS and EdGB gravity. For simplicity, we only considered the leading correction to the homogeneous part of the equation and the leading contribution from the effective stress-energy tensor. In dCS gravity, we found that the correction to the homogeneous part does not break isospectrality. For the correction from the effective stress-energy tensor, we showed that only the odd modes are shifted for non-rotating BHs using our modified Teukolsky formalism, consistent with the results found using metric perturbations in \cite{Cardoso:2009pk, Molina:2010fb, Pani_Cardoso_Gualtieri_2011}. For EdGB gravity, we similarly found no isospectrality breaking in the homogeneous part. For the correction due to the stress-energy tensor, we only focused on the terms driven by the dynamical scalar field. In this case, only the even modes are affected, consistent with the result in \cite{Pani:2009wy, Blazquez-Salcedo:2016enn, Blazquez-Salcedo_Khoo_Kunz_2017}.
	
    There are several future avenues of research that our work enables. First, one can study potential observational signatures of isospectrality breaking in QNM frequencies, the most direct of which would be the branching of the QNM spectrum. One can investigate the extraction of these branching QNMs from real observational data using well tested ringdown analysis frameworks like \texttt{ringdown}~\cite{ringdown_isi_farr:2021} and \texttt{PyRing}~\cite{pyRing:2023}. Additionally, an analysis of how the SNR affects this extraction (similar to resolvability arguments in~\cite{Isi:2021iql}) can aid an understanding of when this effect may be significant. Furthermore, these definite-parity modes of $\Psi_{0,4}$ are also related to other decompositions of gravitational perturbations, such as mass or current quadrupoles, left- or right-circularly polarized modes, and plus or cross polarizations. By studying these relations, one can translate the isospectrality breaking of QNM frequencies to observational effects in other modes or polarizations. For example, different QNM frequencies of even- and odd-parity modes might lead to different frequencies of plus and cross polarizations. In fact, some parity violating theories even feature different propagation speeds for different polarizations --- parameterized theory-agnostic tests for such theories were laid out, for example, in~\cite{Alexander:2017jmt} and more recently in~\cite{Jenks:2023pmk}, and connected to specific parity violating theories in~\cite{Jenks:2023pmk}.
    
    Second, when determining the $\hat{\mathcal{P}}$-transformation of $\Psi_{0,4}^{(1,1)}$, we have made some convenient gauge choices. The choice of certain gauges is fine for Petrov type D spacetimes since $\Psi_{0,4}^{(1,1)}$ are both tetrad- and coordinate-invariant up to $\mathcal{O}(\zeta^1,\epsilon^1)$, which is not the case for Petrov type I spacetimes. The same issue was also encountered in the study of the second-order Teukolsky equation in GR \cite{Campanelli_Lousto_1999, Loutrel_Ripley_Giorgi_Pretorius_2020}, where various authors found that $\Psi_{0,4}^{(0,2)}$ is also not invariant under gauge transformations at $\mathcal{O}(\epsilon^1)$. By either adding quantities at $\mathcal{O}(\epsilon^1)$ to $\Psi_{0,4}^{(0,2)}$ \cite{Campanelli_Lousto_1999}, or by choosing some asymptotically flat coordinate system \cite{Loutrel_Ripley_Giorgi_Pretorius_2020}, one is able to construct gauge-invariant curvature perturbations at $\mathcal{O}(\epsilon^2)$. Due to the connection of our modified Teukolsky formalism to the second-order Teukolsky formalism in GR \cite{Li:2022pcy}, we can then apply a similar procedure to construct gauge-invariant quantities at $\mathcal{O}(\zeta^1,\epsilon^1)$ and extend our definition of definite-parity modes to Petrov type I spacetimes in future work. 
	
    Third, in our examples in dCS and EdGB gravity, we have found the shifts of the QNM frequencies in terms of abstract NP quantities. One can then take the equations here to compute the numerical values of the QNM frequencies directly and compare them to previous results using metric perturbations in \cite{Cardoso:2009pk, Molina:2010fb, Pani_Cardoso_Gualtieri_2011, Wagle_Yunes_Silva_2021, Srivastava_Chen_Shankaranarayanan_2021, Pani:2009wy, Blazquez-Salcedo:2016enn, Blazquez-Salcedo_Khoo_Kunz_2017, Pierini:2021jxd, Pierini:2022eim}. 
    A subset of the authors has already derived the coordinate-based modified Teukolsky equation for slowly-rotating BHs in dCS gravity \cite{dcstyped1} and is currently computing the QNM shifts from it \cite{dcstyped2}. Furthermore, in our examples above, we assumed that the (pseudo)scalar field equation could be inverted, and the EVP method could be applied to these terms. 
    In previous literature, the EVP method has only been shown to be valid for source terms that are directly driven by $\Psi_{0,4}^{(0,1)}$ \cite{Zimmerman:2014aha, Mark_Yang_Zimmerman_Chen_2015, Hussain:2022ins}. It will be worth studying whether the inner product in the EVP method is still well-behaved when integrating the Green's function of the (pseudo)scalar field along with $\Psi_{0,4}^{(0,1)}$ over the contour defined in \cite{Zimmerman:2014aha, Mark_Yang_Zimmerman_Chen_2015, Hussain:2022ins}. In the case without extra non-metric fields, such as higher-derivative gravity, Refs.~\cite{Cano:2023tmv, Cano:2023jbk} have used this modified Teukolsky formalism to compute QNM shifts and found good agreement with their previous results using metric perturbations in \cite{Cano:2021myl}.
	
    Besides comparing QNM frequencies, another way of comparing our results to the approach using metric perturbations is to directly map our definite-parity modified Teukolsky equations [i.e., Eq.~\eqref{eq:definite_parity_eqn}] to the modified ZM and RW equations directly. For non-rotating BHs in GR, this map was found by Chandrasekhar \cite{Chandrasekhar:1975nkd, Chandrasekhar_1983}. Due to isospectrality, one can also find a map between RW and ZM equations \cite{Chandrasekhar_1983}, which was recently extended to slowly-rotating BHs at $\mathcal{O}(\chi^1)$ by \cite{Cano:2021myl}. For Kerr, there has not been a map found between the Teukolsky equation and the RW/ZM equations, since the latter is not known for BHs with arbitrary spin. Some Chandrasekhar-like transformations have been developed, for example, to convert the long-range potential of the Teukolsky equation to a short-range potential \cite{Sasaki:1981sx}. All these transformations are special cases of generalized Darboux transformations \cite{Darboux}, which have been widely used in the study of supersymmetry \cite{Glampedakis:2017rar}. One future avenue is then extending the Chandrasekhar transformation to our modified Teukolsky equation in the slow-rotation expansion and comparing it to the modified ZM and RW equations found in \cite{Cardoso:2009pk, Molina:2010fb, Pani_Cardoso_Gualtieri_2011, Wagle_Yunes_Silva_2021, Srivastava_Chen_Shankaranarayanan_2021, Pani:2009wy, Blazquez-Salcedo:2016enn, Blazquez-Salcedo_Khoo_Kunz_2017, Pierini:2021jxd, Pierini:2022eim, Cano:2020cao, Cano:2021myl}.
	
    Our framework provides a novel approach to studying isospectrality breaking in modified gravity. This work is an intermediate but imperative step in using the modified Teukolsky equations to compute the shifts of QNM frequencies. For BHs with arbitrary spin, our framework is also the only viable analytical approach to study isospectrality breaking since QNMs can only be computed from the (modified) Teukolsky equation, and no (modified) ZM/RW equations are known in this case. Recent works have used spectral methods to investigate quasinormal modes (QNMs) within the context of Schwarzschild~\cite{Chung:2023zdq} and Kerr~\cite{Chung:Kerr} BHs in GR. It is conceivable that these methodologies may be extensible to explore the QNMs of rotating BHs within modified theories of gravity. With this work, we look forward to developing a deeper understanding of these isospectrality-breaking theories of gravity using BH spectroscopy.
	
    \section*{Acknowledgements} \label{sec:Acknowledgements}
	
    We thank Sizheng Ma for the valuable discussions. D.L. and Y.C.'s research is supported by the Simons Foundation (Award No. 568762), the Brinson Foundation, and the National Science Foundation (via Grants No. PHY-2011961 and No. PHY-2011968). P.W. and N.Y. are supported by the Simons Foundation through Award number 896696 and from the National Science Foundation Award PHY-2207650. A.H. and A.Z. are supported by
    the National Science Foundation (via Grant No. PHY-2207594). 
    This work has been assigned preprint number UTWI-14-2023.

    \appendix
    \section{Properties of operators $\hat{P}$, $\hat{\mathcal{C}}$, $\hat{\mathcal{P}}$}
    \label{appendix:parity_operators}
	
    In this appendix, we provide some useful relations for the operators $\hat{P}$, $\hat{\mathcal{C}}$, and $\hat{\mathcal{P}}$ defined in Sec.~\ref{sec:definite_parity_modes_GR}. Let $\alpha,\beta\in\mathbb{C}$, $f,g \in \Lambda^0(\mathcal{U})$, and $\hat{\mathcal{I}}$ be the identity operator. Then $\hat{P}$ defined in Eq.~\eqref{eq:def_parity_trans} satisfies
    \begin{subequations} \label{eq:p_properties}
    \begin{align}
        & \hat{P}^2=\hat{\mathcal{I}}\,, \\
        & \hat{P}[\alpha f +\beta g]
        =\alpha \hat{P}[f]+\beta \hat{P}[g]\,, \\
        & \hat{P}[f\cdot g\cdot h\cdots]
        =\hat{P}[f]\cdot \hat{P}[g]\cdot \hat{P}[h]\cdots\,.
    \end{align}
    \end{subequations}
    The parity operator $\hat{P}$ commutes with all the derivatives:
    \begin{align} \label{eq:p_derivatives}
    [\hat{P},\partial_{t}]=[\hat{P},\partial_{r}]=[\hat{P},\partial_{\phi}]=0
    \end{align}
    except the $\theta$ derivative, with which it anti-commutes:
    \begin{align} \label{eq:p_dtheta}
        \{\hat{P},\partial_{\theta}\}=0\,.
    \end{align}
    
    For the complex conjugate operator $\hat{\mathcal{C}}$, we have
    \begin{subequations} \label{eq:C_properties}
    \begin{align}
        & \hat{\mathcal{C}}^2=\hat{\mathcal{I}}\,, \\
        & \hat{\mathcal{C}}[\alpha f+\beta g]
        =\bar{\alpha}\hat{\mathcal{C}}[f]+\bar{\beta}\hat{\mathcal{C}}[g]\,, \\
        & \hat{\mathcal{C}}[f\cdot g\cdot h\cdots]
        = \hat{\mathcal{C}}[f]\cdot\hat{\mathcal{C}}[g]\cdot\hat{\mathcal{C}}[h]\cdots\,.
    \end{align}
    \end{subequations}
    Since the coordinates are all real, the complex conjugate operator $\hat{\mathcal{C}}$ commutes with all the derivatives:
    \begin{align} \label{eq:C_derivatives}
        [\hat{\mathcal{C}},\partial_{t}]
        =[\hat{\mathcal{C}},\partial_{r}]
        =[\hat{\mathcal{C}},\partial_{\phi}]
        =[\hat{\mathcal{C}},\partial_{\theta}]=0\,.
    \end{align}
    In addition, $\hat{P}$ and $\hat{\mathcal{C}}$ commute with each other. 
    
    In Eq.~\eqref{eq:def_P}, we have combined $\hat{P}$ and $\hat{\mathcal{C}}$ to define another operator $\hat{\mathcal{P}}$, where
    \begin{align}
        \hat{\mathcal{P}}=\hat{\mathcal{C}}\hat{P}\,.
    \end{align}
    Using Eqs.~\eqref{eq:p_properties}--\eqref{eq:C_derivatives}, we find
    \begin{subequations}
    \begin{align}
        & \hat{\mathcal{P}}^2=\hat{\mathcal{I}}\,, \\
        & \hat{\mathcal{P}}[\alpha f +\beta g]
        =\bar{\alpha}\hat{\mathcal{P}}[f]+\bar{\beta}\hat{\mathcal{P}}[g]\,, \\
        & \hat{\mathcal{P}}[f\cdot g\cdot h\cdots]
        =\hat{\mathcal{P}}[f]\cdot\hat{\mathcal{P}}[g]\cdot \hat{\mathcal{P}}[h]\cdots\,.
    \end{align}
    \end{subequations}
    and
    \begin{align} \label{eq:P_properties}
        [\hat{\mathcal{P}},\partial_{t}]
        =[\hat{\mathcal{P}},\partial_{r}]
        =[\hat{\mathcal{P}},\partial_{\phi}]=0\,,\quad
        \{\hat{\mathcal{P}},\partial_{\theta}\}=0\,.
    \end{align}
    
    \section{Reconstruction of NP quantities} \label{appendix:metric_reconstruction_more}
	
    In this appendix, we provide some additional equations of reconstructed NP quantities following \cite{Campanelli_Lousto_1999, Loutrel_Ripley_Giorgi_Pretorius_2020}. Let us first assume a general reconstructed metric $h_{\mu\nu}$ without going to the specific IRG or ORG. In \cite{dcstyped1}, we have only considered the case that the background spacetime is Petrov type D, so the results here are more general. Then to reconstruct NP quantities, the first step is to reconstruct the tetrad. We can first express the reconstructed tetrad in terms of the background tetrad
    \begin{equation} \label{eq:tetrad_expansion}
	    e^{\mu(1)}_{a}=A_{a}{}^{b(1)}e^{\mu(0)}_{b}\,.
    \end{equation}
    As shown in \cite{Campanelli_Lousto_1999, Loutrel_Ripley_Giorgi_Pretorius_2020}, one can always use the six degrees of freedom of tetrad rotations to set some of the $A^{b(1)}_{a}$ coefficients to $0$. Then expanding $h_{\mu\nu}$ in terms of $e^{a(1)}_{\mu}$ and $e^{a(0)}_{\mu}$ using the completeness relation
    \begin{equation} \label{eq:completeness}
	    g_{\mu\nu}=-2l_{(\mu} n_{\nu)}+2 m_{(\mu}\bar{m}_{\nu)}
    \end{equation}
    and its expansion
    \begin{equation} \label{eq:completeness_perturbed}
	    h_{\mu\nu}=-2\left[l_{(\mu}^{(1)}n_{\nu)}^{(0)}
	    -l_{(\mu}^{(0)}n_{\nu)}^{(1)}
	    +m_{(\mu}^{(1)}\bar{m}_{\nu)}^{(0)}
	    +m_{(\mu}^{(0)}\bar{m}_{\nu)}^{(1)}\right]\,,
    \end{equation}
    one can find that \cite{Campanelli_Lousto_1999, Loutrel_Ripley_Giorgi_Pretorius_2020},
    \begin{subequations} \label{eq:perturbed_tetrad}
    \begin{align}
        & l^{\mu(1)}=\frac{1}{2}h_{ll}n^{\mu}\,, \\
        & n^{\mu(1)}=\frac{1}{2}h_{nn}l^{\mu}+h_{ln}n_{\mu}\,, \\
        & m^{\mu(1)}=h_{nm}l^{\mu}+h_{lm}n^{\mu}
        -\frac{1}{2}h_{m\bar{m}}m^{\mu}
        -\frac{1}{2}h_{mm}\bar{m}^{\mu}\,,
    \end{align}
    \end{subequations}
    where we have dropped the superscripts of $e^{\mu(0)}_{a}$ and $h^{(1)}_{ab}$ for simplicity. Notice that the perturbed tetrad in Eq.~\eqref{eq:perturbed_tetrad} has an opposite sign from the one in \cite{Campanelli_Lousto_1999, Loutrel_Ripley_Giorgi_Pretorius_2020} since we used an opposite signature, as one can see in Eqs.~\eqref{eq:completeness} and \eqref{eq:completeness_perturbed}.
	
    To find the spin coefficients, we follow the idea in \cite{Chandrasekhar_1983} to expand the commutation relation defining Ricci rotation coefficients
    \begin{equation} \label{eq:commutation}
	    \left[e_{a}^{\mu},e_{b}^{\mu}\right]
        =\left(\gamma^{c}{}_{ba}-\gamma^{c}{}_{ab}\right)e_{c}^{\mu}
        =C_{ab}{}^{c}e_{c}^{\mu}\,,
    \end{equation}
    and spin coefficients are just linear combinations of Ricci rotation coefficients \cite{Chandrasekhar_1983},
    \begin{align} \label{eq:spin_coefs}
	    & \kappa=\gamma_{131}\,,\quad
        \pi=-\gamma_{241}\,,\quad
        \varepsilon=\frac{1}{2}(\gamma_{121}-\gamma_{341})\,, \nonumber \\
        & \rho=\gamma_{134}\,,\quad
        \lambda=-\gamma_{244}\,,\quad
        \alpha=\frac{1}{2}(\gamma_{124}-\gamma_{344})\,,
        \nonumber \\
        & \sigma=\gamma_{133}\,,\quad
        \mu=-\gamma_{243}\,,\quad
        \beta=\frac{1}{2}(\gamma_{123}-\gamma_{343})\,,
        \nonumber \\
        & \tau=\gamma_{132}\,,\quad
        \nu=-\gamma_{242}\,,\quad
        \gamma=\frac{1}{2}(\gamma_{122}-\gamma_{342})\,.
    \end{align}
    Expanding Eq.~\eqref{eq:commutation} using Eq.~\eqref{eq:tetrad_expansion}, one then finds
    \begin{equation} \label{eq:C_expand}
    \begin{aligned}
	    C_{ab}{}^{c(1)}=& \;\partial_{a}A_{b}{}^{c}-\partial_{b}A_{a}{}^{c} \\
        & \;-\left(A_{a}{}^{d}C_{bd}{}^{c}-A_{b}{}^{d}C_{ad}{}^{c}
        +A_{d}{}^{c}C_{ab}{}^{d}\right)\,,  
    \end{aligned}
    \end{equation}
    where we have dropped the superscript of $C_{ab}{}^{c(0)}$ at the right-hand side. Inserting Eqs.~\eqref{eq:perturbed_tetrad} and \eqref{eq:spin_coefs} into Eq.~\eqref{eq:C_expand} and using the definition in Eq.~\eqref{eq:spin_coefs}, we find the perturbed spin coefficients to be
    \begin{subequations} \label{eq:perturbed_spin_coefs}
    \begin{align}
	    & \begin{aligned}
	         \kappa^{(1)}
            =& \;\frac{1}{2}\delta_{[-2,-2,1,1]}h_{ll}-D_{[-2,0,0,-1]}h_{lm} \\
            & \;-\kappa h_{ln}+\sigma h_{l\bar{m}}
            -\frac{1}{2}\bar{\kappa}h_{mm}-\frac{1}{2}\kappa h_{m\bar{m}}\,, \\
	    \end{aligned} \\
        & \begin{aligned}
            \sigma^{(1)}
            =& \;-\frac{1}{2}D_{[-2,2,1,-1]}h_{m m}
            +(\bar{\pi}+\tau)h_{lm}
            -\frac{1}{2}\bar{\lambda}h_{ll}\,, \\
        \end{aligned} \\
        & \begin{aligned}
            \lambda^{(1)}
            =& \;(\pi+\bar{\tau})h_{n\bar{m}}
            +\frac{1}{2}\boldsymbol{\Delta}_{[-1,1,2,-2]}h_{\bar{m}\bar{m}} \\
            & \;+\lambda h_{ln}-\frac{1}{2}\bar{\sigma}h_{nn}\,, \\
        \end{aligned} \\
        & \begin{aligned}
            \nu^{(1)}
            =& \;-\frac{1}{2}\bar{\delta}_{[2,2,-1,-1]}h_{nn}
            +\boldsymbol{\Delta}_{[0,1,2,0]}h_{n\bar{m}} \\
            & \;+\nu h_{ln}+\lambda h_{nm}-\frac{1}{2}\nu h_{m\bar{m}}
            -\frac{1}{2}\bar{\nu}h_{\bar{m}\bar{m}}\,, \\
        \end{aligned} \\
        & \begin{aligned}
            \epsilon^{(1)}
            =& \;\frac{1}{4}\Big[\boldsymbol{\Delta}_{[-1,1,0,-2]}h_{ll}
            -2D_{[0,0,\frac{1}{2},-\frac{1}{2}]}h_{ln} \\
            & \;-\bar{\delta}_{[-2,0,-3,-2]}h_{lm}
            +\delta_{[-2,0,1,2]}h_{l\bar{m}}
            -(\rho-\bar{\rho})h_{m\bar{m}} \\
            & \;-\bar{\kappa}h_{nm}+\kappa h_{n\bar{m}}
            -\bar{\sigma}h_{mm}+\sigma h_{\bar{m}\bar{m}}\Big]\,, \\
        \end{aligned} \\
        & \begin{aligned}
            \rho^{(1)}
            =& \;\frac{1}{2}\Big[-\mu h_{ll}-(\rho-\bar{\rho}) h_{l n}
            -\bar{\delta}_{[-2,0,-1,0]}h_{lm} \\
            & \;+\delta_{[-2,0,1,2]}h_{l\bar{m}}
            -D_{[0,0,1,-1]}h_{m\bar{m}} \\
            & \;-\bar{\kappa}h_{nm}+\kappa h_{n\bar{m}}\Big]\,, \\
        \end{aligned} \\
        & \begin{aligned}
            \mu^{(1)}
            =& \;\frac{1}{2}\Big[-\rho h_{nn}
            -\bar{\delta}_{[0,2,-2,-1]}h_{nm}
            +\delta_{[0,2,0,1]}h_{n\bar{m}}\\
            & \;+(\mu+\bar{\mu})h_{ln}
            +\boldsymbol{\Delta}_{[-1,1,0,0]}h_{m\bar{m}} \\
            & \;+\nu h_{lm}-\bar{\nu}h_{l\bar{m}}\Big]\,, \\
        \end{aligned} \\
	    & \begin{aligned}
	        \gamma^{(1)}
            =& \;\frac{1}{4}\Big[-D_{[0,2,1,-1]}h_{nn}
            -\bar{\delta}_{[0,2,-2,-1]}h_{nm}\\
            & \;+\delta_{[0,2,2,3]}h_{n\bar{m}}
            -(\mu-\bar{\mu}-4\gamma)h_{ln}
            -(\mu-\bar{\mu})h_{m\bar{m}} \\
            & \;+\nu h_{lm}-\bar{\nu}h_{l\bar{m}}
            +\lambda h_{mm}-\bar{\lambda}h_{\bar{m}\bar{m}}\Big]\,, \\
	    \end{aligned} \\
        & \begin{aligned}
            \alpha^{(1)}
            =& \;\frac{1}{4}\Big[-D_{[-2,0,-1,-2]}h_{n\bar{m}}
            +\delta_{[-2,0,1,1]}h_{\bar{m}\bar{m}}\\
            & \;-\bar{\delta}_{[0,0,-1,-1]}h_{ln}
            +\boldsymbol{\Delta}_{[-2,1,4,-2]}h_{l\bar{m}}
            -\bar{\delta}_{[2,0,-1,-1]}h_{m\bar{m}} \\
            & \;-\nu h_{ll}+3\lambda h_{lm}-\bar{\kappa}h_{nn}
            -\bar{\sigma}h_{nm}\Big]\,, \\
        \end{aligned} \\
        & \begin{aligned}
            \beta^{(1)}
            =& \;\frac{1}{4}\Big[-D_{[-4,2,2,-1]}h_{nm}
            -\bar{\delta}_{[0,2,-1,-1]}h_{mm}\\
            & \;-\delta_{[0,0,-1,-1]}h_{ln}
            +\boldsymbol{\Delta}_{[1,2,2,0]}h_{lm}
            +\delta_{[0,-2,1,1]}h_{m\bar{m}} \\
            & \;-\bar{\nu}h_{ll}-\bar{\lambda}h_{l\bar{m}}
            -\kappa h_{nn}+3\sigma h_{n\bar{m}}\Big]\,,\\
        \end{aligned} \\
        & \begin{aligned}
            \pi^{(1)}
            =& \;\frac{1}{2}\Big[D_{[2,0,-1,0]}h_{n\bar{m}}
            +\tau h_{\bar{m}\bar{m}} \\
            & \;-\delta_{[0,0,-1,-1]}h_{ln}
            +\boldsymbol{\Delta}_{[0,1,0,-2]}h_{l\bar{m}}
            +\bar{\tau}h_{m\bar{m}} \\
            & \;+\lambda h_{lm}-\bar{\sigma}h_{nm}\Big]\,,
        \end{aligned} \\
        & \begin{aligned}
            \tau^{(1)}
            =& \;\frac{1}{2}\Big[-D_{[0,2,0,-1]}h_{nm}
            +\pi h_{mm} \\
            & \;+\delta_{[0,0,1,1]}h_{ln}
            -\boldsymbol{\Delta}_{[1,0,-2,0]}h_{lm}
            +\bar{\pi}h_{m\bar{m}} \\
            & \;-\bar{\lambda}h_{l\bar{m}}+\sigma h_{n\bar{m}}\Big]\,. \\
        \end{aligned}
    \end{align}
    \end{subequations}
    In Eq.~\eqref{eq:perturbed_spin_coefs}, we do not assume anything about the background spacetime, so the background may be Petrov type I, and all the spin coefficients at the background can be nonzero. Thus, we can use the above equations for our analysis in Sec.~\ref{sec:definite_parity_modes_bGR}. For Petrov type D spacetimes in GR, where $\kappa^{(0,0)}=\sigma^{(0,0)}=\lambda^{(0,0)}=\nu^{(0,0)}=0$, our result is the same as the one in \cite{Loutrel_Ripley_Giorgi_Pretorius_2020} up to a minus sign due to the opposite signature we used. However, the result in \cite{Campanelli_Lousto_1999} has some discrepancies with the result here and in \cite{Loutrel_Ripley_Giorgi_Pretorius_2020}, which might be due to errors. If we additionally use the IRG, we can then further set $h_{ll}=h_{ln}=h_{lm}=h_{l\bar{m}}=h_{m\bar{m}}=0$ in Eq.~\eqref{eq:perturbed_spin_coefs}.
	
    To find the perturbed Weyl scalars, one can use Ricci identities to compute Weyl scalars from spin coefficients,
    \begin{subequations} \label{eq:Weyl_scalar_Ricci}
    \begin{align} 
	    & \Psi_0
	    =D_{[-3,1,-1,-1]}\sigma-\delta_{[-1,-3,1,-1]}\kappa\,, \\
        & \Psi_1
        =D_{[0,1,0,-1]}\beta-\delta_{[-1,0,1,0]}\varepsilon
        -(\alpha+\pi)\sigma+(\gamma+\mu)\kappa\,, \\
        & \begin{aligned}
            \Psi_2
            =& \;\frac{1}{3}\Big[\bar{\delta}_{[-2,1,-1,-1]}\beta
            -\delta_{[-1,0,1,1]}\alpha \\
            & \;+D_{[1,1,1,-1]}\gamma
            -\boldsymbol{\Delta}_{[-1,1,-1,-1]}\varepsilon \\
            & \;+\bar{\delta}_{[-1,1,-1,-1]}\tau
            -\boldsymbol{\Delta}_{[-1,1,-1,-1]}\rho \\
            & \;+2(\nu\kappa-\lambda\sigma)\Big]\,, \\
        \end{aligned} \\
        & \Psi_3
        =\bar{\delta}_{[0,1,0,-1]}\gamma
        -\boldsymbol{\Delta}_{[0,1,0,-1]}\alpha
        +(\varepsilon+\rho)\nu-(\beta+\tau)\lambda\,, \\
        & \Psi_4
        =\bar{\delta}_{[3,1,1,-1]}\nu
        -\boldsymbol{\Delta}_{[1,1,3,-1]}\lambda\,.
    \end{align}
    \end{subequations}
    Equation~\eqref{eq:Weyl_scalar_Ricci} works at all order for any spacetime, so we can use them for our analysis in Sec.~\ref{sec:definite_parity_modes_bGR}. Here, we have also followed \cite{Campanelli_Lousto_1999, Loutrel_Ripley_Giorgi_Pretorius_2020} to linearly combine certain Ricci identities such that there are no NP Ricci scalars $\Phi_{ab}$ in the equations, and the equations work for non-vacuum spacetime. Using the NP quantities on the background with the perturbed tetrad in Eq.~\eqref{eq:perturbed_tetrad} and the perturbed spin coefficients in Eq.~\eqref{eq:perturbed_spin_coefs}, one can then write down the perturbed Weyl scalars in terms of metric perturbations directly.
	
    For Petrov type D spacetimes in GR, using Eqs.~\eqref{eq:metric_Hertz_ORG} and \eqref{eq:metric_Hertz_IRG}, one can further write down the perturbed NP quantities in terms of the Hertz potential. In \cite{Kegeles_Cohen_1979, Keidl_Friedman_Wiseman_2007}, they computed the perturbed Weyl scalars directly from the Riemann tensor, and they found in the IRG in Eq.~\eqref{eq:metric_Hertz_IRG},
    \begin{subequations} \label{eq:Weyl_scalar_Hertz}
    \begin{align}
	    & \Psi_0^{(0,1)}
	    =-\frac{1}{2}D_{[-3,1,0,-1]}D_{[-2,2,0,-1]}h_{mm}\,, \\
	    & \begin{aligned}
            \Psi_1^{(0,1)}
            =& \;-\frac{1}{8}\Big[2D_{[-1,1,1,-1]}D_{[0,2,1,-1]}h_{nm} \\
            & \;+D_{[-1,1,1,-1]}\delta_{[-2,2,-2,-1]}h_{mm} \\
            & \;+\bar{\delta}_{[-3,1,-3,-1]}D_{[-2,2,0,-1]}h_{mm}\Big]\,,
	    \end{aligned} \\
	    & \begin{aligned}
            \Psi_2^{(0,1)}
            =& \;-\frac{1}{12}\Big[D_{[1,1,2,-1]}D_{[2,2,2,-1]}h_{nn}^{1} \\
            & \;+2\left(D_{[1,1,2,-1]}\bar{\delta}_{[0,2,-1,-1]}\right. \\
            & \;\left.+\bar{\delta}_{[-1,1,-2,-1]}D_{[0,2,1,-1]}\right)h_{nm} \\
            & \;+\bar{\delta}_{[-1,1,-2,-1]}\bar{\delta}_{[-2,2,-2,-1]}h_{mm}\Big]\,,
	    \end{aligned} \\
	    & \begin{aligned}
            \Psi_3^{(0,1)}
            =& \;-\frac{1}{8}\Big[\left(D_{[3,1,3,-1]}
            \bar{\delta}_{[2,2,0,-1]}\right. \\
            & \;\left.+\bar{\delta}_{[1,1,-1,-1]}D_{[2,2,,2,-1]}\right)h_{nn}^{1} \\
            & \;+\bar{\delta}_{[1,1,-1,-1]}\bar{\delta}_{[0,2,-1,-1]}h_{nm}\Big]\,,
	    \end{aligned} \\
	    & \begin{aligned}
            \Psi_4^{(0,1)}
            =& \;-\frac{1}{2}\Big[\bar{\delta}_{[3,1,0,-1]}
            \bar{\delta}_{[2,2,0,-1]}h_{nn}^{1} \\
            & \;+3\Psi_2\left(\tau\bar{\delta}_{[4,0,0,0]}
            -\rho\boldsymbol{\Delta}_{[0,0,4,0]}\right. \\
            & \;\left.-\mu D_{[4,0,0,0]}+\pi\delta_{[0,4,0,0]}
            +2\Psi_2\right)\bar{\Psi}_{\Hertz}\Big]\,,
        \end{aligned}
    \end{align}
    \end{subequations}
    where $\Psi_{0,4}^{(0,1)}$ reduce to Eq.~\eqref{eq:Hertz_IRG} in the Boyer-Lindquist coordinates of Kerr. We have also defined $h_{nn}^{1}$ to be the piece of $h_{nn}$ proportional to $\bar{\Psi}_{\Hertz}$ in Eq.~\eqref{eq:metric_Hertz_IRG}, i.e., $h_{nn}^{1}=\bar{\delta}_{[1,3,0,-1]}\bar{\delta}_{[0,4,0,3]}\bar{\Psi}_{\Hertz}$.
	
    To compare our results with Eq.~\eqref{eq:Weyl_scalar_Hertz}, we compute the perturbed Weyl scalars using the Ricci identities in Eq.~\eqref{eq:Weyl_scalar_Ricci} in Kerr such that $\varepsilon^{(0,0)}=0$. We also perform a direct calculation by linearizing the Riemann tensor first and then projecting it into the NP basis. For both calculations, we use the tetrad in Eq.~\eqref{eq:perturbed_tetrad} with the IRG, and we find an agreement for $\Psi_{0,1,2,4}^{(0,1)}$. While for $\Psi_{3}^{(0,1)}$, we find a disagreement. This is not very surprising since $\Psi_{3}^{(0,1)}$ is not invariant under both tetrad rotations and infinitesimal coordinate changes at $\mathcal{O}(\epsilon)$. Since both our calculation and Refs.~\cite{Kegeles_Cohen_1979, Keidl_Friedman_Wiseman_2007} use the IRG, we have used the same coordinate freedom. This is also manifested by that our $\Psi_{2}^{(0,1)}$ matches Eq.~\eqref{eq:Weyl_scalar_Hertz}, which is invariant under tetrad rotations at $\mathcal{O}(\epsilon)$ but not invariant under coordinate transformations at $\mathcal{O}(\epsilon)$. Thus, the difference between our result and Eq.~\eqref{eq:Weyl_scalar_Hertz} is due to different tetrad choices, while Refs.~\cite{Kegeles_Cohen_1979, Keidl_Friedman_Wiseman_2007} did not clearly specify their tetrad at $\mathcal{O}(\epsilon)$. 
	
    In the case of Schwarzschild, with the tetrad in Eq.~\eqref{eq:perturbed_tetrad} and the perturbed metric in the IRG in Eq.~\eqref{eq:metric_Hertz_IRG}, we find
    \begin{subequations} \label{eq:Weyl_scalar_Hertz_Schw}
    \begin{align}
        & \Psi_0^{(0,1)}=-\frac{1}{2}D^4\bar{\Psi}_{\Hertz}\,, \\
        & \Psi_1^{(0,1)}=-\frac{1}{2}D^3(\bar{\delta}+4\beta)\bar{\Psi}_{\Hertz}\,, \\
        & \Psi_2^{(0,1)}=-\frac{1}{2}D^2
        (\bar{\delta}+2\beta)(\bar{\delta}+4\beta)\bar{\Psi}_{\Hertz}\,, \\
        & \Psi_3^{(0,1)}=-\frac{1}{2}D\bar{\delta}
        (\bar{\delta}+2\beta)(\bar{\delta}+4\beta)\bar{\Psi}_{\Hertz}
         +\frac{3}{2}\Psi_{2}h_{n\bar{m}}\,, \\
        & \begin{aligned}
            \Psi_4^{(0,1)}
            =& \;-\frac{1}{2}(\bar{\delta}-2\beta) \bar{\delta}
            (\bar{\delta}+2\beta)(\bar{\delta}+4\beta)\bar{\Psi}_{\Hertz} \\
            & \;+\frac{3}{2}\Psi_2\left[\mu D+\rho(\boldsymbol{\Delta}+4\gamma)
            -2 \Psi_2\right]\Psi_{\Hertz}\,,
        \end{aligned}
    \end{align}
    \end{subequations}
    which is the same as Eq.~\eqref{eq:Weyl_scalar_Hertz} in the Schwarzschild limit, except there is an additional $\frac{3}{2}\Psi_{2}h_{n\bar{m}}$ correction to $\Psi_3^{(0,1)}$ due to different tetrad choices.

    \section{Radial Functions in $\mathcal{S}_{\geo}^{(1,1)}$} \label{appendix:radial_functions}
    In this appendix, we provide the radial functions $C_i(r)$ in Eq.~\eqref{eq:H0110dCS} and $D_i(r)$ in Eq.~\eqref{eq:H0100EdGB}. The radial functions $C_i(r)$ are found in \cite{dcstyped1}, where
    \begin{subequations}
    \begin{align}
        \begin{split} \label{eq:C1}
            C_1(r)=& \;57960M^4-39316M^3r-694M^2r^2 \\
            & \;-1050Mr^3+2345r^4 \,,
        \end{split} \\
        C_2(r)=&\;189M^3+120M^2r+70Mr^2\,, \label{eq:C2}\\
        \begin{split} \label{eq:C3}
            C_3(r)=& \;1602M^3-1056M^2r-515Mr^2 \\
            & \;-255r^3\,,
        \end{split} \\
        C_4(r)=& \;954M^2+440Mr+175r^2\,, \label{eq:C4}\\
        C_5(r)=& \;4680M^3-518M^2r-360Mr^2-335r^3\,. \label{eq:C5}
    \end{align}
    \end{subequations}
    The radial functions $D_i(r)$ are given by
    \begin{widetext}
    \begin{subequations}\label{eq:Dfunctions}
    \begin{align}
    & \begin{aligned}\label{eq:D1}
          \frac{1}{(64\pi)^2M}15 r^9 (r-2 M)^3D_1(r)
        = & 168960 M^9+6720 M^8 r-232448 M^7 r^2+129928 M^6 r^3-24108 M^5r^4  \\ & + 13900 M^4 r^5-8090 M^3 r^6+1530 M^2 r^7-150 M r^8+15 r^9\,,
    \end{aligned} \\
    & \begin{aligned}\label{eq:D2}
        \frac{1}{(64\pi)^2M}15 r^7 (r-2 M)^3D_2(r)
        = & 253440 M^8-344992 M^7 r+146720 M^6 r^2-28584 M^5 r^3+16872 M^4r^4 \\& -8240 M^3 r^5+1210 M^2 r^6-75 M r^7+15 r^8\,,
    \end{aligned} \\
    & \begin{aligned}\label{eq:D3}
        \frac{1}{(64\pi)^2M}15 r^8 (r-2 M)^2D_3(r)
        = & 212160 M^8-310624 M^7 r+139352 M^6 r^2-25728 M^5 r^3+14630 M^4r^4 \\& -7720 M^3 r^5+1275 M^2 r^6-120 M r^7+15 r^8\,,
    \end{aligned} \\
    & \begin{aligned}\label{eq:D4}
        & \frac{1}{(64\pi)^2 M^3}15 r^5 (r-2 M)^2D_4(r)
        =400 M^4-96 M^3 r-66 M^2 r^2-130 M r^3-5 r^4\,,
    \end{aligned} \\
    & \begin{aligned}\label{eq:D5}
       & \frac{1}{(64\pi)^2 M^2}15 r^7D_5(r)
        =1840 M^5+48 M^4 r-30 M^3 r^2-260 M^2 r^3-15 M r^4-15 r^5\,.
    \end{aligned}
    \end{align}
    \end{subequations}
    \end{widetext}
    
    \section{$\mathcal{S}$ in the modified Teukolsky equations}
    \label{appendix:source_terms}
		
    In this appendix, we present the source term $\mathcal{S}$ of the modified Teukolsky equations due to the effective stress tensor for non-rotating BHs in dCS and EdGB gravity. Here, we only briefly summarize the procedure in \cite{dcstyped1} and apply it to these two simple non-rotating examples. In addition, for EdGB gravity, we only focus on the source terms with dynamical scalar field $\varphi^{(1,0,1)}$ for simplicity. For a more complete prescription of how to evaluate these source terms, one can refer to \cite{dcstyped1} for slowly-rotating BHs in dCS gravity. The procedure in \cite{dcstyped1} can be extended to BHs with arbitrary spin in dCS and other modified gravity. 
	
    \subsection{dCS} \label{appendix:source_terms_dCS}
	
    As discussed in Sec.~\ref{sec:dCS_S}, for non-rotating BHs in dCS, the only nonzero contribution of Eq.~\eqref{eq:Ricci_dCS} is the term $\left(\nabla^{\sigma}\nabla^{\delta}\vartheta\right)$. In addition, since $\vartheta^{(1,0,0)}$ vanishes, $\vartheta$ only has dynamical contribution $\vartheta^{(1,0,1)}$. For the same reason, all the metric fields in $\mathcal{S}_{\dCS}^{(1,0,1)}$ are evaluated on the stationary Schwarzschild background, so no metric reconstruction is needed. At $\mathcal{O}(\zeta^1,\chi^0,\epsilon^1)$, the only place requiring metric reconstruction is to solve the equation of motion of $\vartheta^{(1,0,1)}$ since it is driven by $\Psi_2^{(0,0,1)}$ [i.e., Eq.~\eqref{eq:R*R_001}]. Given $\vartheta^{(1,0,1)}$ is solved, one can then project the term $\left(\nabla^{\sigma}\nabla^{\delta}\vartheta\right)$ onto the NP tetrad and evaluate all the metric fields using their Schwarzschild values.
	
    Let us present this calculation in more detail. First, inspecting the source terms $S_{1,2}$ in Eq.~\eqref{eq:source_bianchi} of the Bianchi identities in Eqs.~\eqref{eq:BianchiId_Psi0_1_simplify} and \eqref{eq:BianchiId_Psi0_2_simplify}, the only nonzero contributions of $\Phi_{ij}$ are from $\Phi_{00}$, $\Phi_{01}$, and $\Phi_{02}$ since $\kappa^{(1,0,1)}=\sigma^{(1,0,1)}=\lambda^{(1,0,1)}=0$. Then from the definition
    \begin{equation}
	    \Phi_{00}=\frac{1}{2}R_{11}\,,\quad
	    \Phi_{01}=\frac{1}{2}R_{13}\,,\quad
	    \Phi_{02}=\frac{1}{2}R_{33}\,,
    \end{equation}
    we notice the only relevant components of $R_{\mu\nu}$ are $R_{11}$, $R_{13}$, and $R_{33}$. Projecting the equation of $R_{\mu\nu}$ in Eq.~\eqref{eq:Ricci_dCS} onto the NP tetrad, we find
    \begin{widetext}
    \begin{subequations} \label{eq:R_ab_dCS}
    \begin{align}
        & \begin{aligned} \label{eq:R_11_dCS}
            R_{11}^{\dCS}
		  =& \;i\mathcal{R}^{\dCS}_1\bigg\{(D\vartheta)
            \Big[\lambda\Psi_0-\bar{\lambda}
		  \bar{\Psi}_0-(\alpha+\bar{\beta}+\pi)\Psi_1
		  +(\bar{\alpha}+\beta+\bar{\pi})\bar{\Psi}_1
		  +(\varepsilon+\bar{\varepsilon})(\Psi_2-\bar{\Psi}_2)\Big] \\
		  & \;-(\boldsymbol{\Delta}\vartheta)\Big[\bar{\sigma}\Psi_0
            -\sigma\bar{\Psi}_0
		  -\bar{\kappa}\Psi_1+\kappa\bar{\Psi}_1\Big] \\
		  & \;+(\delta\vartheta)\Big[(\bar{\alpha}
            -\beta)\bar{\Psi}_0+\bar{\sigma}
		  \Psi_1+(\varepsilon-\bar{\varepsilon}-\bar{\rho})\bar{\Psi}_1
		  -\bar{\kappa}(\Psi_2-\bar{\Psi}_2)\Big] \\
		  & \;-(\bar{\delta}\vartheta)\Big[(\alpha-\bar{\beta})\Psi_0
            -(\varepsilon-\bar{\varepsilon}+\rho)\Psi_1+\sigma\bar{\Psi}_1
		  +\kappa(\Psi_2-\bar{\Psi}_2)\Big] \\
		  & \;-\frac{1}{2}\Psi_0\{\bar{\delta},\bar{\delta}\}\vartheta
		  +\frac{1}{2}\bar{\Psi}_0\{\delta,\delta\}\vartheta
		  +\Psi_1\{D,\bar{\delta}\}\vartheta
		  -\bar{\Psi}_1\{D,\delta\}\vartheta
		  -\frac{1}{2}(\Psi_2-\bar{\Psi}_2)\{D,D\}\vartheta\bigg\}
		  +\mathcal{R}^{\dCS}_2(D\vartheta)(D\vartheta)\,,
		\end{aligned} \\
		& \begin{aligned} \label{eq:R_13_dCS}
		  R_{13}^{\dCS}
		  =& \;\frac{i}{2}\mathcal{R}^{\dCS}_1\bigg\{(D\vartheta)\Big[\nu\Psi_0
		  -(\gamma+\bar{\gamma}+\mu+\bar{\mu})\Psi_1
		  -2\bar{\lambda}\bar{\Psi}_1+(\bar{\alpha}+\beta+\bar{\pi})
		  (\Psi_2+2\bar{\Psi}_2)
            -2(\varepsilon+\bar{\varepsilon})\bar{\Psi}_3\Big] \\
		  & \;-(\boldsymbol{\Delta}\vartheta)
            \Big[(\alpha+\bar{\beta}+\bar{\tau})\Psi_0
		  -(\varepsilon+\bar{\varepsilon}+\rho+\bar{\rho})\Psi_1
		  -2\sigma\bar{\Psi}_1+\kappa(\Psi_2+2\bar{\Psi}_2)\Big] \\
		  & \;+(\delta\vartheta)\Big[\lambda\Psi_0-(\alpha-\bar{\beta}
		  +\pi-\bar{\tau})\Psi_1+2(\bar{\alpha}-\beta)\bar{\Psi}_1
		  +(\varepsilon-\bar{\varepsilon}-\bar{\rho})
		  (\Psi_2+2\bar{\Psi}_2)+2\bar{\kappa}\bar{\Psi}_3\Big] \\
		  & \;-(\bar{\delta}\vartheta)\Big[(\gamma-\bar{\gamma}
            -\bar{\mu})\Psi_0+(\bar{\alpha}-\beta+\bar{\pi}-\tau)\Psi_1
		  +\sigma(\Psi_2+2\bar{\Psi}_2)-2\kappa\bar{\Psi}_3\Big] \\
		  & \;-\Psi_0\{\boldsymbol{\Delta},\bar{\delta}\}\vartheta
		  +\Psi_1\Big[\{D,\boldsymbol{\Delta}\}
            +\{\delta,\bar{\delta}\}\Big]\vartheta
		  +\bar{\Psi}_1\{\delta,\delta\}\vartheta
		  -(\Psi_2+2\bar{\Psi}_2)\{D,\delta\}\vartheta
		  +\bar{\Psi}_3\{D,D\}\vartheta\bigg\}
		  +\mathcal{R}^{\dCS}_2(D\vartheta)(\delta\vartheta)\,,
        \end{aligned} \\
		& \begin{aligned} \label{eq:R_33_dCS}
            R_{33}^{\dCS}
		  =& \;i\mathcal{R}^{\dCS}_1\bigg\{-(D\vartheta)
            \Big[\bar{\nu}\Psi_1-\bar{\lambda}
            (\Psi_2-\bar{\Psi}_2)-(\bar{\alpha}+\beta+\bar{\pi})\bar{\Psi}_3 +(\varepsilon+\bar{\varepsilon})\bar{\Psi}_4\Big] \\
		  & \;-(\boldsymbol{\Delta}\vartheta)
		  \Big[(\gamma+\bar{\gamma})\Psi_0-(\bar{\alpha}+\beta+\tau)\Psi_1
		  +\sigma(\Psi_2-\bar{\Psi}_2)+\kappa\bar{\Psi}_3\Big] \\
		  & \;+(\delta\vartheta)\Big[\nu\Psi_0
            -(\gamma-\bar{\gamma}+\mu)\Psi_1
		  -(\bar{\alpha}-\beta)(\Psi_2-\bar{\Psi}_2)
		  +(\varepsilon-\bar{\varepsilon}-\bar{\rho})+\bar{\kappa}
		  \bar{\Psi}_4\Big] \\
		  & \;+(\bar{\delta}\vartheta)\Big[\bar{\nu}\Psi_0
            -\bar{\lambda}\Psi_1-\sigma\bar{\Psi}_3+\kappa\bar{\Psi}_4\Big] \\
            & \;-\frac{1}{2}\Psi_0\{\boldsymbol{\Delta},\boldsymbol{\Delta}\}\vartheta+\Psi_1
		  \{\boldsymbol{\Delta},\delta\}\vartheta
            -\frac{1}{2}(\Psi_2-\bar{\Psi}_2)
		  \{\delta,\delta\}\vartheta-\bar{\Psi}_3\{D,\delta\}\vartheta+
		  \frac{1}{2}\bar{\Psi}_4\{D,D\}\vartheta\bigg\}
		  +\mathcal{R}^{\dCS}_2(\delta\vartheta)(\delta\vartheta)\,,\\
            \mathcal{R}^{\dCS}_1\equiv&\;-\left(\frac{1}{\kappa_g}\right)^{\frac{1}{2}}M^2\,,
		  \quad\mathcal{R}^{\dCS}_2\equiv\frac{1}{2\kappa_g\zeta_{\dCS}}\,.
		\end{aligned}
    \end{align}
    \end{subequations}
    \end{widetext}
    The complete procedure of this projection and the projection of other components of $R_{\mu\nu}$ in dCS gravity can be found in \cite{dcstyped1}. Notice, following \cite{dcstyped1}, we have absorbed the coupling constant into the expansion of $\vartheta$ such that its expansion also follows Eq.~\eqref{eq:expansion_Weyl}, so we need to insert an $\zeta^{-1}$ into $\mathcal{R}_2^{\dCS}$ to compensate for this. Although Eq.~\eqref{eq:R_ab_dCS} is complicated, its value at $\mathcal{O}(\zeta^1,\chi^0,\epsilon^1)$ is simple since many Weyl scalars and spin coefficients vanish on the Schwarzschild background. Using
    \begin{align} \label{eq:NP_Schw}
	    & \Psi_{0,1,3,4}^{(0,0,0)}=0\,,\quad
	    \bar{\Psi}_2^{(0,0,0)}=\Psi_2^{(0,0,0)}\,, \nonumber\\
	    & \bar{\alpha}^{(0,0,0)}=\alpha^{(0,0,0)}
	    =-\beta^{(0,0,0)}\,,\quad
	    \bar{\rho}^{(0,0,0)}=\rho^{(0,0,0)}\,,\nonumber \\
	    & \bar{\mu}^{(0,0,0)}=\mu^{(0,0,0)}\,,\quad
	\bar{\gamma}^{(0,0,0)}=\gamma^{(0,0,0)}\,,
    \end{align}
    and other spin coefficients in Schwarzschild vanish, we find
    \begin{align} \label{eq:Phi_dCS}
	    & \Phi_{00,\dCS}^{(1,0,1)}
	    =\Phi_{02,\dCS}^{(1,0,1)}=0\,,\nonumber\\
	    & \Phi_{01,\dCS}^{(1,0,1)}
	    =-\frac{3i}{4}\mathcal{R}^{\dCS}_1\Psi_2(\{D,\delta\}
	    +\rho\delta)\vartheta^{(1,0,1)}\,,
    \end{align}
    where we have dropped the superscripts of terms at $\mathcal{O}(\zeta^0,\chi^0,\epsilon^0)$.
	
    Evaluating $S_{1,2}$ using Eqs.~\eqref{eq:source_bianchi} and \eqref{eq:Phi_dCS}, we find
    \begin{align}
	    & S_{1,\dCS}^{(1,0,1)}
        =\frac{3i}{2}\mathcal{R}^{\dCS}_1\Psi_2
        \left[\delta D^2+3\rho(\delta D+\rho\delta)
	    \right]\vartheta^{(1,0,1)}\,, \nonumber\\
	    & S_{2,\dCS}^{(1,0,1)}
        =-\frac{3i}{2}\mathcal{R}^{\dCS}_1\Psi_2\left[
        \delta^2D+2\alpha\delta D+\rho\delta^2
        +2\alpha\rho\delta\right]\vartheta^{(1,0,1)}\,,
    \end{align}
    where we have used NP equations to make simplifications. Then, inserting $S_{1,2}^{(1,0,1)}$ into the definition of $\mathcal{S}^{(1,1)}$ in Eq.~\eqref{eq:source_def_stress}, we find
    \begin{equation}
    \begin{aligned}
	    \mathcal{S}_{\dCS}^{(1,0,1)}
	    =& \;-3i\mathcal{R}^{\dCS}_1\Psi_2\left[
	    \delta^2D^2+2\alpha\delta D^2+2\rho\delta^2D \right.\\
        & \;\left.+4\alpha\rho\delta D+2\rho^2\delta^2
        +4\alpha\rho^2\delta\right]\vartheta^{(1,0,1)} \\
        \equiv& \;i\mathcal{Q}^{\dCS}\vartheta^{(1,0,1)}\,.
    \end{aligned}
    \end{equation}
    Using the transformation properties in Eq.~\eqref{eq:spin_coeff_00_10_parity}, one can easily show that $\hat{\mathcal{P}}\mathcal{Q}^{\dCS}=\mathcal{Q}^{\dCS}$. Following the definition in Eq.~\eqref{eq:source_dCS_2}, we can write 
    \begin{equation} \label{eq:F_dCS}
	    \mathcal{F}^{\dCS}=-24\pi^{-\frac{1}{2}}\mathcal{Q}^{\dCS}\square^{-1}
	    \left[\left(\frac{M}{r}\right)^3D^2\right]\,,
    \end{equation}
    where $D^2$ comes from converting Hertz potential $\bar{\Psi}_{\Hertz}^{(0,0,1)}$ to $\Psi_2^{(0,0,1)}$, and $\square^{-1}$ comes from inverting the equation of motion of $\vartheta^{(1,0,1)}$ in Eq.~\eqref{eq:EOM_theta_11_dCS}. One can easily check that $\hat{\mathcal{P}}\mathcal{F}^{\dCS}=\mathcal{F}^{\dCS}$, so non-rotating dCS BHs admit definite-parity modes as expected.
	
    \subsection{EdGB} \label{appendix:source_terms_EdGB}
	
    For EdGB, as discussed in Sec.~\ref{sec:EdGB_S}, we choose to focus on the terms in $\mathcal{S}^{(1,0,1)}$ proportional to $\varphi^{(1,0,1)}$ or its derivatives, so all the metric fields are stationary. To compute the terms in $\mathcal{S}^{(1,0,1)}$ driven by GW perturbations in GR, one can follow similar procedures in \cite{dcstyped1}. Following the same argument in Appendix~\ref{appendix:source_terms_dCS}, one only needs to evaluate $\Phi_{00}$, $\Phi_{01}$, and $\Phi_{02}$, or alternatively $R_{11}$, $R_{13}$, and $R_{33}$ for this contribution. Projecting Eq.~\eqref{eq:tracericci-sgb} onto the NP tetrad, we find
    \begin{widetext}
    \begin{subequations}
    \begin{align}
        & \begin{aligned} \label{eq:R_11_EdGB}
            R_{11}^{\EdGB}
            =& \;\frac{1}{2}\mathcal{R}_1^{\EdGB}\bigg\{
            -(D\varphi)\Big[\lambda\Psi_0+\bar{\lambda}\bar{\Psi}_0
            -(\alpha+\bar{\beta}+\pi)\Psi_1
            -(\bar{\alpha}+\beta+\bar{\pi})\bar{\Psi}_1
            +(\varepsilon+\bar{\varepsilon})(\Psi_2+\bar{\Psi}_2)\Big]\\
            & \;+(\boldsymbol{\Delta}\varphi)
            \Big[\bar{\sigma}\Psi_0+\sigma\bar{\Psi}_0
            -\bar{\kappa}\Psi_1-\kappa\bar{\Psi}_1\Big] \\
            & \;+(\delta\varphi)\Big[(\bar{\alpha}-\beta)\bar{\Psi}_0
            +(\varepsilon-\bar{\varepsilon}-\bar{\rho})\bar{\Psi}_1
            -\bar{\sigma}\Psi_1+\bar{\kappa}(\Psi_2+\bar{\Psi}_2)\Big] \\
            & \;+(\bar{\delta}\varphi)\Big[(\alpha-\bar{\beta})\Psi_0
            -\sigma\bar{\Psi}_1-(\varepsilon-\bar{\varepsilon}+\rho)\Psi_1
            +\kappa(\Psi_2+\bar{\Psi}_2)\Big] \\
            & \;+\frac{1}{2}\Psi_0\{\bar{\delta},\bar{\delta}\}\varphi
            +\frac{1}{2}\bar{\Psi}_0\{\delta,\delta\}\varphi
            -\Psi_1\{D,\bar{\delta}\}\varphi-\bar{\Psi}_1\{D,\delta\}\varphi
            +\frac{1}{2}(\Psi_2+\bar{\Psi}_2)\{D,D\}\varphi\bigg\}
            +\mathcal{R}^{\EdGB}_2(D\varphi)(D\varphi)\,,
	    \end{aligned} \\
	    & \begin{aligned} \label{eq:R_13_EdGB}
            R_{13}^{\EdGB}
            =& \;\frac{1}{4}\mathcal{R}_1^{\EdGB}\bigg\{
            -(D\varphi)\Big[\nu\Psi_0
            -(\gamma+\bar{\gamma}+\mu+\bar{\mu})\Psi_1+2\bar{\lambda}\bar{\Psi}_1
            +(\bar{\alpha}+\beta+\bar{\pi})(\Psi_2-2\bar{\Psi}_2)
            +2(\varepsilon+\bar{\varepsilon})\bar{\Psi}_3\Big] \\
            & \;+(\boldsymbol{\Delta}\varphi)
            \Big[(\alpha+\bar{\beta}+\bar{\tau})\Psi_0
            -(\varepsilon+\bar{\varepsilon}+\rho+\bar{\rho})\Psi_1
            +2\sigma\bar{\Psi}_1+\kappa(\Psi_2-2\bar{\Psi}_2)\Big] \\
            & \;-(\delta\varphi)\Big[\lambda\Psi_0
            -(\alpha-\bar{\beta}+\pi-\bar{\tau})\Psi_1
            -2(\bar{\alpha}-\beta)\bar{\Psi}_1
            +(\varepsilon-\bar{\varepsilon}-\bar{\rho})(\Psi_2-2\bar{\Psi}_2)
            -2\bar{\kappa}\bar{\Psi}_3\Big] \\
            & \;+(\bar{\delta}\varphi)\Big[(\gamma-\bar{\gamma}-\bar{\mu})\Psi_0
            +(\bar{\alpha}-\beta+\bar{\pi}-\tau)\Psi_1
            +\sigma(\Psi_2-2\bar{\Psi}_2)+2\kappa\bar{\Psi}_3\Big] \\
            & \;+\Psi_0\{\boldsymbol{\Delta},\bar{\delta}\}\varphi
            -\Psi_1\left[\{D,\boldsymbol{\Delta}\}
            +\{\delta,\bar{\delta}\}\right]\varphi
            +\bar{\Psi}_1\{\delta,\delta\}\varphi
            +(\Psi_2-2\bar{\Psi}_2)\{D,\delta\}\varphi
            +\bar{\Psi}_3\{D,D\}\varphi\bigg\}
            +\mathcal{R}^{\EdGB}_{2}(D\varphi)(\delta\varphi)\,,
	    \end{aligned} \\
	    & \begin{aligned} \label{eq:R_33_EdGB}
            R_{33}^{\EdGB}
            =& \;\frac{1}{2}\mathcal{R}_1^{\EdGB}\bigg\{
            (D\varphi)\Big[\bar{\nu}\Psi_1-\bar{\lambda}(\Psi_2+\bar{\Psi}_2)
            +(\bar{\alpha}+\beta+\bar{\pi})\bar{\Psi}_3
            -(\varepsilon+\bar{\varepsilon})\bar{\Psi}_4\Big] \\
            & \;+(\boldsymbol{\Delta}\varphi)\Big[
            (\gamma+\bar{\gamma})\Psi_0-(\bar{\alpha}+\beta+\tau)\Psi_1
            +\sigma(\Psi_2+\bar{\Psi}_2)-\kappa\bar{\Psi}_3\Big] \\
            & \;-(\delta\varphi)\Big[\nu\Psi_0-(\gamma-\bar{\gamma}+\mu)\Psi_1
            -(\bar{\alpha}-\beta)(\Psi_2+\bar{\Psi}_2)
            -(\varepsilon-\bar{\varepsilon}-\bar{\rho})\bar{\Psi}_3
            -\bar{\kappa}\bar{\Psi}_4\Big] \\
            & \;-(\bar{\delta}\varphi)\Big[\bar{\nu}\Psi_0
            -\bar{\lambda}\Psi_1+\sigma\bar{\Psi}_3-\kappa\bar{\Psi}_4\Big] \\
            & \;+\frac{1}{2}\Psi_0\{\boldsymbol{\Delta},\boldsymbol{\Delta}\}\varphi
            -\Psi_1\{\boldsymbol{\Delta},\delta\}\varphi
            +\frac{1}{2}(\Psi_2+\bar{\Psi_2})\{\delta,\delta\}\varphi
            -\bar{\Psi}_3\{D,\delta\}\varphi
            +\frac{1}{2}\bar{\Psi}_4\{D,D\}\varphi\bigg\}
            +\mathcal{R}^{\EdGB}_{2}(\delta\varphi)(\delta\varphi)\,,
	    \end{aligned} \\
	    & \mathcal{R}_1^{\EdGB}=-\kappa_g^{\frac{1}{2}}M^2\,,\quad
	    \mathcal{R}_2^{\EdGB}=\frac{1}{2\zeta_{\EdGB}}\,,
    \end{align}
    \end{subequations}
    \end{widetext}
    where one can refer to \cite{dcstyped1} for more details of this projection in dCS gravity. Similarly, we have absorbed one coupling constant into the expansion of $\varphi$ to be consistent with Eq.~\eqref{eq:expansion_Weyl}, so $\mathcal{R}_2^{\EdGB}$ contains an extra factor of $\zeta^{-1}$. Using Eq.~\eqref{eq:NP_Schw}, one can find that the $\mathcal{O}(\zeta^1,\chi^0,\epsilon^1)$ contributions to $\Phi_{ij}$ with dynamical $\varphi$ are
    \begin{widetext}
    \begin{align} \label{eq:Phi_EdGB}
	    & \Phi_{00,\EdGB}^{(1,0,1)}
	    =\frac{1}{2}\mathcal{R}_1^{\EdGB}\Psi_2D^2\varphi^{(1,0,1)}
	    +\mathcal{R}_2^{\EdGB}D\varphi^{(1,0,0)}D\varphi^{(1,0,1)}\,, \nonumber\\
	    & \Phi_{01,\EdGB}^{(1,0,1)}
	    =-\frac{1}{8}\mathcal{R}_1^{\EdGB}
	    \Psi_2\left(\{D,\delta\}+\rho\delta\right)\varphi^{(1,0,1)}
	    +\frac{1}{2}\mathcal{R}_2^{\EdGB}D\varphi^{(1,0,0)}
        \delta\varphi^{(1,0,1)}\,, \nonumber\\
	    & \Phi_{01,\EdGB}^{(1,0,1)}
	    =\frac{1}{2}\mathcal{R}_1^{\EdGB}\Psi_2
	    \left(\delta^2+2\alpha\delta\right)\varphi^{(1,0,1)}\,, 
    \end{align}   
    \end{widetext}
    where we have also used that $\delta^{(0,0,0)}\varphi^{(1,0,0)}=\bar{\delta}^{(0,0,0)}\varphi^{(1,0,0)}=0$ since $\varphi^{(1,0,0)}$ is a pure radial function \cite{Ayzenberg:2016ynm}. For simplicity, we have also dropped the superscripts of terms at $\mathcal{O}(\zeta^0,\chi^0,\epsilon^0)$. Using Eqs.~\eqref{eq:source_bianchi} and \eqref{eq:Phi_EdGB}, we find
    \begin{widetext}
    \begin{align}
	    S_{1,\EdGB}^{(1,0,1)}
	    =& \;\frac{3}{4}\mathcal{R}_1^{\EdGB}\Psi_2\left[\delta D^2
	    +\rho\left(\delta D+\rho\delta\right)\right]\varphi^{(1,0,1)}
	    +\frac{1}{2}\mathcal{R}_2^{\EdGB}
	    \left[D\varphi^{(1,0,0)}\delta D
	    -(D^2-\rho D)\varphi^{(1,0,0)}\delta\right]\varphi^{(1,0,1)}\,, \\
	    S_{2,\EdGB}^{(1,0,1)}
	    =& \;-\frac{3}{4}\mathcal{R}_1^{\EdGB}\Psi_2
        \left(\delta^2D+2\alpha\delta D
        +3\rho\delta^2+6\alpha\rho\delta\right)\varphi^{(1,0,1)}
        +\frac{1}{2}\mathcal{R}_2^{\EdGB}D\varphi^{(1,0,0)}
	    (\delta^2+2\alpha\delta)\varphi^{(1,0,1)}\,,
    \end{align}
    and using Eq.~\eqref{eq:source_def_stress}, we find
    \begin{equation}
    \begin{aligned}
        \mathcal{S}_{\EdGB}^{(1,0,1)}
        =& \;-\frac{3}{2}\mathcal{R}_1^{\EdGB}\Psi_2
        \left(\delta^2D^2+2\alpha\delta D^2+2\rho\delta^2D
        +4\alpha\rho\delta D+2\rho^2\delta^2
        +4\alpha\rho^2\delta\right)\varphi^{(1,0,1)} \\
        & \;+\mathcal{R}_2^{\EdGB}\left\{
        \left(D^2-2\rho D\right)\varphi^{(1,0,0)}\delta^2
        +\left[2\alpha\left(D^2-2\rho D\right)
        +\frac{1}{2}\delta D^2\right]\varphi^{(1,0,0)}
        \delta\right\}\varphi^{(1,0,1)} \\
        =& \;\mathcal{Q}^{\EdGB}\varphi^{(1,0,1)}\,.
    \end{aligned}
    \end{equation} 
    \end{widetext}
    Using the transformation properties in Eq.~\eqref{eq:spin_coeff_00_10_parity} and that $\varphi^{(1,0,0}$ is purely radial, we find that $\hat{\mathcal{P}}\mathcal{Q}^{\EdGB}=\mathcal{Q}^{\EdGB}$. Follow the definition in Eq.~\eqref{eq:source_EdGB}, we can write 
    \begin{equation}
	    \mathcal{F}^{\EdGB}
        =\mathcal{Q}^{\EdGB}\square^{-1}\,,
    \end{equation}
    where $\square^{-1}$ comes from inverting the equation of motion of $\varphi^{(1,0,1)}$ in Eq.~\eqref{eq:EdGB_EOM_scalar_11}. One can check that $\hat{\mathcal{P}}\mathcal{F}^{\EdGB}=\mathcal{F}^{\EdGB}$, so non-rotating EdGB BHs also admit definite-parity modes.

    \bibliographystyle{apsrev4-1}
    \bibliography{reference}
	
\end{document}